%% file: paper.tex
\documentclass[acmtog,nonacm]{acmart}

\acmSubmissionID{579}

\input{commands}

\input{acronyms}

\input{math}
\myexternaldocument{supplemental}
\begin{document}

\title{\name: Editing Point Patterns by Image Manipulation}

\author{Xingchang Huang}
\affiliation{%
	\institution{Max-Planck-Institut für Informatik}
    \country{Germany}
}
\email{xhuang@mpi-inf.mpg.de}

\author{Tobias Ritschel}
\affiliation{%
	\institution{University College London}
    \country{United Kingdom}
}
\email{t.ritschel@ucl.ac.uk}

\author{Hans-Peter Seidel}
\affiliation{%
	\institution{Max-Planck-Institut für Informatik}
    \country{Germany}
}
\email{hpseidel@mpi-sb.mpg.de}

\author{Pooran Memari}
\affiliation{%
	\institution{CNRS, LIX, \'Ecole Polytechnique, IP Paris, INRIA}
    \country{France}
}
\email{memari@lix.polytechnique.fr}

\author{Gurprit Singh}
\affiliation{%
	\institution{Max-Planck-Institut für Informatik}
    \country{Germany}
}
\email{gsingh@mpi-inf.mpg.de}

\begin{teaserfigure}
  \input{images/teaser}
  \vspace{-.4cm}
  \caption{
  Our framework facilitate point pattern design by representing both density and correlation as a three-channel raster image (a). 
  These images can be edited (c) in terms of their density or correlation using off-the-shelf image manipulation software.
  The resulting point patterns are shown before (b) and after the edits (d). Please see the accompanied supplemental material for vector graphic images.
}
  \label{fig:Teaser}
\end{teaserfigure}
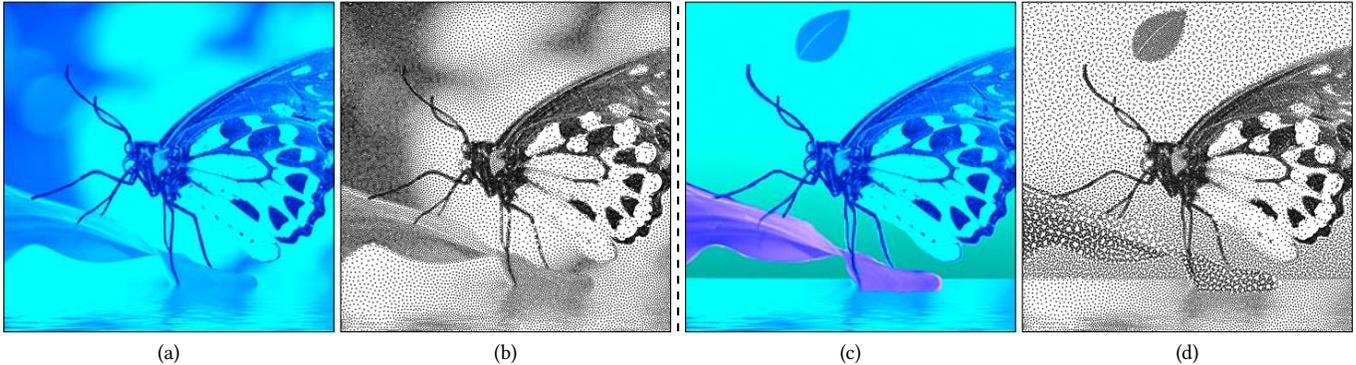

\begin{abstract}
Point patterns are characterized by their density and correlation. 
While spatial variation of density is well-understood, analysis and synthesis of spatially-varying correlation is an open challenge.
No tools are available to intuitively edit such point patterns, primarily due to the lack of a compact representation for spatially varying correlation.
We propose a low-dimensional perceptual embedding for point correlations. 
This embedding can map point patterns to common three-channel raster images, enabling manipulation with off-the-shelf image editing software.
To synthesize back point patterns, we propose a novel edge-aware objective that carefully handles sharp variations in density and correlation.
The resulting framework allows intuitive and backward-compatible manipulation of point patterns, such as recoloring, relighting to even texture synthesis that have not been available to 2D point pattern design before.
Effectiveness of our approach is tested in several user experiments.
Code is available at \href{https://github.com/xchhuang/patternshop}{https://github.com/xchhuang/patternshop}.
\end{abstract}

\ccsdesc[500]{Computing methodologies~Point-based Models; Non-photorealistic rendering; Image manipulation; Neural networks}

\keywords{Blue noise; Point pattern editing and synthesis; Image stippling}

\acmJournal{TOG}
\acmYear{2023} \acmVolume{42} \acmNumber{4} \acmArticle{1} \acmMonth{8} \acmPrice{}\acmDOI{10.1145/3592418}

\maketitle

\acresetall

\section{Introduction}
\label{sec:introduction}

Point patterns are characterized by their underlying density and correlations.
Both properties can vary over space (\refFig{Teaser}), but two key challenges limit the use of such patterns: first, a reliable representation and, second, the tools to manipulate this representation.

While algorithms \cite{zhou2012point,oztireli2012analysis} are proposed in the literature to generate point patterns with specific correlations (\eg blue-, green-, pink-, etc. noise), designing specific correlation requires understanding of the power spectrum or \ac{PCF}~\cite{heck2013blue} only a handful of expert users might have. 
Further, the space spanned by these \emph{colored} noises (correlations) is also limited to a handful of noises studied in the literature~\cite{zhou2012point,oztireli2012analysis}.
Addressing this, we embed point correlations in a 2D space in a perceptually optimal way.
In that space, two point correlations have similar 2D coordinates if they are perceptually similar.
We simply sample {all realizable point correlations} and perform \ac{MDS} on a perceptual metric, without the need for user experiments, defined on the pairs of all samples.
Picking a 2D point in that space simply selects a point correlation.
Moving a point changes the correlation with perceptual uniformity.

The next difficulty to adopt point patterns of spatially varying density and correlations is the lack of tools for their intuitive manipulation.
Modern creative software gives unprecedented artistic freedom to change images by relighting \cite{pellacini2010envylight}, painting~\cite{strassmann1986hairy,hertzmann2001image},  material manipulation \cite{pellacini2007appwand,renzo2014appim}, changing canvas shape~\cite{rottshaham2019singan}, completing missing parts~\cite{bertalmio2000inpainting}, as well as transferring style ~\cite{gatys2016styletransfer} or textures~\cite{efros2001image,zhou2018nonstationary,sendik2017deep}.
Unfortunately, no such convenience exists when aiming to manipulate point patterns as previous approaches are specifically designed to process three-channel (RGB) raster images.
Our core idea is to convert point patterns into off-the-shelf three-channel raster image.
We use the \cielab color space to encode density as one-dimensional luminance (L) and two-dimensional chroma (AB) as correlation.
The resulting images can the be manipulated harnessing all the power of  typical raster image editing software.
 
While spatially-varying density is naturally mapped to single-channel images, this is more difficult for correlation.
Hence, it is often assumed spatially-invariant and represented by either the power spectrum~\cite{ulichney1987digital,lagae2008} or the \ac{PCF}~\cite{wei2011differential,oztireli2012analysis}. 
Some work also handles spatially-varying density or/and correlation \cite{chen2013bilateral,roveri2017general}, but with difficulties in handling sharp variation of density and/correlation.
We revisit bilateral filtering to handle sharp variation both in density and correlation.

Finally, we show how to generate detailed spatial \cielab maps of correlation and density given an input point pattern using a learning-based system, trained  by density and correlation supervision on a specific class of images, \eg human faces.

In summary, we make the following contributions:
\begin{itemize}[leftmargin=3.4mm]
\setlength\itemsep{0.5mm}
\item Two-dimensional perceptual embedding of point correlations,
\item Spatially-varying representation of point pattern density and perceptually-embedded correlation as LAB raster images,
\item A novel definition of edge-aware correlation applicable to hard edges in density and/or correlation,
\item An optimization-based system to synthesize point patterns defined by density and embedded correlation maps according to said edge-aware definition,
\item Intuitive editing of point pattern correlation and density by recoloring, painting, relighting, etc. of density and embedded correlation maps in legacy software such as Adobe Photoshop.
\item A learning-based system to predicts density and embedded correlation maps from an input point pattern.
\end{itemize}

\section{Previous Work}
\label{sec:previous_work}

\paragraph{Sample correlations}
Correlations among stochastic samples are widely studied in computer graphics. 
From halftoning \cite{ulichney1987digital}, reconstruction \cite{yellott1983spectral}, anti-aliasing \cite{cook1986stochastic,dippe1985antialiasing} to Monte Carlo integration~\cite{durand2011frequency,singh19analysis}, correlated samples have made a noticeable impact.
Recent works~\cite{xu2020ladybird,zhang2019active} have also shown direct impact of correlated samples on training accuracy in machine learning.
Among different sample correlations, \emph{blue} noise~\cite{ulichney1987digital} is the most prominent in literature. 
Blue noise enforces point separation and is classically used for object placement~\cite{kopf2006recursive,reinert2013interactive} 
 and point stippling~\cite{deussen2000floating,secord2002weighted,schulz2021multiclass}.  
However, modern approaches do not insist on point separation in faithful stippling~\cite{martin2017survey,kim2009stippling,deussen2013halftoning,paul2012image}.
Different kind of \emph{colored} noises (\eg green, red) are studied in literature~\cite{lau99digital,zhou2012point} for half-toning and stippling purposes. 
But the space spanned by these point correlations is limited to a few bases~\cite{oztireli2012analysis}.
We propose an extended space of point correlations that helps express large variety of correlations. 
Our framework embeds these correlations in a two-dimensional space, allowing representation of different correlations with simple two-channel color maps. 
This makes analysis and manipulation of correlations easier by using off-the-shelf image editing software.

\paragraph{Analysis}
To analyze different sample correlations, various spectral~\cite{ulichney1987digital,lagae2008} and spatial~\cite{wei2011differential,oztireli2012analysis} tools are developed. 
For spectral methods, the Fourier power spectrum and it's radially averaged version are used to analyze sample characteristics. 
In the spatial domain, \ac{PCF} is used for analysis which evaluates pairwise distances between samples that are then binned in a 1D or a 2D histogram. 

\paragraph{Synthesizing blue-noise correlation}
Blue noise sample correlation is most commonly used in graphics.
There exist various algorithms to generate blue noise sample distributions (see~\citet{yan2015survey}). 
Several optimization-based methods \cite{lloyd1982least,balzer2009capacity,liu2009centroidal, schmaltz2010electrostatic,fattal2011blue,degoes2012blue, heck2013blue,kailkhura2016stair,qin2017wasserstein}, as well as tiling-based \cite{ostromoukhov2004fast,kopf2006recursive,wachtel2014fast} and number-theoretic approaches \cite{ahmed2015aa,ahmed2016low,ahmed2017simple} are developed over the past decades to generate blue noise samples.

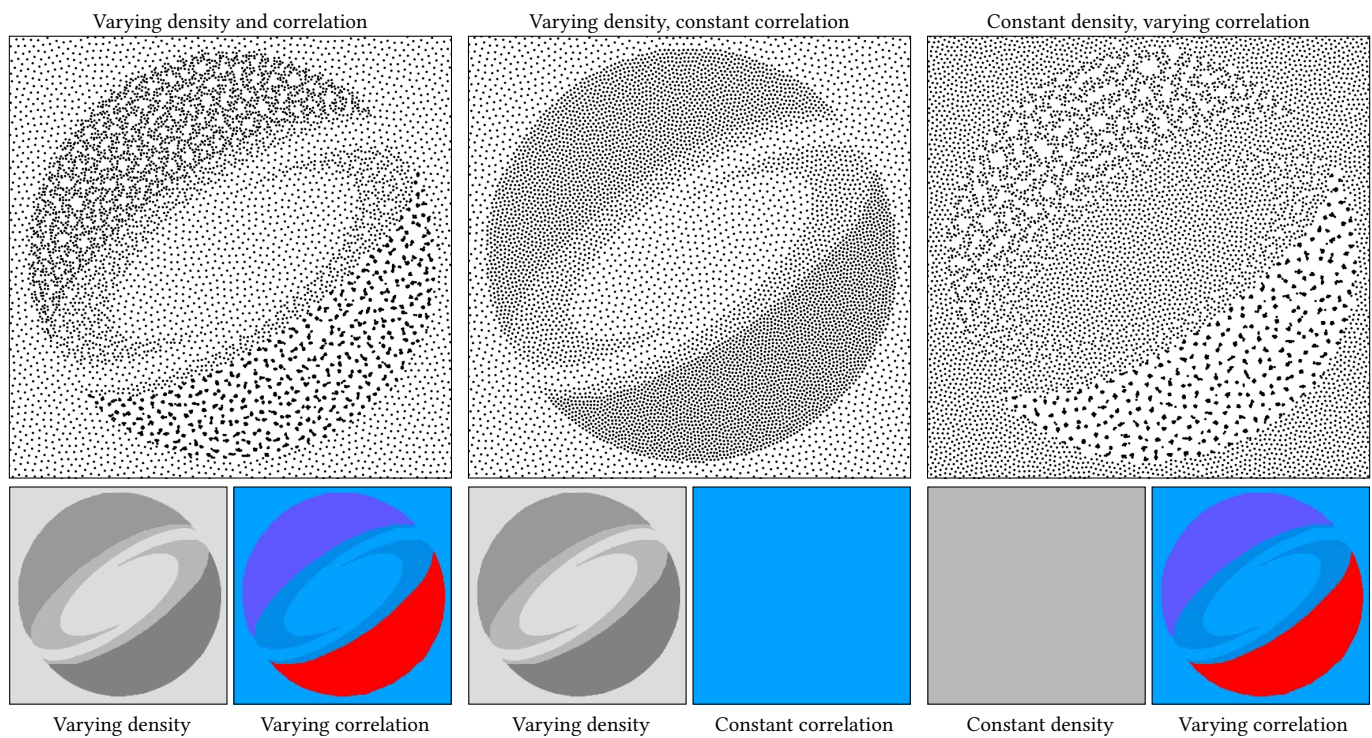
\begin{figure*}[t!]
    \centering
    \input{images/siggraph_logo/fig_siggraph_logo_ours}
    \caption{Our framework allows manipulating density and correlations independently. 
    Top row shows the point set synthesized using the density (left) and the correlation map (right) shown in the bottom row.
    The first column shows a point set when both density and correlation are spatially varying. 
    Second column has constant correlation but spatially varying density. 
    Third column point set reproduces the Siggraph logo just from spatially varying correlations.}
    \label{fig:siggraph_logo}
\end{figure*}

\paragraph{Synthesizing other correlations}
There exist methods to generate samples with different target correlations. 
For example, \citet{zhou2012point} and \citet{oztireli2012analysis} proposed to synthesize point correlations defined from a user-defined target \ac{PCF}. 
\citet{wachtel2014fast} proposed a tile-based optimization approach that can generate points with a user-defined target power spectrum.
All these methods require heavy machinery to ensure the optimization follows the target.
\citet{leimkuhler2019deep}  simplified this idea  and proposed a neural network-based optimization pipeline. 
All these approaches, however, require the user to know how to design a \emph{realizable} \ac{PCF} or a power spectrum~\cite{heck2013blue}. 
This can severely limit the usability of point correlations to only handful of expert users.  
Our framework lifts this limitation and allows us to simply use a two-dimensional space to define correlations. 
Once a user picks a 2D point ---which we visualize as color of different chroma--- our framework automatically finds the corresponding \ac{PCF} from the embedded space and synthesizes the respective  point correlations.
It is also straightforward to generate spatially varying correlations using our framework by defining a range of colors as correlations. 
So far, only~\citet{roveri2017general} have synthesized spatially varying correlations but remain limited by how well a user can design \acp{PCF}.

\paragraph{Image and point pattern editing}
Editing software allows artists to tailor the digital content to their needs.
Since \emph{color image} is the easily available data representation, today's software are specifically designed to process three-channel (RGB) images.
Modern pipelines allow effects like relighting~\cite{sun2019single}, recoloring, image retargeting~\cite{rottshaham2019singan}, inpainting~\cite{bertalmio2000inpainting}, style transfer~\cite{gatys2016styletransfer} and texture synthesis~\cite{efros2001image,zhou2018nonstationary,sendik2017deep} to be performed directly in the three-channel RGB representation. 
However, most digital assets \eg materials, color pigments, light fields, patterns, sample correlations, are best captured in high-dimensional space.
This makes it  hard for image-based software to facilitate editing these components. 
A lot of research has been devoted in the past to support editing materials~\cite{pellacini2007appwand,an2011appwarp,renzo2014appim}, light fields~\cite{jarabo2014how}, color pigments~\cite{sochorova2021practical}, natural illumination~\cite{pellacini2010envylight} with workflows similar to images.

Synthesizing textures with elements~\cite{ma2011discrete,reinert2013interactive,emilien2015worldbrush,reddy2020discovering,hsu2020autocomplete} and patterns with different variations~\cite{guerrero2016patex} using 2D graphical primitives has also been proposed. 
These work focus on updating point and element locations to create patterns for user-specific goals and designing graphical interface for user interactions.
However, none of the previous work allows editing spatially-varying point correlations in an intuitive manner. 
In this work, we propose a pipeline to facilitate correlation editing using simple image operations. 
Instead of directly working with points, we encode their corresponding spectral or spatial statistics in a low-dimensional (3-channel) embedding. 
This low-dimensional latent space can then be represented as an image 
in order to allow artists to manipulate point pattern designs using any off-the-shelf image editing software. 
There exists previous work that encode point correlations~\cite{leimkuhler2019deep} for a single target or point pattern structures~\cite{tu2019pointpattern} using a neural network. But these representations do not disentangle the underlying density and correlation, thereby, not facilitating editing operations. 

\paragraph{Latent and perceptual spaces}
Reducing high-dimensional signals into low-dimensional ones has several benefits.
Different from latent spaces in generative models which are still high-dimensional, such as for StyleGAN \cite{abdal2019image2stylegan}, the application we are interested in here is a usability one, where the target space is as low as one-, two- or three- dimensional, so users can actively explore it, \eg on a display when searching \cite{duncan1989visual}.
This is common to do for color itself \cite{fairchild2013color,nguyen2015data}.
Ideally, the embedding into a lower dimension is done such, that distance in that space is proportional to perceived distances \cite{lindow2012perceptually}.
This was pioneered by \citet{pellacini2000toward} for BRDF, with a methodology close to ours (\ac{MDS}).
Other modalities, such as acoustics \cite{pols1969perceptual}, materials \cite{wills2009toward}, textures \cite{henzler2020neuraltexture} and even fabricated objects \cite{piovarci2018modelling} were successfully organized in latent spaces.

\section{Overview}
\label{sec:overview}
\begin{figure}[htb]
    \centering
    \includegraphics[width=\linewidth]{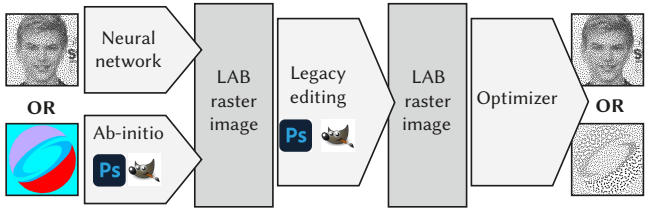}
    \caption{The workflow of \name. \revision{A user can select a point pattern and run it through a \ac{NN} to produce a three-channel raster image, or draw that image directly, ab-initio in a program like Adobe Photoshop or GIMP.
    The edited pattern can be synthesized using our optimization.}}
    \label{fig:workflow}
\end{figure}%
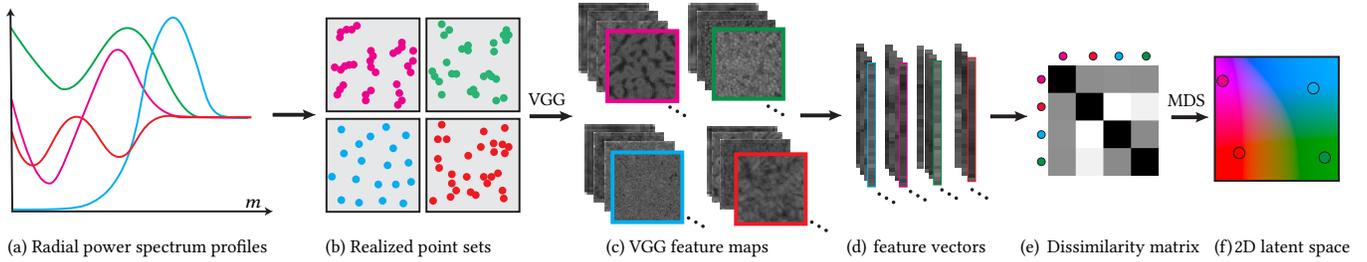
\begin{figure*}[htb]
\input{images/our_approach/embedding_overview_powspectra}
\caption{
\label{fig:embedding_overview}
We illustrate the embedding from \pcfSampleCount-dimensional space to a 2D space. 
(a) Random power spectra are generated and their corresponding realizable point patterns in (b) are synthesized using gradient descent. 
The point patterns are then rasterized and pass through a pre-trained VGG network~\cite{simonyan2014vgg19} to generate (c) feature maps, which are flattened into feature vectors (d).
A dissimilarity matrix (e) computed from these VGG feature vectors are used to bring \pcfSampleCount-dimensional representation to a 2D space (f) using \ac{MDS} (\cref{sec:embedding}).
}
\end{figure*}
\cref{fig:workflow} shows the workflow of \name:
A user selects a point pattern with spatially varying density and correlation and runs this through a \ac{NN} to produce a three-channel raster image.
{Alternatively, they can draw that image directly, ab-initio in a program like Adobe Photoshop or GIMP.}
This raster image can then be edited in any legacy software.
Density can be manipulated by editing luminance.
To change correlation, we provide a perceptual 2D space of point correlations (\cref{fig:mds_space}).
This space is perceptually uniform, \ie distances are proportional to human perception and {covers a wide range}, \ie it contains most realizable points correlations.
\Cref{fig:siggraph_logo} demonstrates one such example edit.
A user may edit 
spatially varying density (\cref{fig:siggraph_logo}b) 
or correlation (\cref{fig:siggraph_logo}c) 
or both (\cref{fig:siggraph_logo}a)
using our three-channel representation.
A final optimizer produces the edited point pattern.

Correlations are generally characterized by a \ac{PCF} or power spectrum. 
A \ac{PCF} encodes spatial statistics---\eg the pairwise distances between points---distributed over an \pcfSampleCount-bin histogram.
Handling such an \pcfSampleCount-dimensional correlation space is neither intuitive nor supported by existing creative software.
To tackle this problem, in~\cref{sec:embedding}, we embed \acf{PCF} into a two-dimensional space.
We optimize the distance between all pairs of latent coordinates in that space to be proportional to the perceived visual distance between their point pattern realizations.

Moreover, synthesizing such patterns requires support for edge-aware transitions for both density and correlation.
In~\cref{sec:density_aware_pcfs}, we outline our formalism to handle such edge-aware density and correlations.
\Cref{sec:synthesis_and_editing} defines the optimization objective, that enables finding a point pattern to have the desired density and correlation.
\Cref{subsec:implementation} provides details on the implementation.
\Cref{sec:results} outlines various application scenarios supported by our pipeline, followed by comparative analyses, concluding discussions and future works.

\section{Latent embedding of point correlations}
\label{sec:embedding}

\paragraph{Overview}
In order to encode the correlation of a point pattern as a 2D coordinate, we first generate a corpus of basic point patterns (each with constant correlation).
Next, we extract perceptual features of these patterns.
The pairwise distance between all features forms a perceptual metric.
This metric is used to embed a discrete set of basis correlations into a 2D space.
From this discrete mapping, we can further define a continuous mapping from any continuous correlations into a latent space as well as back from any continuous latent space coordinate to correlations.
This process is a pre-computation, only executed once, and its result can be used to design many point patterns, same as one color space can be used to work with many images.
We detail each step in the following:

\paragraph{Data generation} First, we systematically find a variety of point patterns to span a gamut of correlations $\{\pcf_i\}$.
We start by defining a power spectrum as a Gaussian mixture of either (randomly) one or two modes with means randomly uniform across the domain and uniform random variance of up to a quarter of the unit domain (\cref{fig:embedding_overview}a).
Not all such power spectra are realizable, therefore, we run a gradient descent optimization \cite{leimkuhler2019deep} to obtain realizable point patterns (\cref{fig:embedding_overview}b).
We finally use the \ac{PCF} of that realization $\pointPattern_i$ as $\pcf_i$.
Please see Supplemental Sec. 1.1 for details.

\paragraph{Metric}
A perceptual metric $\perceptualDistance_{i,j}$ assigns a positive scalar to every pair of stimuli  $i$ and $j$ (here: {point patterns}) which is low only if the pair is perceived similar.
As point patterns are stationary textures, their perception is characterized by the spatially aggregated statistics of their visual features \cite{portilla2000parametric}.
Hence, the perceived distance between any two point patterns with visual features $\visualFeatureStats_i$ and $\visualFeatureStats_j$ is their $\mathcal L_1$ distance $\perceptualDistance_{i,j}=|\visualFeatureStats_i-\visualFeatureStats_i|_1$.

To compute visual feature statistics $\visualFeatureStats_i$ for one pattern $\pointPattern_i$, we 
first rasterize $\pointPattern_i$ three times with point splats of varying Gaussian splat size of 0.015, 0.01 and 0.005 \cite{tu2019pointpattern}.
Second, each image of that triplet is converted into VGG feature activation maps \cite{simoncelli2001natural}.
Next, we compute the Gram matrices \cite{gatys2016styletransfer} for the \texttt{pool\_2} and \texttt{pool\_4} layers in each of the three images.
Finally, we stack the triplet of pairs of Gram matrices into a vector $\visualFeatureStats_i$ of size $3\times(64^2+128^2)=61440$.

\Cref{fig:embedding_overview} shows four example patterns leading to a $4\times4$ dissimilarity matrix. 

\paragraph{Dimensionality reduction}
To first map the basis set of $\{\pointPattern_i\}$ to latent 2D codes, we apply \ac{MDS} (\cref{fig:embedding_overview}f).
\ac{MDS} assigns every pattern $\pointPattern_i$ a latent coordinate $\latentCoordinate_i$ so that the joint stress of all assignments 
\begin{align}
\label{eq:mds}
\embeddingCost(\{\latentCoordinate_i\}) &= 
\sum_{i,j}
(\perceptualDistance_{i,j} - ||\latentCoordinate_i - \latentCoordinate_j||)^2
\end{align}
is minimized across the set $\{\latentCoordinate_i\}$ of latent codes.

After optimizing with \cref{eq:mds}, we rotate the latent coordinates, so that blue noise is indeed distributed around a bluish color of the chroma plane and then normalize the 2D coordinates to $[0, 1]^2$.

\paragraph{Encoding}
Once the latent space is constructed, we can sample correlation \pcf at any continuous coordinate \latentCoordinate in the latent space using the \ac{IDW} method:
\begin{align}
\label{eq:idw}
\encoding(\latentCoordinate) &= 
\sum_i
\pcf_i
w_i(\latentCoordinate)
/
\sum_i
w_i(\latentCoordinate)
\quad\text{with}
\\
w_i(\latentCoordinate) &= 1/\max((||\latentCoordinate-\latentCoordinate_i||_2)^{\locality(\latentCoordinate)}, 10^{-10})
.
\end{align}
The idea is to first compute the distance between the current location \latentCoordinate and the existing locations $\latentCoordinate_i$ in the \ac{MDS} space. 
The inverse of these distances, raised to a power $\locality(\latentCoordinate)$, is used as the weight to interpolate the correlations $\pcf_i$.
$\locality\in\mathbb R^2\mapsto\mathbb R$ is a function that is low in areas of the latent space with a low density of examples and high for areas where \ac{MDS} has embedded many exemplars.
We implement \locality by kernel density estimation on the distribution $\{\latentCoordinate_i\}$ itself with a Parzen window of size 0.01 and clamping and scaling this density linearly to the $(3, 6)$-range.

\paragraph{Decoding}
We can now also embed any correlation \pcfTarget, that is not in the discrete basis set $\{\pcf_i\}$.
Let \targetVisualFeatureStats be the  visual features stats of \pcfTarget.
We simply look up the visual-feature-nearest training exemplar and return its latent coordinate
\begin{equation}
\label{eq:decoding}
\decoding(\pcf)=
\argmin_{\latentCoordinate_i}
||
\targetVisualFeatureStats
-
\visualFeatureStats_i
||_2
.
\end{equation}

\begin{figure*}[t!]
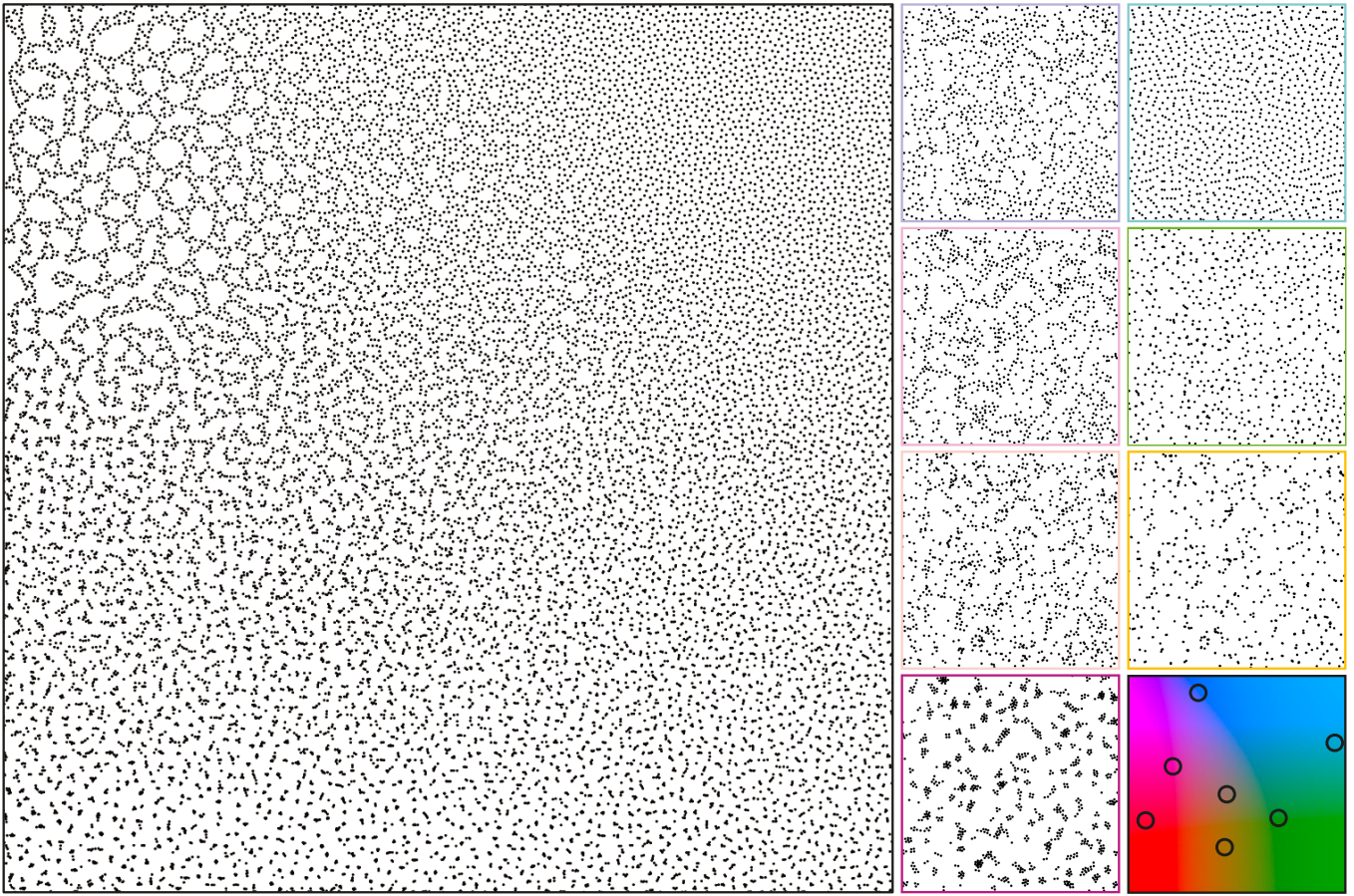

\begin{overpic}[scale=1.09,grid=false]{images/mds_space}
\end{overpic}
\caption{
\label{fig:mds_space}
The left image shows the resynthesized point pattern with a continuously varying correlations in the 2D latent space $\encoding(\latentCoordinate)$. 
The bottom-right (black) square shows  different coordinates and their color in \cielab. 
Point patterns with the corresponding spatially-invariant correlations for each of these coordinates are shown with respective color coding in the right halve. 
Blue noise can be spotted on the top-right side of the space.
}
\end{figure*}%
\paragraph{Latent space visualizations}
The latent space $\encoding(\latentCoordinate)$ of correlations is visualized in~\cref{fig:mds_space} and more visualizations can be found in Supplemental Fig. 5.
Every point in $\encoding(\latentCoordinate)$ is a \ac{PCF} and we use the machinery that will be defined in \cref{subsec:synthesis} to generate a point pattern that has such a spatially-varying correlation. 

\section{Edge-aware density and correlations}
\label{sec:density_aware_pcfs}

\paragraph{Assumptions}
Input to this step is a discrete point pattern \pointPattern, a continuous density \revision{map} \density and a continuous guidance \revision{map} \guide \revision{to handle spatially-varying correlations}.
The aim is to estimate correlation at one position so it is unaffected by points that fall onto locations on the other side of an edge.
Output is a continuous spatially-varying correlation \revision{map}.

\paragraph{Definition}
We define the edge-aware radial density function as
\begin{equation}
    \label{eq:pcf}
    \pcfEstimate(\pointPattern,\density,\guide)
    (\point,\radius)
    =
    \sum_{i=1}^\pointCount
    \kernel(\point,\point_i,\radius,\density,\guide),
\end{equation}
a mapping from location \point, radius \radius to density, conditioned on the point pattern \pointPattern \revision{with \pointCount points} subject to the kernel
\begin{equation}
    \label{eq:DensityEstimator}
    \kernel(\point, \point_i, \radius, \density, \guide)
    =
    \frac{
    \hphantom{\sum_{\location\in\mathbb R^2}}
    \spatial(\point_i, \point, \radius, \density)
    \cdot
    \guidance(\point_i, \point, \guide)
    }{
    \sum_{\location\in\mathbb R^2}
    \spatial(\location, \point, \radius, \density)
    \cdot
    \guidance(\location, \point, \guide)
    },
\end{equation}
that combines a \emph{spatial} term \spatial and a \emph{guidance} term \guidance, both to be explained next.
Intuitively, this expression visits all discrete points $\point_i$ and soft-counts if they are in the ``relevant distance'' and on the ``relevant side'' of the edge relative to a position \point and a radius \radius. 
\revision{The \location locations represent a dense grid  used to compute the normalization term as explained next.}

\paragraph{Spatial term}
The spatial \spatial-term is a Gaussian \normal with mean \radius and standard deviation (bandwidth) \bandwidth.
It is non-zero for distances similar to \radius and falls of with bandwidth \bandwidth:
\begin{equation}
    \label{eq:PCFEstimator}
    \spatial(\point_i, \point, \radius, \density) = 
    \normal
    \left(
    \frac{||\point_i-\point||_2}{\density(\point)};
    \radius,
    \bandwidth
    \right)
\end{equation}


The distance between two points is scaled by the density at the query position.
As suggested by \citet{zhou2012point}, this assures that the same pattern at different scales, \ie densities, indeed maps to the same correlation.
Bandwidth \bandwidth is chosen proportional to the number of points \pointCount.

\paragraph{{Guidance} term}
The \guidance-term establishes if two points $\point$ and $\point_i$ in the domain are related.
This is inspired by the way Photon Mapping adapts its kernels to the guide by external information such as normals \cite{jensen2001realistic} or joint image processing makes use of guide images \cite{petschnigg2004digital}.
If $\point$ is related to $\point_i$, $\point_i$ is used to compute correlation or density around $\point$, otherwise, it is not.
Relation of $\point$ and $\point_i$ is measured as the pair similarity
\begin{equation}
    \guidance(\point, \point_i, \guide)=
    ||
    \guide(\point)^\mathsf T
    \cdot\guideBandwidth\cdot
    \guide(\point_i)
    ||
    ,
\end{equation}
where \guide is a {guidance} map and \guideBandwidth is the (diagonal) bandwidth matrix, controlling how guide dimensions are discriminating against each other.
The {guidance} map can be anything with distances defined on it, \revision{but our results will use correlation itself.}

\paragraph{Normalization}
The denominator in~\cref{eq:DensityEstimator} makes the kernel sum to 1 when integrated over \location for a fixed \point as in normalized convolutions \cite{knutsson1993normalized}.
Also, if \kernel extends outside the domain for points close to the boundary, this automatically re-scales the kernel.

\paragraph{Example}
\mywfiguretext{Kernel}{0.45}{Density estimation kernel.}
\refFig{Kernel} shows this kernel in action.
We show the upper left corner of the domain (white) as well as some area outside the domain (grey with stripes).
Correlation is computed at the yellow point \revision{\point}.
The kernel's spatial support is the blue area.
The area outside the domain (stripes) will get zero weight.
Also, the \revision{grey} area inside the domain which is different in the guide (due to different correlation) is ignored, controlled by the \guidance-term.

\paragraph{Discussion}
Our formulation is different from the one used by \citet{chen2013bilateral}, who warp the space to account for guide differences, as \citet{wei2011differential} did for density variation.
This does not work for patterns with varying correlation: a point on an edge of a correlation should not contribute to the correlation of another point in a scaled way.
Instead of scaling space,  we embed points  into a higher dimensional space \cite{chen2007real,jensen2001realistic} only to perform density estimation on it.

\mywfigure{Bilateral}{0.42}{Bilateral.}
Consider estimating the \ac{PCF} at the yellow point, close to an edge between an area of blue and an area of green noise as shown on \refFig{Bilateral}.
The dark gray point lies on the other side of the edge.
In common bilateral handling, it would contribute to a different bin (orange horizontal movement, distance in \ac{PCF} space) in the formulation of \citet{wei2010multi}, which is adequate for density, but not for correlation.
Our approach suppresses those points (blue arrow and vertical movement, density in \ac{PCF} space).

\section{Synthesis and Editing}
\label{sec:synthesis_and_editing}
We will now derive a method to synthesize a point pattern with a desired spatially varying target density and correlation (\cref{subsec:synthesis}).
Users can then provide these target density and correlation as common raster images (\cref{subsec:Encoding}) and edit them in legacy software (\cref{subsec:Editing}).

\subsection{Synthesizing with desired correlation and density}
\label{subsec:synthesis}
Given \cref{eq:pcf} we can estimate $\pcfEstimate(\pointPattern,\density,\guide)$, the spatially-varying \ac{PCF} field for a triplet of point pattern \pointPattern, density map \density and \revision{guidance map} \guide.
Recall, we can also compare this spatially-varying \ac{PCF} to another one, we shall call \pcfTarget, \eg by spatially and radially integrating their point-wise norms.
Hence, we can then optimize for a point pattern \pointPattern to match a given density map \density, a given guidance map \guide and a given target \ac{PCF} \pcfTarget by
\begin{equation}
\label{eq:synthesis_cost}
\argmin_\pointPattern
\synthesisCost(\pointPattern)
=
\int_x
\int_r
(\pcfEstimate(\pointPattern,\density,\guide)(\point,\radius)
 -\pcfTarget(\point,\radius))^2
\,
\mathrm d \point
\,
\mathrm d \radius
.
\end{equation}
%
%
\revision{We use \pcfTarget as our guidance map \guide to evaluate \pcfEstimate}.
Note, that we do not explicitly ask the result \pointPattern to have a specific density.
This happens implicitly:
recall, that our edge-aware correlation estimate \cref{eq:PCFEstimator} will scale point-wise distances according to density before computing the \ac{PCF}.
Hence, the only way for the optimization to produce the target \pcfTarget is to scale the distances proportional to density.

In practice, we Monte Carlo-estimate this objective \revision{using $10\times10$ jittered samples $\{\jitterlocation_i\} \in [0, 1]^2$ along the spatial dimension and \pcfSampleCount regular samples $\{\radius_j\}$ along the radius dimension (ranging from $0.1$ to $2\pcfRmax$ as detailed in \cref{subsec:implementation})}, as in
\begin{equation}
\synthesisCostEstimate(\pointPattern)
=
\sum_{i=1}^{10\times 10}
\sum_{j=1}^\pcfSampleCount
(\pcfEstimate
(\pointPattern,\density,\pcfTarget)
(\jitterlocation_i,\radius_j)
-
\pcfTarget(\jitterlocation_i,\radius_j))^2
.
\end{equation}
and optimize over \pointPattern using ADAM \cite{kingma2014adam}.

To have this synthesis produce a point pattern \pointPattern, a user now only needs to provide a spatially-varying  density \density and correlation \pcfTarget.

\subsection{Encoding into raster images}
\label{subsec:Encoding}
{Users provide} density \density, and correlation \pcfTarget as common discrete raster images on which we assume suitable sampling operations to be defined (\eg bilinear interpolation) to evaluate them at continuous positions.

For density \density, this is trivial and common practice in the point pattern literature.
For correlation, \pcfTarget, we use our embedding \encoding from \cref{sec:embedding} to decode the complex \ac{PCF} from a latent 2D coordinate, from only two numbers per spatial location.
Hence, density and correlation require only three numbers per spatial position.
We pack those three number into the three image color channels.
More specifically, density into the  \lspace and latent correlation into the \abspace channel of a \cielab color image, we call \featureMap.

\subsection{Editing point patterns as raster images}
\label{subsec:Editing}
Any editing operation that is defined on a three-channel image in any legacy image manipulation software can now be applied to the point pattern feature image \featureMap.

Working in \cielab color space, users have freedom to select the first-channel to edit density, the two latter channels to edit the correlations, or edit both at the same time.
Since \lspace and \abspace channels do not impact each other, \cielab color space is ideal for manipulating the density or the correlation independently as luminance and chrominance are perceptually decorrelated by-design.

While this is in a sense the optimal space to use when editing point patterns in legacy image editing software, it is not necessarily an intuitive user interface.
In a fully-developed software system, on top of legacy software, a user would not be shown colors or pick colors to see, respectively, edits, correlations.
Instead, they would only be presented the point pattern, and select from an image similar to \cref{fig:mds_space}, and all latent encoding would be hidden.
We will give examples of such edits in \cref{sec:results}.
For the case of ab-initio design, no input point pattern is required and a user would freely draw correlation and density onto a blank canvas.  

\begin{figure*}[t!]
    \centering
    \input{images/manual_edits/fig_manual_edits}
    \vspace{-2.5mm}
    \caption{
    We start with a given density and correlation map (a), and perform edits directly on this three-channel raster image (c) to obtain the point pattern in (d) after resynthesis. 
    In (c), we edit the density map (\lspace-channel) with gradients from left to right and edit correlations in the \abspace-channel to enhance different segments of the image.
    No neural network is used here. Source image credit: Artmajeur user Bianchini Jr. Used with permission under Media Licence.
    }
    \label{fig:manual_edits}
\end{figure*}
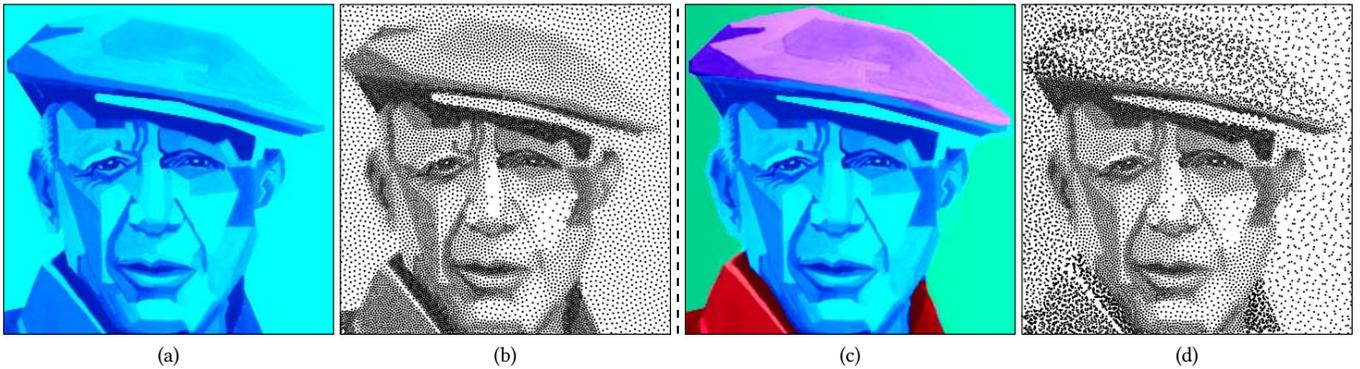

\begin{figure*}[t!]
    \centering
    \input{images/face_editing_final/fig_face_editing_manual}
    \vspace{-2.5mm}
    \caption{
    Input points (first column) generated by our synthesis and mapped to raster images (second column).
    An edit is performed in Adobe Photoshop (third column), and the a new point patterns is resynthesized (fourth column). 
    In the first row, we edit the background correlation using the \abspace channels and add some flowers on the hair using the \lspace channel.
    The second row shows another edit where we change the correlation in the background using the \abspace channels. 
    In the third row we change the correlation in the background (\abspace channels) and add a density gradient (\lspace channel). To generate the results in the second column from the first column, we use the neural networks trained on our human faces, animal faces and churches datasets, respectively.
    }
    \label{fig:face_editing_manual}
\end{figure*}
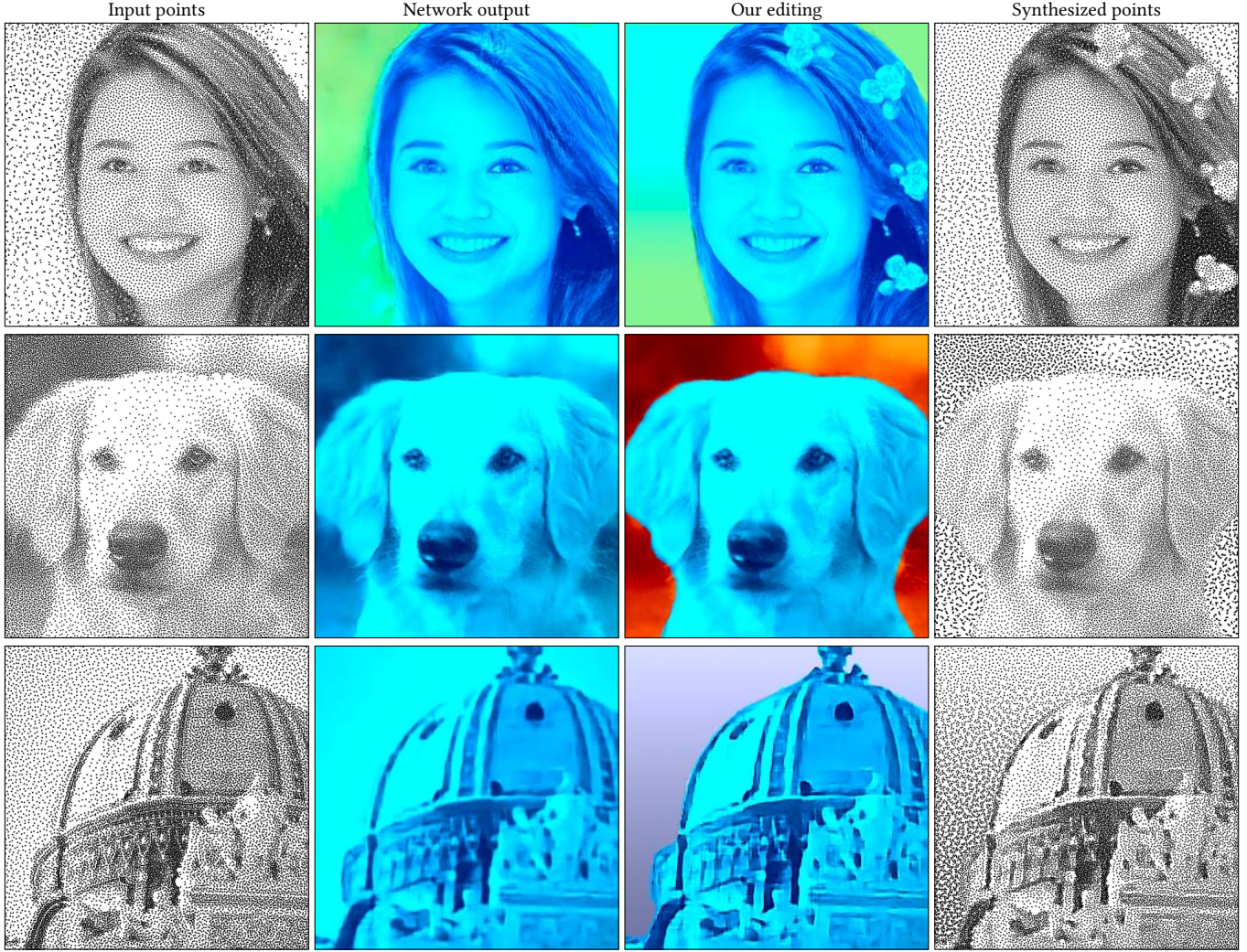

\section{Implementation}
\label{subsec:implementation}
We implement our framework mainly in PyTorch~\cite{paszke2017automatic}.
All experiments run on a workstation with an NVIDIA GeForce RTX 2080 graphics card and an Intel(R) Core(TM) i7-9700 CPU @ 3.00GHz.

\paragraph{Embedding}
In total, we collect 1,000 point patterns and each of them has 1,024 point samples.
To perform \ac{MDS}, the 2D latent coordinates $\{\latentCoordinate_i\}$ are initialized randomly in $[0,1]^2$.
The \ac{MDS} optimization~\cref{eq:mds} runs in batches of size $20$, using the ADAM optimizer~\cite{kingma2014adam} with a learning rate of $0.005$ for $1,000$ iterations.

If latent coordinates are quantized to 8 bit, there is only $256^2$ many different possible correlations $\{\pcf_i\}$.
We pre-compute these and store them in a $256\times256\times \pcfSampleCount$ look-up table $\mathtt{lut}$ w.r.t.\ each latent 2D coordinate to be used from here on.

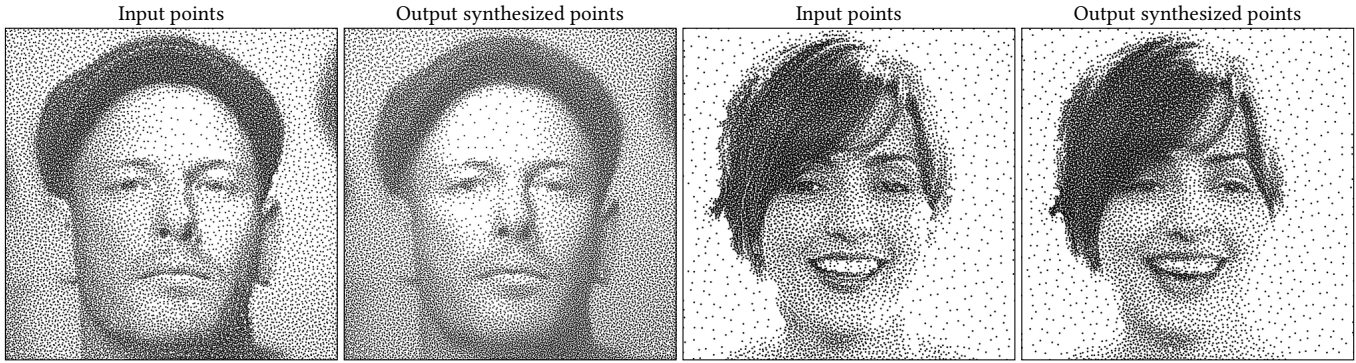
\begin{figure*}[t!]
    \centering
    \input{images/face_editing_final/fig_face_editing_neural}
    \vspace{-1.0mm}
    \caption{
    Our \ac{NN} maps the input point pattern to a density and correlation map. 
    To obtain different point pattern editing effects, we apply different advanced filters to the output density map. 
    From left, the second column shows the relighting effect using the method by~\citet{ClipDrop}.
    Fourth column shows the change in the facial expression and the eyes direction performed using a neural filter from Adobe Photoshop.
    }
    \label{fig:face_editing_neural}
\end{figure*}

\paragraph{Edge-aware \ac{PCF} estimator}
To compute the pair similarity between two guide values in \guide, the bandwidth matrix \guideBandwidth is set to be $0.005$-diagonal.
We use $\pcfSampleCount=20$ bins to estimate the \ac{PCF}. 
The binning is performed over the distance range from 0.1 to 2 $\pcfRmax$, where $\pcfRmax = 2\sqrt{\frac{1}{2\sqrt{3}n}}$. 
The point count \pointCount is chosen as the product between a constant and the total inverse intensity in the \lspace-channel of the given feature image~\featureMap, such that an image with dark pixels has more points. 
To compute local \ac{PCF} for each pixel in \featureMap, we consider only the $k$-nearest neighbor points, and not all points, where $k=50$ and $\pcfBandwidth = 0.26$. 

With each \ac{PCF}, $\pcf_i$ we also pre-compute and store $\learnignRate_i$, the best \ac{LR} (see Supplemental Sec. 1.2 for a definition) as we found different spectra to require different \acp{LR}.
During synthesis, we find the \ac{LR} for every point, by sampling the correlation map at that point position and using the \ac{LR} of the manifold-nearest exemplar of that correlation.

\paragraph{Synthesis}
The initial points when minimizing \cref{eq:synthesis_cost} are randomly sampled \revision{in $[0, 1]^2$} proportional to the density map \density and the optimization runs for $5,000$ iterations.
We note that the denominator in \cref{eq:pcf} is a sum over many points which could be costly to evaluate, but as it does not depend on the point pattern \pointPattern itself, it can be pre-computed before optimizing \cref{eq:synthesis_cost}. We use \texttt{C++} and \texttt{Pybind11} to accelerate this computation and the whole initialization stage takes around $5$ seconds.

To faithfully match the intensity of point stipples with the corresponding density map~\cite{spicker2017quantifying}, we also optimize for dot sizes as a post processing step.

\paragraph{Editing}
We use Adobe Photoshop 2022 for editing which has native  \cielab support. 
We devised two simple interfaces scripted into Adobe Photoshop 2022, one for interactive visualization of colors and point correlations and the other for editing and synthesis.

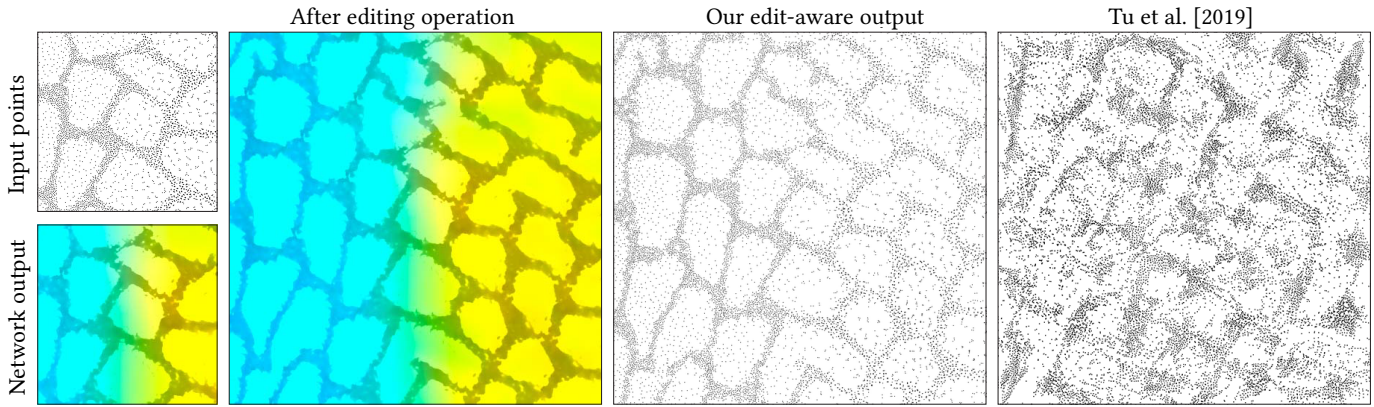
\begin{figure*}[t!]
\input{images/texture_synthesis/fig_texture_synthesis_comparison}
\caption{
\label{fig:point_pattern_expansion}
Starting from an input pattern of tree cover density (left, top), we use a network specialized on geomatic data to estimate correlation and density (left, bottom).
We can then apply existing image operations, such as Adobe Photoshop's ``Content-Aware Fill'' (second column) to achieve ``Point Pattern Expansion'' (third column), which compares favorably to direct optimization of points to match the inputs' VGG-based Gram matrices, deep correlation matrices and histograms \cite{tu2019pointpattern} (last column).
}
\end{figure*}

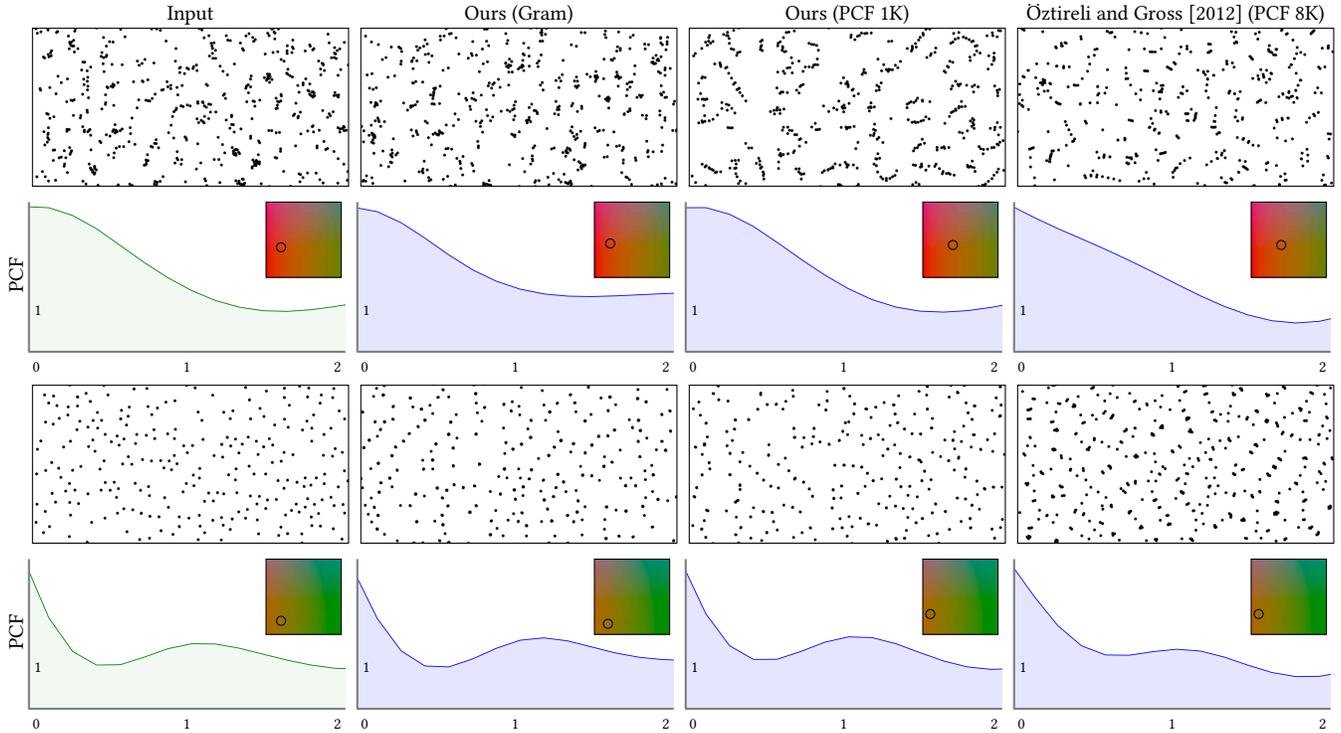
\begin{figure*}[t]
\input{images/comparison_cengiz/fig_comparison_with_cengiz}
\caption{
    For an input point pattern, we query its nearest neighbor in the latent space.
    Using Gram matrices (middle-left) the nearest neighbor quality is closer to the input than using \acp{PCF} (middle-right). 
    The inset images alongside PCFs show the latent space coordinates.  
    The correlation space from~\citet{oztireli2012analysis} uses 8K \acp{PCF} but is still not diverse enough to provide good nearest neighbor for the input.
    Our latent space has only 1000 base point patterns.
    We also perform a user study on this as detailed  in~\cref{subsec:user_study}.
    }
     \vspace{-2.5mm}
     \label{fig:comparison_with_cengiz}
\end{figure*}

\section{Results}
\label{sec:results}
In this section, we demonstrate our design choices and compare our pipeline with existing methods through some application scenarios. 
We also perform a user study to evaluate the effectiveness and usability of our framework.
More results can be found in the accompanied supplemental material.

\subsection{Ab-initio point pattern design}
\label{subsec:point_pattern_design}

We demonstrate examples of point pattern edits in~\cref{fig:Teaser} and~\cref{fig:manual_edits}. 
We start with a given density and correlation map (1st and 3rd columns). 
We choose blue noise for the correlation map to start with and start editing both the density (\lspace-channel) and the correlation (\abspace-channel). 
For~\cref{fig:Teaser}, we add density gradient transition on the background and add a simple leaf by editing the density channel. We also assign different correlations to different segments like the butterfly, leaf and background.
For~\cref{fig:manual_edits}, the  correlation edits are done in the \abspace channel of the Picasso image to assign different correlations to different image segments, \eg background, face, cap and the shirt. We also add a gradient density in the background from left to right by editing the \lspace-channel of the image.

We show more results in Supplemental Fig. 4 to demonstrate that our framework provides a straightforward way to edit spatially-varying point correlations by picking correlations from our correlation space (\cref{fig:mds_space}), instead of by designing \acp{PCF} or power spectra.

\mysubsection{Neural network-aided point pattern design}{Stippling}
Besides manually drawing correlation and density, we propose a second alternative: a \ac{NN},  based on \texttt{pix2pix} \cite{isola2017imagetoimage}, which automatically produces density and correlation maps from legacy point patterns.
We curated paired synthetic datasets for class-specific training from three categories, including human faces from \texttt{CelebA} by~\citet{CelebAMask-HQ}, animal faces from~\citet{choi2020stargan} and churches from~\citet{yu2015lsun}.
As the \ac{NN} maps point patterns to raster images, training data synthesis proceeds in reverse:
For each of the three-channel raster image, the gray-scale image of each original image is directly used as the \lspace-channel. We generate the correlation map (\abspace-channel) by assigning them random chroma \ie latent correlation coordinates.
Next, our synthesis from \cref{subsec:synthesis} is used to instantiate a corresponding point pattern.
Finally the patterns is rasterized, as \texttt{pix2pix} works with raster images.
For further data generation, network architecture and training details, please see Supplemental Sec. 1.4. We also perform an ablation study on the network architecture and training in Supplemental Fig. 6.

This pipeline enables freely changing local density or correlation in point patterns of different categories as seen in ~\cref{fig:face_editing_manual}. 
As shown in~\cref{fig:face_editing_neural}, this also allows advanced filtering such as relighting or facial expression changes on point patterns. In Supplemental Fig. 7, we show additional results where the input point patterns, generated by other image stippling methods~\cite{zhou2012point}~\cite{salaun2022scalable}, can be edited using our framework.

\mysubsection{Point pattern expansion}{Expansion}
Here we train our network on density from the Tree Cover Density \cite{buettner2017european} dataset in combination with random spatially-varying correlation maps using anisotropic Gaussian kernel with varying kernel size and rotation as detailed in Supplemental Sec. 1.3. 
Similar works can be found in~\citet{kapp2020data}, \citet{ tu2019pointpattern} and \citet{ huang2022point} for sketch-based or example-based point distribution synthesis.

\Cref{fig:point_pattern_expansion} illustrates one such representative example of point-based texture synthesis using our method.
Our network reconstructs the density and correlation map that captures the gradient change of correlation and spatially-varying density.
By using the content-aware filling tool in Adobe Photoshop, we can perform texture expansion (second column) based on the network output. 
More specifically, we first expand the canvas size of the map, select the unfilled region, and use content-aware filling to automatically fill the expanded background.
We also compare our method with a state-of-the-art point pattern synthesis method from~\citet{tu2019pointpattern} which takes an exemplar point pattern as input and uses VGG-19~\cite{simonyan2014vgg19} features for optimization. 

\subsection{Realizability}
\label{sec:realizability}
We construct the manifold \cref{fig:mds_space} from exemplars that are realizable, but an interpolation of realizable \acp{PCF} is not guaranteed to be realizable.
We test if realizability still holds in practice as follows.
For each \ac{PCF} $\pcf(\latentCoordinate_i)$, we generate a point set instance using our method and compute the resulting \ac{PCF} $\pcf'(\latentCoordinate_i)$.
We then compute
$A = \mathbb E_i[(|\pcf'(\latentCoordinate_i) - \pcf(\latentCoordinate_i)|)]$,
the average error between \ac{PCF} and realized PCF and
$B = \max_{ij} (|\pcf(\latentCoordinate_j) - \pcf(\latentCoordinate_i)|)$
the maximum difference between any two \acp{PCF}.
The relative error A/B is $3\times10^{-5}$, indicating that the realization error is five order of magnitude smaller than the \ac{PCF} signal itself.

\subsection{Comparisons}

\paragraph{Comparisons with~\citet{oztireli2012analysis}}
To the best of our knowledge,~\citet{oztireli2012analysis} is the most relevant related work which studies the space of point correlations using \acp{PCF}. 
However, we observe that \acp{PCF} are not the best way to characterize the perceptual differences between different point correlations. 
In~\cref{fig:comparison_with_cengiz}, we show that the Gram matrices (our input proximity to MDS) better encode the perceptual similarity between neighboring point correlations.

\begin{figure}[t!]
    \centering
    \input{images/comparison_roveri/fig_comparison_roveri}
    \vspace{-1.0mm}
    \caption{
    We compare the synthesis quality of our optimization against \citet{roveri2017general}. 
    To run \citeauthor{roveri2017general} method (left), we use the stored \acp{PCF} within each pixel of \featureMap. 
    The zoom-ins in the bottom two rows show that \citeauthor{roveri2017general} cannot handle well the sharp transitions in the correlations.
    }
    \label{fig:comparison_roveri}
\end{figure}
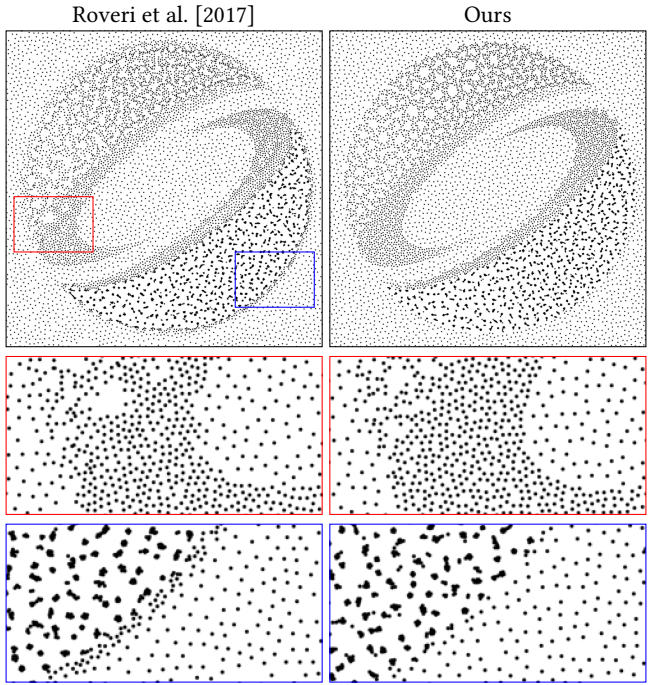

\paragraph{Comparisons with~\citet{roveri2017general}}

\Cref{fig:comparison_roveri} shows an example of synthesizing point patterns using our synthesis method and the synthesis method proposed by \citet{roveri2017general}. To the best of our knowledge, \citet{roveri2017general} is the only competitor that supports point pattern synthesis from spatially-varying density and correlation. We demonstrate that their method may synthesize point patterns with artifacts around the sharp transitions between two correlations (bottom-right and top-left of the logo). Our method, by taking the bilateral term into account, can handle sharp transition between correlations more accurately.

This relation can be further quantified as follows:
Let \pointPattern be a point pattern, \editedPointPattern be an edited version of that and $\pcf(\pointPattern)$, respectively, $\pcf(\editedPointPattern)$ be their correlations.
Now, first, let $\pcf_\mathrm{PCA}(\editedPointPattern)$ be the correlations of \editedPointPattern, projected into the space spanned by the PCA of the correlations in and only in \pointPattern,
and, second, $\pcf_\mathrm{Ours}(\editedPointPattern)$ be the correlations of \editedPointPattern, projection into our palette (\cref{fig:mds_space}).
The error of those projections is 
$e_\mathrm{PCA} = |
\pcf_\mathrm{PCA}(\editedPointPattern) - 
\pcf(\editedPointPattern)
|$
and
$e_\mathrm{Ours} = |
\pcf_\mathrm{Ours}(\editedPointPattern) - 
\pcf(\editedPointPattern)
|$
, respectively.
We evaluate these error values for different figures.
\cref{fig:Teaser} shows 96$\times$ improvement,  
\cref{fig:manual_edits} shows 130$\times$ improvement and the three rows in~\cref{fig:face_editing_manual} shows 471$\times$, 461$\times$ and 59$\times$ improvement using our approach ($e_\mathrm{Ours}$ vs. $e_\mathrm{PCA}$). 
This indicates that our latent space can preserve one to two order of magnitudes more edit details than \citet{roveri2017general} approach which restricts itself to the input exemplar.

\subsection{User study}
\label{subsec:user_study}
We performed a set of user experiments to verify i) the perceptual uniformity of our embedding ii) the ability of users to navigate in that space iii) the usefulness of the resulting user interface.

\paragraph{Embedding user experiment}
34 subjects (S) were shown a self-timed sequence of 9 four-tuples of {point patterns} with constant density and correlation in a horizontal arrangement (\cref{fig:comparison_with_cengiz}).
The leftmost pattern was a reference.
The three other patterns were nearest neighbors to the reference in a set of:
i) our basis patterns using our perceptual metric,
ii) our basis patterns using \ac{PCF} metric, and
iii) patterns from the \ac{PCF} space suggest by \citet{oztireli2012analysis} using \ac{PCF} metric.
Ss were instructed to rate the similarity of each of the three leftmost patterns to the reference on a scale from 1 to 5 using radio buttons.

\begin{figure}[ht]
\centering
\includegraphics*[width = \linewidth]{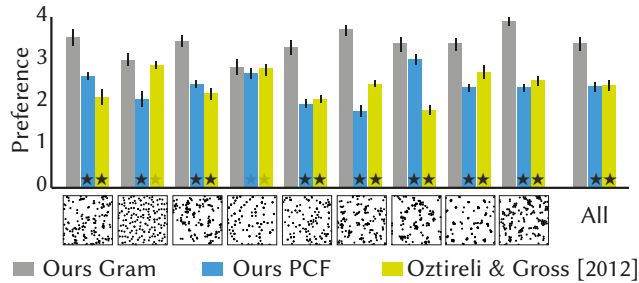}
\vspace{-.6cm}
\caption{Results of the embedding user experiment. 
Errors bars are standard errors of the mean.
A black or gray star is significance against our method at the $p=.001$ or $.005$-level.}
\label{fig:study}
\end{figure}

The mean preferences, across the 10 trials, were a favorable 3.59, 2.52 and 2.55 which is significant ($p<.001$, two-sided $t$-test) for ours against both other methods.
A per-pattern breakdown is seen in \refFig{study}.
This indicates our metric is perceptually more similar to user responses than two other published ones. 
\Cref{fig:comparison_with_cengiz} shows two examples where the nearest-neighbor of the query point pattern is perceptually different using different metrics.
The \ac{PCF} of two point patterns can be similar even when they are perceptually different.

\begin{figure}[t!]
    \centering
    \input{images/user_study/fig_user_study_points_vs_chroma}
    \vspace{-1.0mm}
    \caption{
    Results of the navigation user experiment. 
    Users are shown the latent space visualized as spatially-varying points (\cref{fig:mds_space}) on the left and as spatially-varying chroma on the right.
    The large dots represent the locations of our selected $8$ reference point patterns. 
    The small dots are the locations that the users chose as perceptually similar point patterns wrt.\ the reference.
    }
    \label{fig:user_study_points_vs_chroma}
\end{figure}
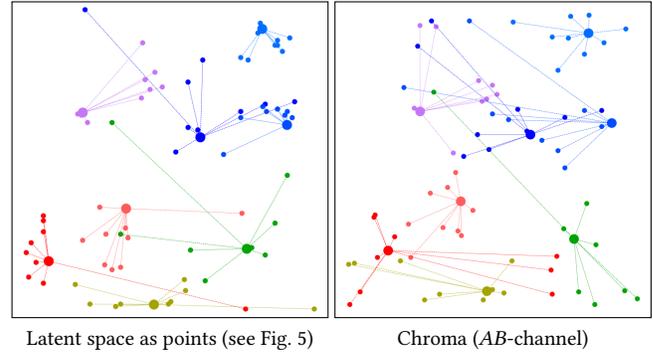

\paragraph{Navigation user experiment}
In this experiment, $N=9$ Ss were shown, $8$ reference point correlations and, second, the palette of all correlations covered by our perceptual embedding (\cref{fig:mds_space}).
Ss were asked to pick coordinates in the second image by clicking locations in the embedding, so that the corresponding point correlations of the picked coordinates perceptually match the reference correlations.

The users' locations were off by 14.9\,\% of the embedding space diagonal.
We would not be aware of published methods for intuitive correlation design to compare this number to.
Instead, we have compared to the mistakes users make when picking colors using the LAB color picker in Adobe Photoshop.
Another  $N=9$ Ss, independent to the ones shown the palette of correlations, made an average mistake of 21.3\% in that case.
We conclude, that our embedding of pattern correlation into the chroma plane is significantly more intuitive ($p<0.2$, $t$ test) than picking colors.
\cref{fig:user_study_points_vs_chroma} shows the points users clicked (small dots), relative to the true points (large dots).

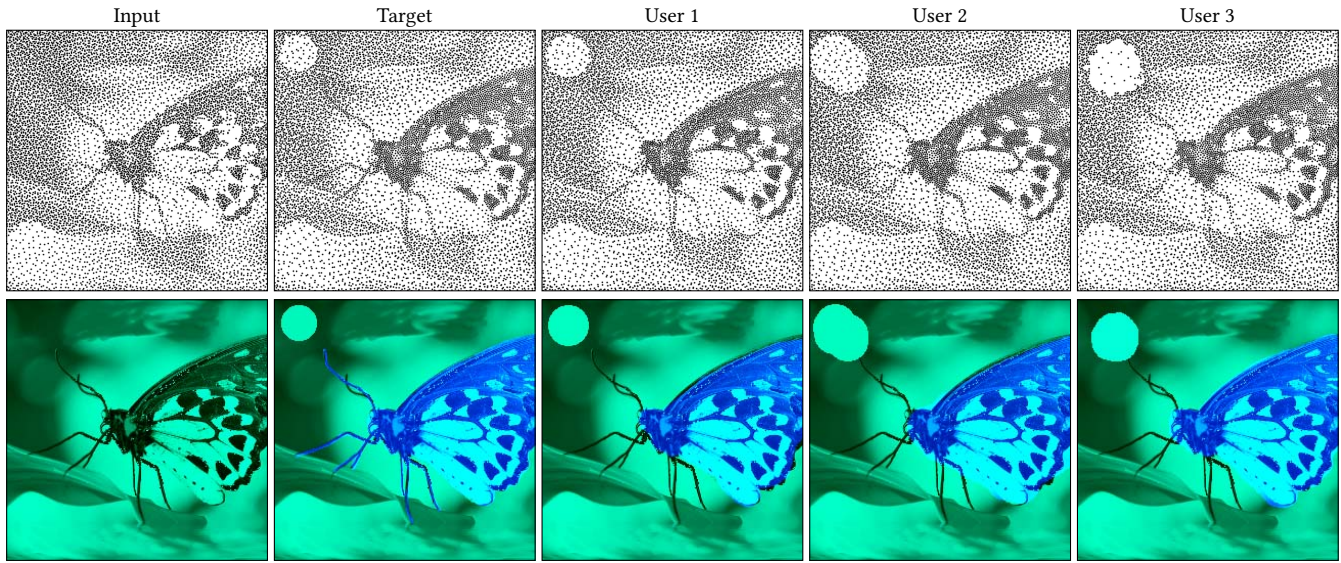
\begin{figure*}[t!]
    \centering
    \input{images/user_study/fig_user_study_edits}
    \vspace{-2.5mm}
    \caption{
    User edits from the usefulness experiment. 
    We show that users can use our system to design point patterns (first row) by editing \lspace-channel and \abspace-channel (second row) to match the target one from an initialized input.
    }
    \label{fig:user_study_edits}
\end{figure*}

\paragraph{Usefulness experiment}
Ss were asked to reproduce a target stippling with spatially varying correlation and density by means of Adobe Photoshop that was enhanced to handle point correlation as LAB space using our customized interface.
Ss were shown the point patterns of our perceptual latent space, and asked to edit the density and correlation separately to reproduce the reference from an initialized LAB image.
After each editing, generating the point pattern incurred a delay of one minute. 
Note that we intentionally reduce the number of iterations to optimize the point patterns from user edits to offer faster feedback in around a minute with around 10,000 points.
Details on this experiment are found in Supplemental Sec. 2.4.
There is no objective metric to measure the result quality, so we report three user-produced point patterns in \cref{fig:user_study_edits}.
The whole process takes 15 minutes on average for all three users, as they usually require multiple trials on picking the correlations and our synthesis method does not run interactively.

\mysubsection{Performance}{Performance}
\mywfigurevspace{PointCountVsTime}{0.33}{Timing.}{-1.2cm}
We summarize the run-time statistics of our synthesis method w.r.t. the number of synthesized points in \cref{fig:PointCountVsTime}.
Editing time is not reported as it can be biased by the editing skills of the users.

\mysection{Conclusion and Future Work}{Conclusion}
We propose a novel framework to facilitate point pattern design, by introducing a perceptual correlation space embedded in a two-channel image using a dimensionality reduction method (\ac{MDS}).
This allows users to design or manipulate density and correlation by simply editing a raster image using any off-the-shelf image editing software.
Once edited, the new features can be resynthesized to the desired point pattern using our optimization (\cref{subsec:synthesis}).
To better handle sharp transitions in density and correlation during synthesis, we formulate a novel edge-aware \ac{PCF} (\cref{sec:density_aware_pcfs}) estimator. 
The resulting framework allows wide range of manipulations using simple image editing tools that were not available to point patterns before. 
Users can design novel spatially varying point correlations and densities without any expert knowledge on \ac{PCF} or power spectra or can use a \ac{NN} to get correlation and density for a specific class, such as faces.

\paragraph{Limitations}
The latent space spanned by the bases point correlations in~\cref{fig:mds_space} is by no means perfect.
Synthesizing point patterns with smooth transitions from two extreme locations in the latent space may result in some unwanted artifacts.
Our synthesis method takes minutes to synthesize points which is far from interactive rate that is more friendly to artists and users. 

\paragraph{Future work}
A neural network with such latent space is a promising direction where the correlations within the geomatic data can be predicted and edited according to user-defined conditions (environment, pollution, global warming, etc).
Our approach happens to rasterize points, which ideally is to be replaced by directly operating on points \citet{qi2017pointnet,hermosilla2018montecarlo}.

The editing operations can also be improved by proper artistic guidance or by interactive designing tools that intuitively manipulate the desired density and correlations for desired goals.
Extending our pipeline for visualization purposes is another fascinating future direction.
Another interesting future direction is to extend the current pipeline to multi-class and multi-featured point patterns, which appear in real-world patterns.
Designing a latent pattern space that spans a larger gamut of correlations (also anisotropic and regular ones) present in nature (\eg sea shells, tree, stars, galaxy distributions) is a challenging future problem to tackle.
There is an exciting line of future works that can be built upon our framework. 
Many applications like material appearance, haptic rendering, smart-city design, re-forestation, planet colonization can benefit from our framework.

\section*{Acknowledgments}
We would like to thank the anonymous reviewers for their detailed and constructive comments. We thank artist Bianchini Jr. for the Pablo Piccaso artwork which we use under the Media Licence.

\bibliographystyle{ACM-Reference-Format}
\bibliography{paper}

\end{document}


\title{Supplemental Material\\\name: Editing Point Patterns by Image Manipulation}

\author{Xingchang Huang}
\affiliation{%
	\institution{Max-Planck-Institut für Informatik}
    \country{Germany}
}
\email{xhuang@mpi-inf.mpg.de}

\author{Tobias Ritschel}
\affiliation{%
	\institution{University College London}
    \country{United Kingdom}
}
\email{t.ritschel@ucl.ac.uk}

\author{Hans-Peter Seidel}
\affiliation{%
	\institution{Max-Planck-Institut für Informatik}
    \country{Germany}
}
\email{hpseidel@mpi-sb.mpg.de}

\author{Pooran Memari}
\affiliation{%
	\institution{CNRS, LIX, \'Ecole Polytechnique, IP Paris, INRIA}
    \country{France}
}
\email{memari@lix.polytechnique.fr}

\author{Gurprit Singh}
\affiliation{%
	\institution{Max-Planck-Institut für Informatik}
    \country{Germany}
}
\email{gsingh@mpi-inf.mpg.de}

\acmJournal{TOG}
\acmYear{2023} \acmVolume{42} \acmNumber{4} \acmArticle{1} \acmMonth{8} \acmPrice{}\acmDOI{10.1145/3592418}

\maketitle

\acresetall

\section*{Introduction}
In this supplemental, we provide further details on the implementation (\refSec{ImplementationDetails}), as well as additional results (\refSec{Results}).

\mysection{Additional Implementation Details}{ImplementationDetails}

\begin{figure}
\includegraphics[width=\linewidth]{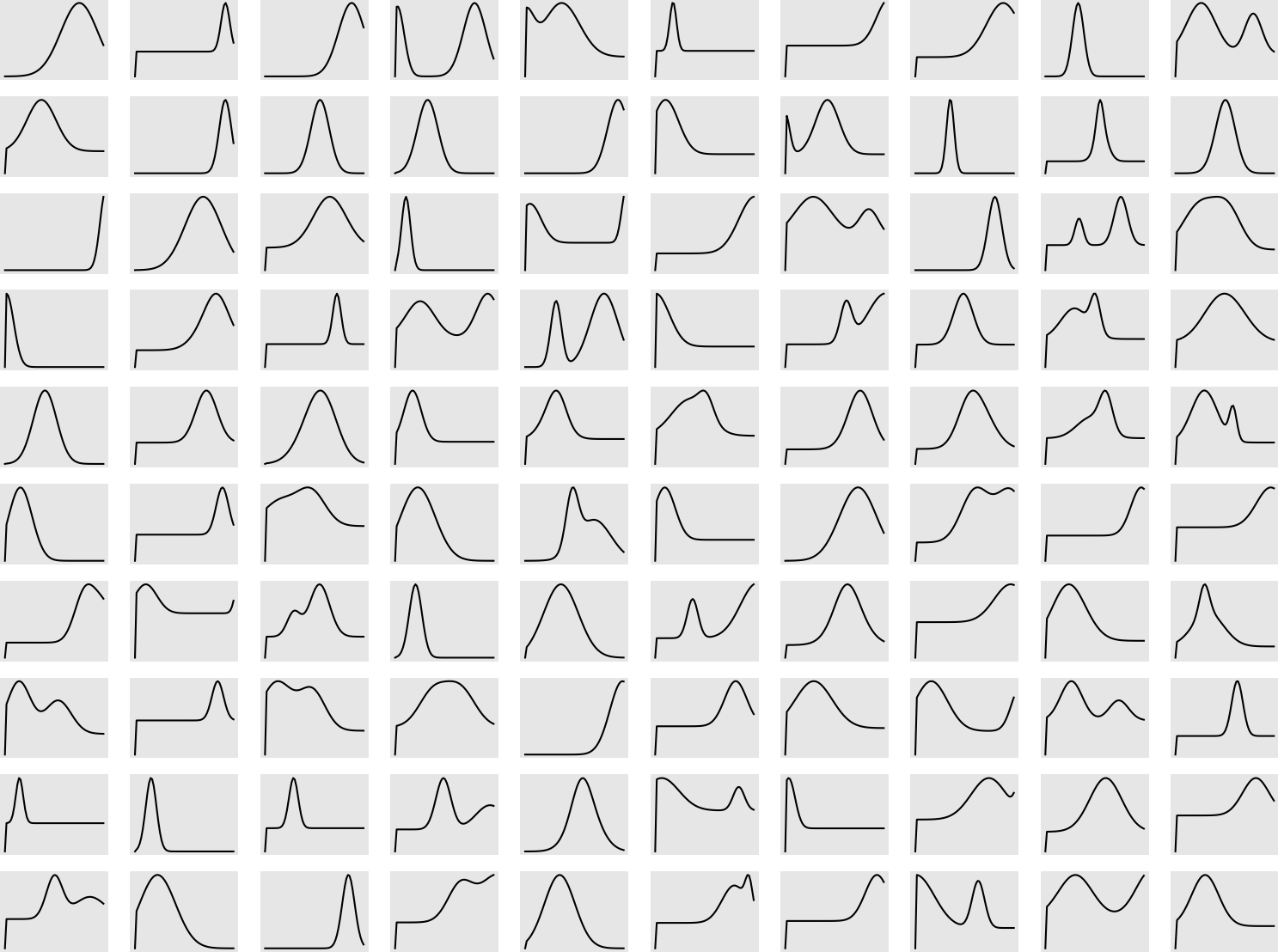}
\caption{
\label{fig:point_correlations_visualization}
Example power spectra used to learn the perceptual embedding.
}
\end{figure}

\mysubsection{Point correlations generation}{PointCorrelationGeneration}
We use a mixture of Gaussians to sample the gamut of possible power spectra.
We use power spectra here, as they allow us to directly work with different range of frequencies unlike \acp{PCF}.
The value of a power spectrum bin $\PowerSpectrum_{\PowerSpectrumBin}$ is computed as
\begin{equation}
    \label{eq:PointPatternGeneration}
    \PowerSpectrum_\PowerSpectrumBin = 
    \sum_{i=1}^{\NumGMM}
    \left(
    \PowerSpectrumScale_i
    \cdot
    \exp(-\frac{(\PowerSpectrumBin-\mu_i)^2}{2\sigma_i^2}
    )
    \right)
    + \PowerSpectrumGamma
 .
\end{equation}
We use a mixture with, randomly, either  
$\NumGMM=1$ or $\NumGMM=2$ Gaussians and sample the parameters from the range 
$\PowerSpectrumGamma \in \left\{0,1\right\},$ 
$\mu_i \in [0, 68]$, 
$\sigma_i \in [2, 12]$, and 
$\PowerSpectrumScale_i \in [1, 3]$, respectively, so as to cover the required range of frequencies, including blue, green and red noises.
We vary \PowerSpectrumBin from $0$ to $\pcfSampleCount=63$.
The DC $\PowerSpectrum_0$ is set to $0$.
A sample of 100 random power spectra produced by this approach is seen in \cref{fig:point_correlations_visualization}.

The generated power spectra are only used to run the method of~\citet{leimkuhler2019deep} to produce a point pattern that is used in the following steps, while the power spectrum can be discarded.

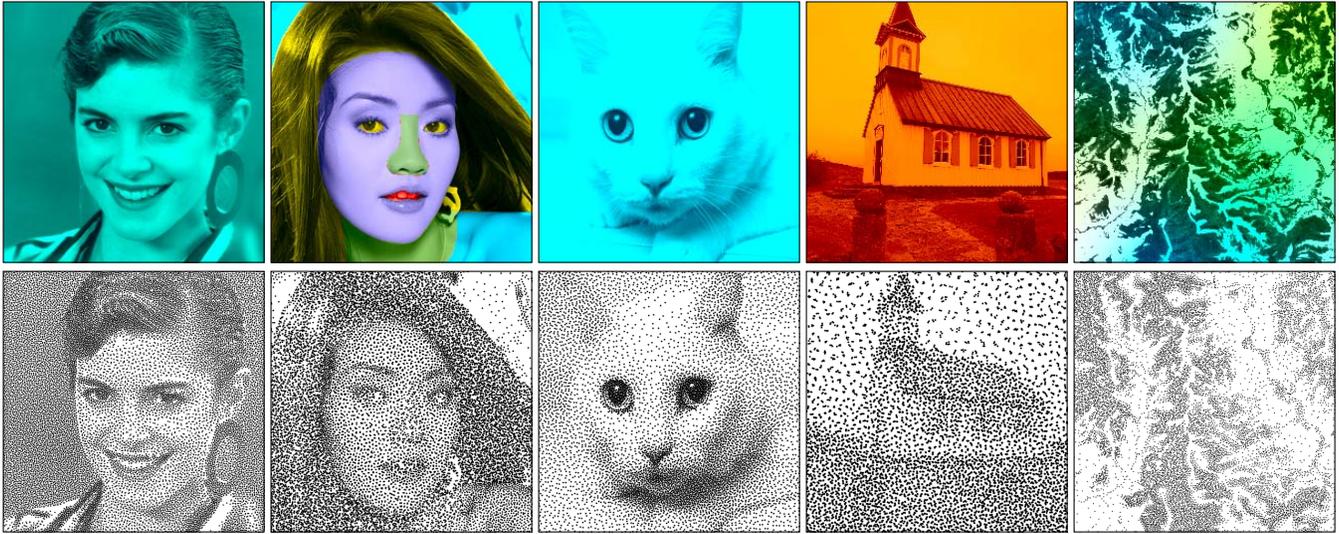
\begin{figure*}[t!]
    \centering
    \input{images/training_data/training_data}
    \caption{
    Examples of paired training data from human faces, animal faces, churches and tree cover density datasets, respectively.
    }
    \label{fig:training_data}
\end{figure*}

\begin{figure*}[t!]
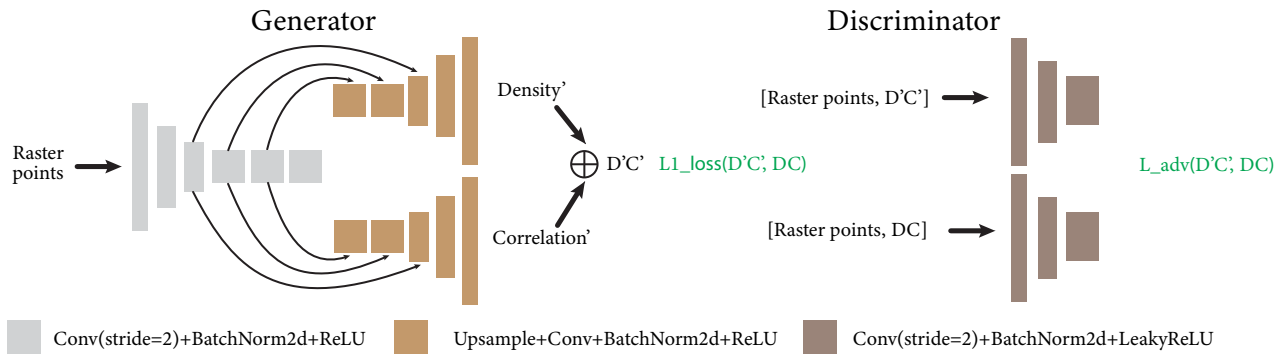

\begin{overpic}[scale=3.8,grid=false]{images/our_approach/network_architecture}
\end{overpic}
\caption{
\label{fig:network_architecture}
Our adapted cGAN architecture.
}
\end{figure*}
\mysubsection{Optimal learning rate details}{learning_rates}
The best \ac{LR} \learnignRate for a correlation \pcf is the one out of 0.02, 0.01, 0.005, 0.001, 0.0005, 0.0001 and 0.00005 that, when using a fixed number of 1000 iterations to minimize \ac{PCF} error of a randomly initialized pattern, leads to the lowest VGG error.
We find these $\{\learnignRate_i\}$ in a grid search pre-process for the set of all training \acp{PCF} $\{\pcf_i\}$.

\mysubsection{Data generation for neural network-aided point pattern design}{DataGeneration}

We achieve this by utilizing our proposed latent space and a large set of images to generate a dataset with varying density maps (represented by gray-scale images) and varying correlation maps (represented by different colors).
All the density and correlation maps have the same resolution $\pixelResolutionWidth \times \pixelResolutionHeight$. 

For each density and correlation map of the following datasets, our synthesis is used to generate a point pattern with spatially varying density and correlation to get pairs (a point pattern, a density and correlation map) for training our networks. 
The number of points is computed as $\pointCount = 50,000 \times \mathbb E_\location(\mathbf{1}-\density(\location))$. \cref{fig:training_data} shows examples of paired training data from different datasets.

\paragraph{Human faces stippling dataset}
We generate the human faces stippling dataset using the face images from~\citet{CelebAMask-HQ}. 
We use 10,000 gray-scale face images as the density maps. 
Each of the face images are used to generate two correlation maps, one for uniform correlation with a random chroma assigned to all pixels, another for spatially-varying correlations. 
To generate a spatially-varying correlation map, we utilize the facial segmentation masks including skin, hair, background and generate a correlation map by assigning random chroma to each of the segment.
In total, we generate 20,000 density and correlation maps.

\paragraph{Animal faces and outdoor churches stippling datasets}
Similarly, we use the gray-scale images of animal faces~\cite{choi2020stargan}.  and outdoor churches~\cite{yu2015lsun} as density maps for the two datasets. 
Different from the human faces dataset~\citet{CelebAMask-HQ}, no segmentation masks are provided for animals faces and churches. Therefore, for each density map, we generate a uniform correlation map by randomly sampling a color and assign it to all pixels. 
In total, we generate around 15,000 density and correlation maps.

\paragraph{Tree Cover Density dataset for point pattern expansion}
We use the Tree Cover Density data~\cite{buettner2017european} as our density maps. 
To generate a correlation map for each density map, we generate either a uniform correlation maps or a spatially-varying correlation maps using anisotropic Gaussian kernels with varying locations, kernel sizes and orientations. 
To generate a uniform correlation map, we randomly sample a color and assign it to all pixels. 
To generate a spatially-varying correlation map, the number of Gaussian kernels is randomly sampled from [2, 16]. 
For each of the Gaussian kernel, the location (mean) is randomly sampled in [0, 1], the orientation is randomly sampled from [-180, 180] degrees and the kernel size (variance) for x-, y-axis are randomly sampled from [0.15, 0.25]. 
The Gaussian kernels are summed with equal weight to generate a correlation map.
In total, we generate around 20,000 density and correlation maps.

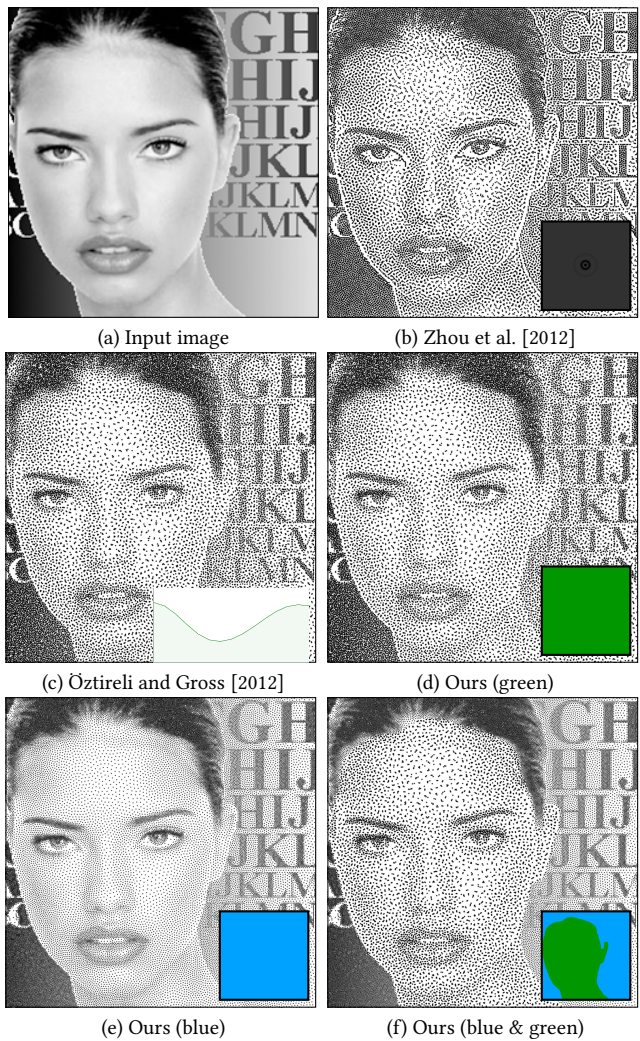
\begin{figure}[tbh]
\input{images/point_correlation_design/fig_point_correlation_design.tex}
\caption{
        Our framework provides a straightforward way to design spatially varying point correlations by picking correlations from our correlation palette. The picked correlations are visualized as chroma in (d), (e) and (f), which automatically find the corresponding \ac{PCF} from the embedded space.
     }
     \label{fig:point_correlation_design}
\end{figure}

\mysubsection{Network architecture and training}{NetworkDetails}

\cref{fig:network_architecture} shows the details of our network architecture, adapted from the cGAN~\cite{isola2017imagetoimage} framework. In the generator side, we take rasterized points as input and output the density and correlation maps in separate branches. The density and correlation predictions are concatenated in the output layer followed by a Sigmoid function. We use a U-Net architecture with 6 downsampling and upsampling layers with skip connections. For the discriminator, we use a three-layer convolutional architecture to extract patch-based features.

Point patterns are rasterized to a resolution of $\pixelResolutionWidth \times \pixelResolutionHeight$ as input. 
Output is the three-channel image for each rasterized point pattern. Compared with the original framework~\cite{isola2017imagetoimage}, the major change is to use two branches for regressing density and correlation map separately for better quality.
Predicted density and correlation are concatenated at the last layer, followed by a Sigmoid function to normalize output values between $[0, 1]$.
The network is trained with a combination of \networkLossPixel loss and adversarial loss \networkLossGAN between the output and the ground truth. 
The total loss $\networkLossTotal = \networkLossPixel + 0.001\networkLossGAN$ is minimized to update network weights during training.
We use the ADAM optimizer~\cite{kingma2014adam}, with an initial learning rate of $0.0001$ for both generator and discriminator and a batch size of $8$. 
Learning rate decays by half after every 100 epochs. 
The network is trained for $400$ epochs in $24$ hours. With each $\pixelResolutionWidth \times \pixelResolutionHeight$ rasterized point image as input, the network inference time is about $0.005$ seconds per frame to get the density and correlation map with a resolution of $\pixelResolutionWidth \times \pixelResolutionHeight \times 3$.

\mysection{Additional Results}{Results}

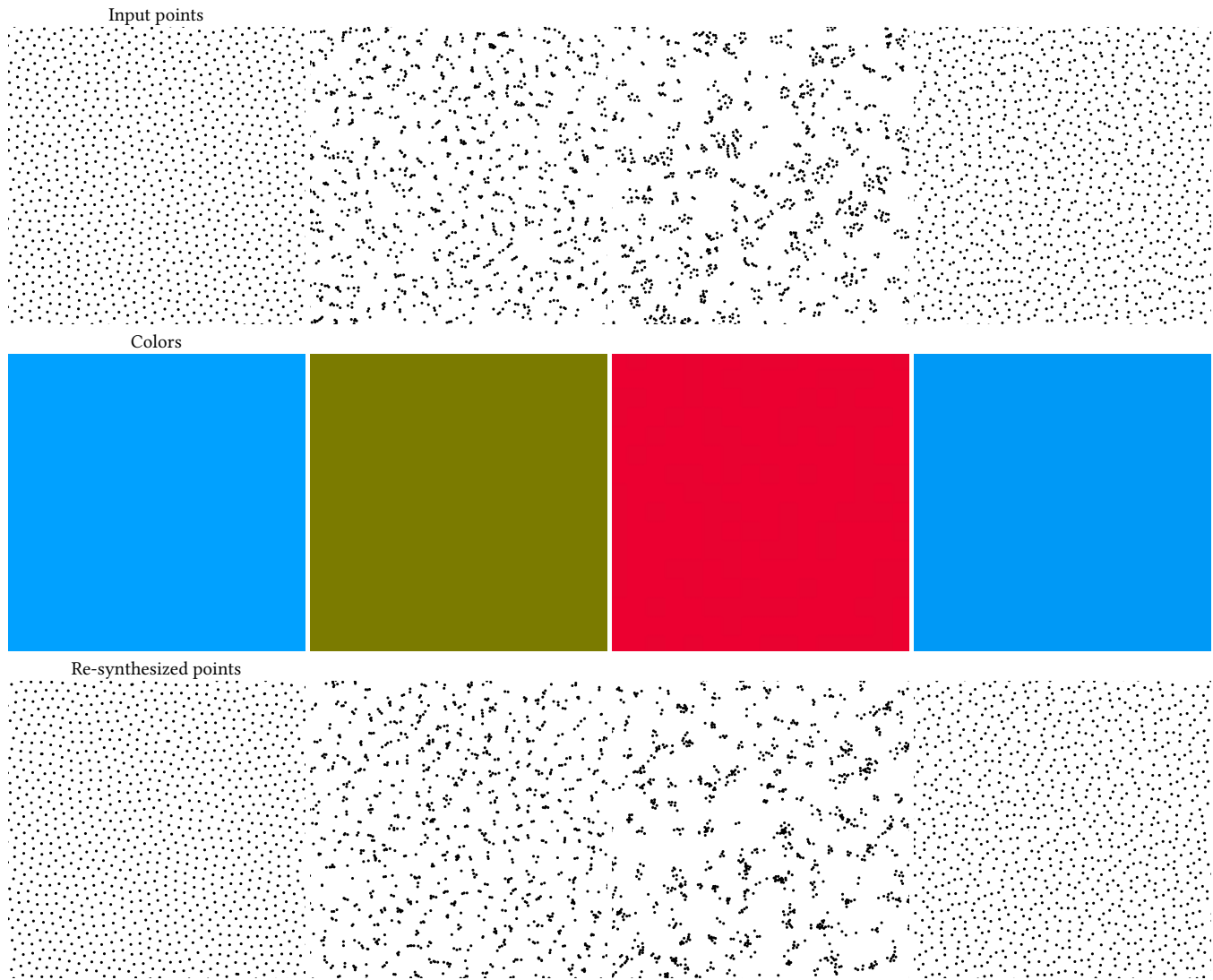
\begin{figure*}[t!]
\centering
\input{images/knownnoises}
\caption{
We search four known noises (blue, green, red, step noises) generated by \citet{leimkuhler2019deep} their nearest-neighbors in our embedding space using VGG16~\citet{simonyan2014very} based gram metric as shown in main paper equation 1. The color of their nearest-neighbors are shown in the second row. Lastly, we can use our synthesis method to realize back similar point patterns.
}
\label{fig:known_noises}
\end{figure*}

\mysubsection{Latent space}{Latent}
One property about our point correlation embedding space is that we locate some known point correlations to their corresponding semantic colors including blue, green, pink/red and step noises generated by \citet{leimkuhler2019deep}.
The found colors can roughly match the semantic meaning, as shown in \cref{fig:known_noises}.
More specifically, we search those four noises (blue, green, red, step noises) in our embedding space using Eq. 4 in the main paper.
The colors of their nearest-neighbors are shown in the second row. The colors are then used to re-synthesize the point sets as shown in the third row.

\subsection{Ab-initio point pattern design}

In~\cref{fig:point_correlation_design}, we compare our \cielab space representation to traditional approaches~\cite{zhou2012point,oztireli2012analysis}. 
Unlike these methods, we do not need to tailor a specific power spectrum or a \ac{PCF} to represent point correlations which requires expert knowledge from the end users. 
Instead, we can simply pick correlations from our correlation palette to design the point correlations and use that to synthesize point patterns.
We start with a given density map in~\cref{fig:point_correlation_design}(a).
\Cref{fig:point_correlation_design}(b) and (c) uses traditional \ac{PCF} representation.
\Cref{fig:point_correlation_design}(d) and (e) shows our
green and blue noise stippling which require painting the \abspace-channel with the specific color (latent coordinate).
green and blue noise stippling which require painting the \abspace-channel with the specific color (latent coordinate).
We create spatially varying point pattern \cref{fig:point_correlation_design}(f) by simply assigning green color to the face and blue color to the background Previous methods~\cite{zhou2012point,oztireli2012analysis} are not able to create point patterns with spatially-varying correlations like ours.

\mysubsection{Neural network-aid point pattern design}{NetworkResults}

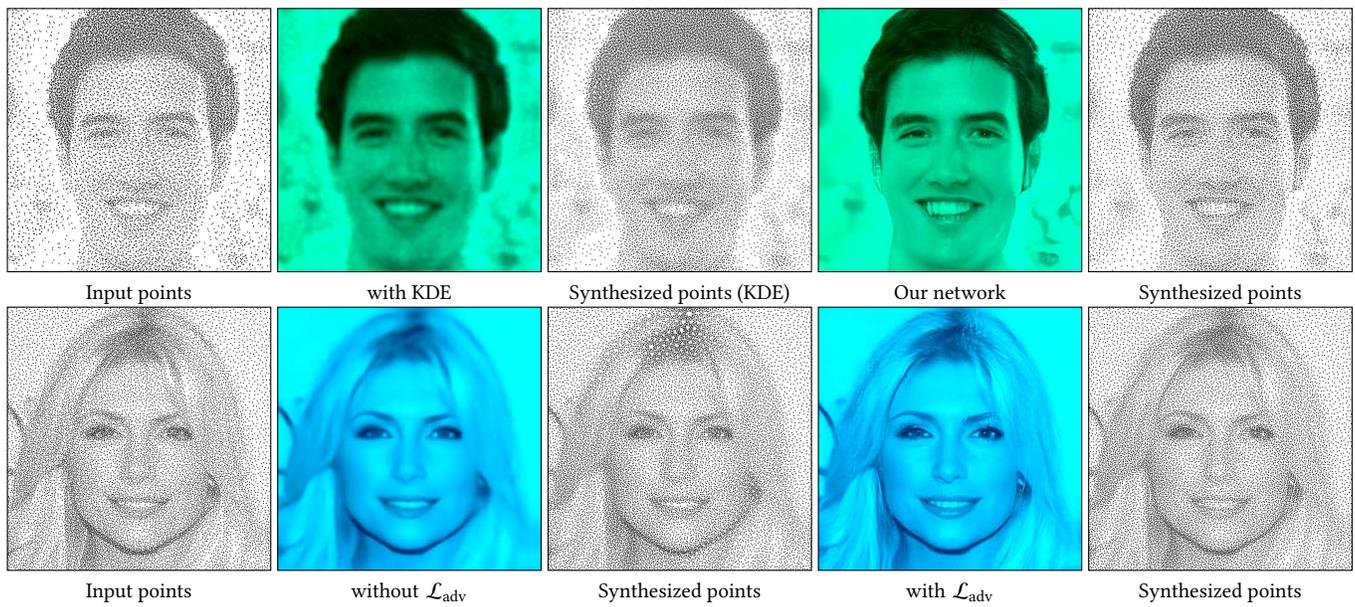
\begin{figure*}[t!]
    \centering
    \input{images/ablation_study/fig_ablation_study}
    \caption{
    In the first row, we analyze the impact of density estimation from given point pattern. 
    We compare traditional kernel density estimation (KDE) with our network density reconstruction.
    In the second row, we study the impact of using different losses during training.
    }
    \label{fig:ablation_study}
\end{figure*}

\paragraph{Ablation study}
Here we study the impact of our network (trained on faces) components. Firstly, we demonstrate that our network is important in terms of density estimation. 
As shown in~\cref{fig:ablation_study} (first row), one way to estimate density from points is to perform traditional kernel density estimation. 
However, this can lead to blurry results given the number of points is finite. 
Our network trained on face images, on the other hand, can reconstruct higher-quality density map. 
Note that no existing method can perform correlation estimation from points, both density and correlation estimation branches are important in our network.
Secondly, we study the impact of using \networkLossGAN during training. 
The second row shows that training with \networkLossGAN can produce sharper density and correlation map that is used to synthesize points closer to the input compared with training without \networkLossGAN.

\paragraph{Input points from existing methods}

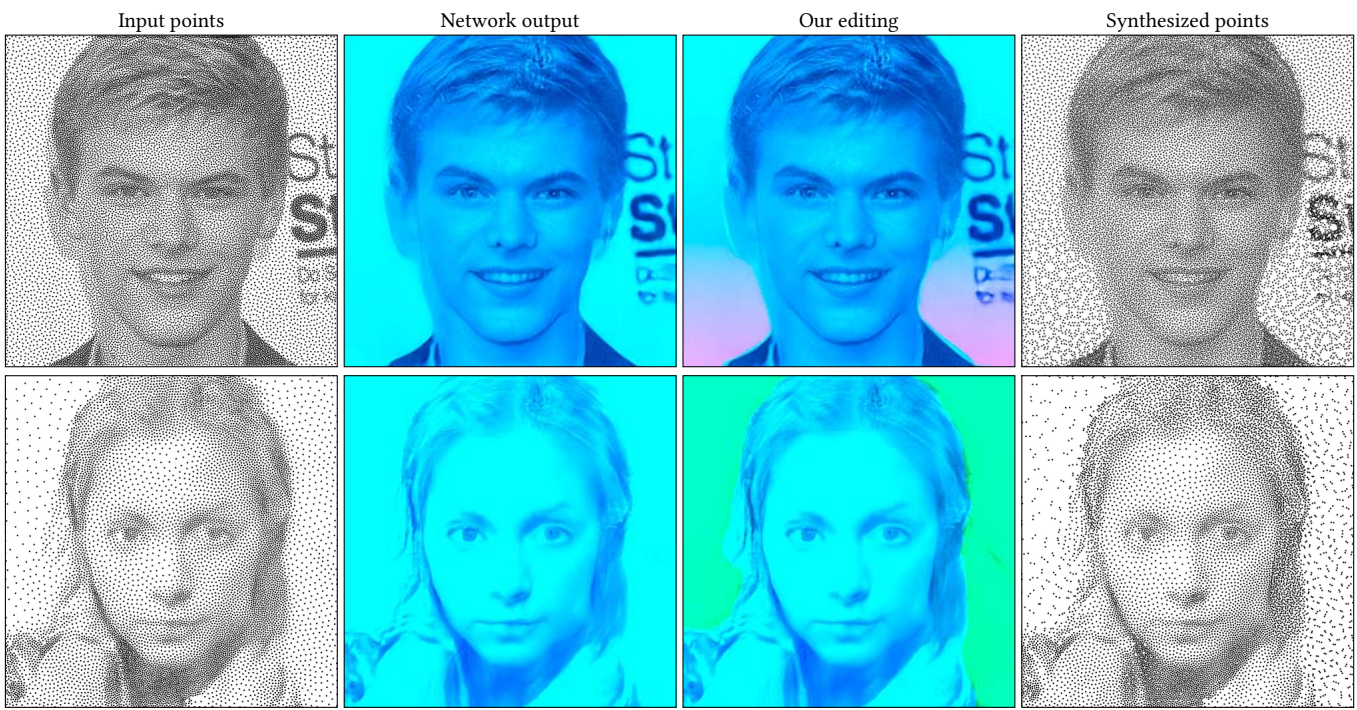
\begin{figure*}[t!]
    \centering
    \input{images/face_editing_existing_methods/fig_face_editing_existing_methods}
    \caption{
    We generate the input points using~\citet{salaun2022scalable} (first row) and \citet{zhou2012point} (second row).
    These point patterns goes as input to our network to obtain the underlying density and correlation map (second column). 
    We change the background correlation in both rows (third column), sharpen the density map and change the hair correlation in the second row, to get edited point patterns (fourth column).
    }
    \label{fig:face_editing_existing_methods}
\end{figure*}

In~\cref{fig:face_editing_existing_methods}, the input points are synthesized using existing methods. 
We use \citet{zhou2012point} to generate the input points with CCVT profile (in the first row) and \citet{salaun2022scalable} to generate blue noise face stippling (in the second row).
Our network reconstructs the underlying correlation and density which can be edited to obtain new synthesized points with spatially-varying correlation. 
Note that our network can only faithfully reconstruct the correlations which are covered by the latent space.

\subsection{User study}
\paragraph{Usefulness experiment}
\label{sec:usefulness_study}
In this experiment, we first explain the concept of density (\lspace-channel) and correlation (\abspace-channel) to the users so that they are able to pick correlation from our correlation palette and density as otherwise they are able to pick color by switching between the \lspace- and \abspace-channels, a built-in function of Photoshop that can natively work in LAB mode.

\bibliographystyle{ACM-Reference-Format}
\bibliography{paper}

%% file: commands.tex
\usepackage{overpic}
\usepackage{pict2e} 
\usepackage{stmaryrd}
\usepackage{newfloat}
\usepackage{savesym}
\usepackage{wrapfig}
\usepackage{bbm}
\usepackage{flushend}
\usepackage{tabularx}
\usepackage{mathrsfs}
\usepackage[most]{tcolorbox}
\usepackage[a]{esvect}
\usepackage[outline]{contour}
\usepackage{lipsum,enumitem}
\usepackage{soul}
\usepackage{microtype}
\usepackage{amsmath}
\usepackage{amsfonts}
\usepackage{booktabs}
\usepackage{hyperref}
\usepackage[capitalize]{cleveref}
\usepackage{multirow}
\usepackage{graphicx}
\usepackage{tikz}
\usepackage{dsfont}
\usepackage{pifont}
\usepackage{xcolor}
\usepackage{wrapfig}
\usepackage{overpic}  
\usepackage{nicefrac}
\usepackage{ctable}
\usepackage{natbib}
\usepackage{xspace}
\usepackage{algorithm}
\usepackage[noend]{algpseudocode} 
\usepackage{siunitx}

\newcommand{\revision}[1]{\textcolor[rgb]{0.0, 0.0, 0.0}{{#1}}}

\newcommand{\mywfigurevspace}[4]{%
\begin{wrapfigure}{r}{#2\columnwidth}%
 \vspace{#4}%
    \centering
    \includegraphics[width=#2\columnwidth]{#1}%
    \vspace{-.2cm}%
    \caption{#3}%
    \label{fig:#1}%
    \vspace{-1.0cm}%
\end{wrapfigure}%
}

\newcommand{\mywfigure}[3]{%
\begin{wrapfigure}{r}{#2\columnwidth}%
 \vspace{-.2cm}%
    \centering
    \includegraphics[width=#2\columnwidth]{#1}%
    \vspace{-.2cm}%
    \caption{#3}%
    \label{fig:#1}%
    \vspace{-.2cm}%
\end{wrapfigure}%
}

\newcommand{\mywfiguretext}[3]{%
\begin{wrapfigure}{r}{#2\columnwidth}%
 \vspace{-.3cm}%
    \centering
    \begin{overpic}[width=#2\columnwidth]{images/#1}
        \put(42, 32){\point} 
    \end{overpic}
    \vspace{-.5cm}%
    \caption{#3}%
    \label{fig:#1}%
    \vspace{-.3cm}%
\end{wrapfigure}%
}

\soulregister\cref7
\soulregister\cite7
\soulregister\cite7
\soulregister\shortcite7
\soulregister\eg0
\soulregister\ie0
\soulregister\etal0

\newcommand{\mysection}[2]{\section{#1}\label{sec:#2}}
\newcommand{\mysubsection}[2]{\subsection{#1}\label{sec:#2}}

\newcommand{\refSec}[1]{Sec.~\ref{sec:#1}}
\newcommand{\refFig}[1]{Fig.~\ref{fig:#1}}
\newcommand{\refEq}[1]{Eq.~\ref{eq:#1}}

\newcommand{\ie}{i.e.,\ }
\newcommand{\eg}{e.g.,\ }

\newcommand{\mymath}[2]{\newcommand{#1}{\TextOrMath{$#2$\xspace}{#2}}}

\newcommand{\unsure}[1]{{\sethlcolor{yellow}\hl{#1}}}

\usepackage{xr}
\makeatletter
\newcommand*{\addFileDependency}[1]{
  \typeout{(#1)}
  \@addtofilelist{#1}
  \IfFileExists{#1}{}{\typeout{No file #1.}}
}
\makeatother

\newcommand*{\myexternaldocument}[1]{%
    \externaldocument[supplemental-]{#1}%
    \addFileDependency{#1.tex}%
    \addFileDependency{#1.aux}%
}

\setcitestyle{square,nosort}
\makeatletter
\renewcommand*{\NAT@spacechar}{~}
\makeatother

\crefname{section}{Sec.}{Secs.}
\Crefname{section}{Sec.}{Secs.}

\graphicspath{{images/}}

\soulregister\ref7
\soulregister\cite7
\soulregister\citeetal7
\soulregister\refSec7
\soulregister\refFig7
\soulregister\refTbl7
\soulregister\refEq7
\soulregister\cite7
\soulregister\ref7
\soulregister\pageref7
\soulregister\shortcite7
\soulregister\eg7
\soulregister\ie7
\soulregister\etal7
\soulregister\unsure7

\DeclareGraphicsExtensions{.png,.jpg,.pdf,.ai,.psd}
\usepackage[nolist]{acronym}

%% file: acronyms.tex
\begin{acronym}
\acro{NN}{Neural Network}
\acro{CNN}{Convolutional Neural Network}
\acro{MC}{Monte Carlo}
\acro{MLP}{Multi-Layered Perceptron}
\acro{PCF}{Pair Correlation Function}
\acro{PCA}{Principal Component Analysis}
\acro{MDS}{Multidimensional scaling}
\acro{IDW}{Inverse Distance Weighted}
\acro{LR}{Learning Rate}
\end{acronym}

\newcommand{\name}{Patternshop}

%% file: math.tex
 \newcommand{\estimate}[1]{\hat{#1}}
 
\mymath{\density}d
\mymath{\pcf}g
\mymath{\pcfEstimate}{\estimate\pcf}
\mymath{\pcfTarget}{\bar\pcf}
\mymath{\pointPattern}P
\mymath{\kernel}{\kappa}\mymath{\densithKernel}{\kernel_\mathrm d}
\mymath{\correlationKernel}{\kernel_\mathrm g}
\mymath{\point}{\mathbf x}
\mymath{\location}{\mathbf y}
\mymath{\jitterlocation}{\mathbf q}
\mymath{\locationSpaced}{\location_{\hphantom{i}}}
\mymath{\guidance}{\mathcal H}
\mymath{\spatial}{\mathcal S}
\mymath{\pcfBandwidth}{\sigma}
\mymath{\editedPointPattern}{{\bar{\pointPattern}}}
\mymath{\pointCount}n
\mymath{\pcfSampleCount}m
\mymath{\radius}r
\mymath{\pcfRmax}{r_\mathrm{max}}
\mymath{\normal}{\mathcal N}
\mymath{\bandwidth}{\sigma}
\mymath{\intensity}{I}
\mymath{\inverseIntensity}{s}
\mymath{\densityScaling}{\alpha}
\mymath{\guide}h
\mymath{\guideBandwidth}{\Sigma}
\mymath{\densityMap}D
\mymath{\pcfMap}G
\mymath{\latentMap}Z
\mymath{\featureMap}F
\mymath{\editedFeatureMap}{\bar{\featureMap}}
\mymath{\edit}{\mathtt{edit}}
\mymath{\editedDensityMap}{\bar\densityMap}
\mymath{\editedPCFMap}{\bar\pcfMap}
\mymath{\editedLatentMap}{\bar\latentMap}
\mymath{\synthesisCost}{c_\mathrm{Syn}}
\mymath{\synthesisCostEstimate}{\estimate{c}_\mathrm{Syn}}
\mymath{\visualFeatureStats}{\mathbf v}
\mymath{\targetVisualFeatureStats}{\bar{\mathbf v}}
\mymath{\perceptualDistance}{\mathcal D}
\mymath{\latentCoordinate}{\mathbf z}
\mymath{\networkLossPixel}{\mathcal L_1}
\mymath{\networkLossGAN}{\mathcal L_\mathrm{adv}}
\mymath{\networkLossTotal}{\mathcal L_\mathrm{tot}}
\mymath{\locality}{\phi}
\mymath{\embeddingCost}{c_\mathrm{Emb}}
\mymath{\encoding}{f}
\mymath{\decoding}{f^{-1}}
\mymath{\pixelResolutionWidth}{256}
\mymath{\pixelResolutionHeight}{256}
\mymath{\learnignRate}{\lambda}

\mymath{\PowerSpectrum}{p}
\mymath{\PowerSpectrumBin}{b}
\mymath{\PowerSpectrumScale}{w}
\mymath{\PowerSpectrumGamma}{\gamma}
\mymath{\NumGMM}{n_\mathrm{GMM}}
\mymath{\mean}{\mu}
\mymath{\stddev}{\sigma}

\mymath{\cielab}{CIELAB}
\mymath{\lspace}{L}
\mymath{\aspace}{A}
\mymath{\bspace}{B}
\mymath{\abspace}{AB}

\mymath{\argmin}{\operatorname{arg\,min}}

%% file: images/teaser.tex

\newcommand{\PlotSingleImage}[1]{%
    \begin{tikzpicture}[scale=1.05]
        \begin{scope}
            \clip (0,0) -- (4.175,0) -- (4.175,4.175) -- (0,4.175) -- cycle;
            \path[fill overzoom image=images/manual_edits/#1] (0,0) rectangle (4.175cm,4.175cm);
        \end{scope}
        \draw (0,0) -- (4.175,0) -- (4.175,4.175) -- (0,4.175) -- cycle;
        \begin{scope}
        \end{scope}
    \end{tikzpicture}%
}

\small
\hspace*{-2.5mm}
\begin{tabular}{c@{\;}c@{\;}c@{\;}c@{\;}c@{}}
\PlotSingleImage{art_00056_ours_dencorr.jpg}
&
\PlotSingleImage{art_00057_ours_points_n29501_lowres.png}
&
\begin{tikzpicture}
    \draw[black, thick, dashed] (0,0) -- (0,4.4);
\end{tikzpicture}
&
\PlotSingleImage{art_00061_ours_dencorr.jpg}
&
\PlotSingleImage{art_00061_ours_points_n23629_lowres.png}
\\
(a) & (b) & & (c) & (d)
\end{tabular}

%% file: images/siggraph_logo/fig_siggraph_logo_ours.tex

\newcommand{\PlotSingleImage}[1]{%
        \begin{scope}
            \clip (0,0) -- (2.5,0) -- (2.5,2.5) -- (0,2.5) -- cycle;
            \path[fill overzoom image=images/siggraph_logo/#1] (0,0) rectangle (2.5cm,2.5cm);
        \end{scope}
        \draw (0,0) -- (2.5,0) -- (2.5,2.5) -- (0,2.5) -- cycle;
}

\newcommand{\TwoColumnFigure}[2]{%
        \hspace*{-1.75mm}
        \begin{tikzpicture}[scale=1.155]
            \PlotSingleImage{#1}
        \end{tikzpicture}
         & 
         \begin{tikzpicture}[scale=1.155]
            \PlotSingleImage{#2}
        \end{tikzpicture}
}

\small
\hspace*{-4mm}
\begin{tabular}{c@{\;}c@{\;}c@{}}
Varying density and correlation
&
Varying density, constant correlation
&
Constant density, varying correlation
\\
\begin{tikzpicture}[scale=2.35]
    \PlotSingleImage{logo_varyingDensityCorrelation_ours_points_n8126_lowres.png}
\end{tikzpicture}
&
\begin{tikzpicture}[scale=2.35]
    \PlotSingleImage{logo_varyingDensity_ours_points_n8126_lowres.png}
\end{tikzpicture}
&
\begin{tikzpicture}[scale=2.35]
    \PlotSingleImage{logo_varyingCorrelation_ours_points_n8352_lowres.png}
\end{tikzpicture}
\\
\begin{tabular}{c@{\;}c@{}}
    \TwoColumnFigure{logo_varyingDensityCorrelation_l.png}{logo_varyingDensityCorrelation_ab.jpg}
    \\
    Varying density & Varying correlation
\end{tabular}
&
\begin{tabular}{c@{\;}c@{}}
    \TwoColumnFigure{logo_varyingDensity_l.png}{logo_varyingDensity_ab.jpg}
    \\
    Varying density & Constant correlation
\end{tabular}
&
\begin{tabular}{c@{\;}c@{}}
    \TwoColumnFigure{logo_varyingCorrelation_l.png}{logo_varyingCorrelation_ab.jpg}
    \\
    Constant density & Varying correlation
\end{tabular}
\\
\end{tabular}

%% file: images/our_approach/embedding_overview_powspectra.tex
\begin{overpic}[scale=1.46,grid=false]{images/our_approach/embedding_overview_powspectra_four}
    \footnotesize
    \put(18,2.3){\pcfSampleCount} 
    \put(0.25,-1){(a) Radial power spectrum profiles}
    \put(24,-1){(b) Realized point sets}
    \put(45,-1){(c) VGG feature maps}
    \put(65,-1){feature vectors}
    \put(78,-1){Dissimilarity matrix}
    \put(92,-1){2D latent space}
    \footnotesize
    \put(39.2, 10){VGG}
    \put(87.0, 9.9){MDS}
    \put(63,-1){(d)}
    \put(76,-1){(e)}
    \put(90.5,-1){(f)}
\end{overpic}

%% file: images/manual_edits/fig_manual_edits.tex

\newcommand{\PlotSingleImage}[1]{%
    \begin{tikzpicture}[scale=1.05]
        \begin{scope}
            \clip (0,0) -- (4.175,0) -- (4.175,4.175) -- (0,4.175) -- cycle;
            \path[fill overzoom image=images/manual_edits/#1] (0,0) rectangle (4.175cm,4.175cm);
        \end{scope}
        \draw (0,0) -- (4.175,0) -- (4.175,4.175) -- (0,4.175) -- cycle;
        \begin{scope}
        \end{scope}
    \end{tikzpicture}%
}

\small
\hspace*{-2.5mm}
\begin{tabular}{c@{\;}c@{\;}c@{\;}c@{\;}c@{}}
%
\\
\PlotSingleImage{art_00020_ours_dencorr.jpg}
&
\PlotSingleImage{art_00020_ours_points_n11246_lowres.png}
&
\begin{tikzpicture}
    \draw[black, thick, dashed] (0,0) -- (0,4.4);
\end{tikzpicture}
&
\PlotSingleImage{art_00021_ours_dencorr.jpg}
&
\PlotSingleImage{art_00021_ours_points_n11463_lowres.png}
\\
(a) & (b) & & (c) & (d)

\end{tabular}

%% file: images/face_editing_final/fig_face_editing_manual.tex

\newcommand{\PlotSingleImage}[1]{%
    \begin{tikzpicture}[scale=1.05]
        \begin{scope}
            \clip (0,0) -- (4.2,0) -- (4.2,4.2) -- (0,4.2) -- cycle;
            \path[fill overzoom image=images/#1] (0,0) rectangle (4.2cm,4.2cm);
        \end{scope}
        \draw (0,0) -- (4.2,0) -- (4.2,4.2) -- (0,4.2) -- cycle;
        \begin{scope}
        \end{scope}
    \end{tikzpicture}%
}

\small
\hspace*{-2mm}
\begin{tabular}{c@{\;}c@{\;}c@{\;}c@{}}
Input points & Network output & Our editing & Synthesized points
\\
\PlotSingleImage{face_editing_final/face5_00105_input_points_lowres.png}
&
\PlotSingleImage{face_editing_final/face5_00105_edit00001_ours_dencorr.jpg}
&
\PlotSingleImage{face_editing_final/face5_00105_edit00003_ours_dencorr.jpg}
&
\PlotSingleImage{face_editing_final/face5_00105_edit00003_ours_points_lowres.png}
\\
\PlotSingleImage{editing_other_categories/afhq_00050_input_points_lowres.png}
&
\PlotSingleImage{editing_other_categories/afhq_00050_edit00001_ours_dencorr.jpg}
&
\PlotSingleImage{editing_other_categories/afhq_00050_edit00002_ours_dencorr.jpg}
&
\PlotSingleImage{editing_other_categories/afhq_00050_edit00002_ours_points_lowres.png}
\\
\PlotSingleImage{editing_other_categories/church_00035_input_points_lowres.png}
&
\PlotSingleImage{editing_other_categories/church_00035_edit00001_ours_dencorr.jpg}
&
\PlotSingleImage{editing_other_categories/church_00035_edit00004_ours_dencorr.jpg}
&
\PlotSingleImage{editing_other_categories/church_00035_edit00004_ours_points_lowres.png}
\end{tabular}

%% file: images/face_editing_final/fig_face_editing_neural.tex

\newcommand{\PlotSingleImage}[1]{%
    \begin{tikzpicture}[scale=1.05]
        \begin{scope}
            \clip (0,0) -- (4.2,0) -- (4.2,4.2) -- (0,4.2) -- cycle;
            \path[fill overzoom image=images/face_editing_final/#1] (0,0) rectangle (4.2cm,4.2cm);
        \end{scope}
        \draw (0,0) -- (4.2,0) -- (4.2,4.2) -- (0,4.2) -- cycle;
        \begin{scope}
        \end{scope}
    \end{tikzpicture}%
}

\small
\hspace*{-2mm}
\begin{tabular}{c@{\;}c@{\;}c@{\;}c@{}}
Input points & Output synthesized points & Input points & Output synthesized points
\\
\PlotSingleImage{face5_00104_input_points_lowres.png}
&
\PlotSingleImage{face5_00104_edit00003_ours_points_lowres.png}
&  
\PlotSingleImage{face5_00103_input_points_lowres.png}
&
\PlotSingleImage{face5_00103_edit00006_ours_points_lowres.png}
\\
\end{tabular}

%% file: images/texture_synthesis/fig_texture_synthesis_comparison.tex

\begin{overpic}[scale=2.55,grid=false]{images/texture_synthesis/texture_synthesis_comparison_tu_lowres}
    \put(-2,16){\rotatebox{90}{Input points}}
    \put(-2,1){\rotatebox{90}{Network output}}
    \put(19,28.7){After editing operation}
    \put(50,28.7){Our edit-aware output}
    \put(80,28.7){\citet{tu2019pointpattern}}
\end{overpic}

%% file: images/comparison_cengiz/fig_comparison_with_cengiz.tex

\newcommand{\PlotSingleImage}[1]{%
    \begin{tikzpicture}[scale=1]
        \begin{scope}
            \clip (0,0) -- (4.2,0) -- (4.2,4.2) -- (0,4.2) -- cycle;
            \path[fill overzoom image=images/comparison_cengiz/#1] (0,0) rectangle (4.2cm,4.2cm);
        \end{scope}
        \draw (0,0) -- (4.2,0) -- (4.2,4.2) -- (0,4.2) -- cycle;
        \begin{scope}
        \end{scope}
    \end{tikzpicture}%
}

\newcommand{\PlotSingleHalfImage}[1]{%
    \begin{tikzpicture}[scale=1]
        \begin{scope}
            \clip (0,0) -- (4.2,0) -- (4.2,2.1) -- (0,2.1) -- cycle;
            \path[fill overzoom image=images/comparison_cengiz/#1] (0,0) rectangle (4.2cm,4.2cm);
        \end{scope}
        \draw (0,0) -- (4.2,0) -- (4.2,2.1) -- (0,2.1) -- cycle;
        \begin{scope}
        \end{scope}
    \end{tikzpicture}%
}

    \newcommand{\PlaceImage}[5]{
    \path[fill overzoom image={#1}] (#2,#3) rectangle ++(#4,#5);
    \draw[black] (#2,#3) rectangle ++(#4,#5);
}

\newcommand{\PlotPCFWithInset}[4]{
    \!\!
    \begin{tikzpicture}[scale=1]
        \begin{scope}
            \clip (0.0,0) -- (4.2,0) -- (4.2,2.1) -- (0,2.1) -- cycle;
            \path[fill overzoom image=images/comparison_cengiz/#1] (0,0) rectangle (6.2cm,2.1cm);
        \end{scope}
        \begin{scope}
         \draw[gray, thick] (0,0.01) -- (4.2,0.01);
         \draw[gray, thick] (0,0.01) -- (0,2.0);
        \end{scope}
        \begin{scope}
        \scriptsize
        \filldraw[thick] (0.1, -0.25) circle (0pt) node[anchor=base,rotate=0] {0};
        \filldraw[thick] (2.1, -0.25) circle (0pt) node[anchor=base,rotate=0] {1};
        \filldraw[thick] (4.1, -0.25) circle (0pt) node[anchor=base,rotate=0] {2};
        \filldraw[thick] (0.12, 0.5) circle (0pt) node[anchor=base,rotate=0] {1};
        \end{scope}
        \begin{scope}
            \PlaceImage{images/comparison_cengiz/#2}{3.15}{1}{1}{1}
            \draw[black](#3,#4) circle (0.06);
        \end{scope}
    \end{tikzpicture}
}

\small
\hspace*{-2mm}
\begin{tabular}{c@{\;}c@{\;}c@{\;}c@{\;}c@{}}
& Input & Ours (Gram) & Ours (PCF 1K) & \citet{oztireli2012analysis} (PCF 8K)
\\
&
\PlotSingleHalfImage{/pattern171_target_n1024_points_lowres.png}
&
\PlotSingleHalfImage{/pattern171_gram_n1024_points_lowres.png}
&
\PlotSingleHalfImage{/pattern171_pcf_n1024_points_lowres.png}
&
\PlotSingleHalfImage{/pattern171_pcfcengiz_n1024_points_lowres.png}
\\
\rotatebox{90}{\qquad\qquad PCF}
&
\PlotPCFWithInset{/pattern171_target_n1024_pcf.png}{/pattern171_coordinate_reference.jpg}{3.35}{1.4}
&
\PlotPCFWithInset{/pattern171_gram_n1024_pcf.png}{/pattern171_coordinate_ours_gram.jpg}{3.36}{1.45}
&
\PlotPCFWithInset{/pattern171_pcf_n1024_pcf.png}{/pattern171_coordinate_ours_pcf.jpg}{3.55}{1.43}
&
\PlotPCFWithInset{/pattern171_pcfcengiz_n1024_pcf.png}{/pattern171_coordinate_ours_pcf.jpg}{3.55}{1.43}
\\
&
\PlotSingleHalfImage{/pattern68_target_n1024_points_lowres.png}
&
\PlotSingleHalfImage{/pattern68_gram_n1024_points_lowres.png}
&
\PlotSingleHalfImage{/pattern68_pcf_n1024_points_lowres.png}
&
\PlotSingleHalfImage{/pattern68_pcfcengiz_n1024_points_lowres.png}
\\
\rotatebox{90}{\qquad\qquad PCF}
&
\PlotPCFWithInset{/pattern68_target_n1024_pcf.png}{/pattern68_coordinate_reference.jpg}{3.35}{1.18}
&
\PlotPCFWithInset{/pattern68_gram_n1024_pcf.png}{/pattern68_coordinate_ours_gram.jpg}{3.33}{1.14}
&
\PlotPCFWithInset{/pattern68_pcf_n1024_pcf.png}{/pattern68_coordinate_ours_pcf.jpg}{3.25}{1.27}
&
\PlotPCFWithInset{/pattern68_pcfcengiz_n1024_pcf.png}{/pattern68_coordinate_ours_pcf.jpg}{3.25}{1.27}
\\
\end{tabular}

%% file: images/comparison_roveri/fig_comparison_roveri.tex

\newcommand{\PlotSingleImage}[1]{%
    \begin{tikzpicture}[scale=1.05]
        \begin{scope}
            \clip (0,0) -- (4.0,0) -- (4.0,4.0) -- (0,4.0) -- cycle;
            \path[fill overzoom image=images/comparison_roveri/#1] (0,0) rectangle (4.0cm,4.0cm);
        \end{scope}
        \draw (0,0) -- (4.0,0) -- (4.0,4.0) -- (0,4.0) -- cycle;
    \end{tikzpicture}%
}

\newcommand{\ImageWithZoomInBox}[3]{%
    \begin{tikzpicture}[scale=1.05]
        \begin{scope}
            \clip (0,0) -- (4.0,0) -- (4.0,4.0) -- (0,4.0) -- cycle;
            \path[fill overzoom image=images/comparison_roveri/#1] (0,0) rectangle (4.0cm,4.0cm);
        \end{scope}
        \draw (0,0) -- (4.0,0) -- (4.0,4.0) -- (0,4.0) -- cycle;
        \begin{scope}
            \draw[#2] (0.1,1.2) rectangle (1.1,1.9);
            \draw[#3] (2.9,0.5) rectangle (3.9,1.2);
        \end{scope}
    \end{tikzpicture}%
}

\newcommand{\ZoominImage}[2]{%
    \begin{tikzpicture}[scale=1.05]
        \begin{scope}
            \clip (0,0) -- (4,0) -- (4,2.0) -- (0,2.0) -- cycle;
            \path[fill overzoom image=images/comparison_roveri/#1] (0,0) rectangle (4.0cm,2.0cm);
        \end{scope}
        \draw[#2] (0,0) -- (4.0,0) -- (4.0,2.0) -- (0,2.0) -- cycle;
        
    \end{tikzpicture}%
}

\hspace*{-3mm}
\begin{tabular}{c@{\;}c@{}}
\citet{roveri2017general} & Ours 
\\
\ImageWithZoomInBox{logo_roveri_points_n10000_lowres.png}{red}{blue}
&
\PlotSingleImage{logo_ours_points_n10000_lowres.png}
\\
\ZoominImage{roveri_points_n10000_lowres_crop1.jpg}{red}
&
\ZoominImage{ours_points_n10000_lowres_crop1.jpg}{red}
\\
\ZoominImage{roveri_points_n10000_lowres_crop5.jpg}{blue}
&
\ZoominImage{ours_points_n10000_lowres_crop5.jpg}{blue}

\end{tabular}

%% file: images/user_study/fig_user_study_points_vs_chroma.tex

\newcommand{\PlotSingleImage}[1]{%
    \begin{tikzpicture}[scale=1.0]
        \begin{scope}
            \clip (0,0) -- (4.2,0) -- (4.2,4.2) -- (0,4.2) -- cycle;
            \path[fill overzoom image=images/user_study/#1] (0,0) rectangle (4.2cm,4.2cm);
        \end{scope}
        \draw (0,0) -- (4.2,0) -- (4.2,4.2) -- (0,4.2) -- cycle;
        \begin{scope}
        \end{scope}
    \end{tikzpicture}%
}

\small
\hspace*{-2mm}
\begin{tabular}{c@{\;}c@{\;}c@{\;}c@{\;}c@{}}
\PlotSingleImage{user_study_2_pointset.png}
&
\PlotSingleImage{user_study_2_chroma.png}
\\
Latent space as points (see \cref{fig:mds_space}) & Chroma (\abspace-channel)

\end{tabular}

%% file: images/user_study/fig_user_study_edits.tex

\newcommand{\PlotSingleImage}[1]{%
    \begin{tikzpicture}[scale=1.05]
        \begin{scope}
            \clip (0,0) -- (3.3,0) -- (3.3,3.3) -- (0,3.3) -- cycle;
            \path[fill overzoom image=images/user_study/#1] (0,0) rectangle (3.3cm,3.3cm);
        \end{scope}
        \draw (0,0) -- (3.3,0) -- (3.3,3.3) -- (0,3.3) -- cycle;
        \begin{scope}
        \end{scope}
    \end{tikzpicture}%
}

\small
\hspace*{-2mm}
\begin{tabular}{c@{\;}c@{\;}c@{\;}c@{\;}c@{}}
Input & Target  & User 1 & User 2 & User 3
\\
\PlotSingleImage{init_points_lowres.png}
&
\PlotSingleImage{reference_points_lowres.png}
&
\PlotSingleImage{user1_points_lowres.png}
&
\PlotSingleImage{user2_points_lowres.png}
&
\PlotSingleImage{user3_points_lowres.png}
\\
\PlotSingleImage{init_edit.jpg}
&
\PlotSingleImage{reference_edit.jpg}
&
\PlotSingleImage{user1_edit.jpg}
&
\PlotSingleImage{user2_edit.jpg}
&
\PlotSingleImage{user3_edit.jpg}
\end{tabular}

%% file: images/training_data/training_data.tex

\newcommand{\PlotSingleImage}[1]{%
    \begin{tikzpicture}[scale=1.05]
        \begin{scope}
            \clip (0,0) -- (3.3,0) -- (3.3,3.3) -- (0,3.3) -- cycle;
            \path[fill overzoom image=images/training_data/#1] (0,0) rectangle (3.3cm,3.3cm);
        \end{scope}
        \draw (0,0) -- (3.3,0) -- (3.3,3.3) -- (0,3.3) -- cycle;
        \begin{scope}
        \end{scope}
    \end{tikzpicture}%
}

\small
\hspace*{-2mm}
\begin{tabular}{c@{\;}c@{\;}c@{\;}c@{\;}c@{}}
%
\PlotSingleImage{face5_07433_dencorr.jpg}
&
\PlotSingleImage{face5_15980_dencorr.jpg}
&
\PlotSingleImage{afhq_00041_dencorr.jpg}
&
\PlotSingleImage{church_00036_dencorr.jpg}
&
\PlotSingleImage{tcd2_04883_dencorr.jpg}
\\
\PlotSingleImage{face5_07433_points_lowres.png}
&
\PlotSingleImage{face5_15980_points_lowres.png}
&
\PlotSingleImage{afhq_00041_points_lowres.png}
&
\PlotSingleImage{church_00036_points_lowres.png}
&
\PlotSingleImage{tcd2_04883_points_lowres.png}
\end{tabular}

%% file: images/point_correlation_design/fig_point_correlation_design.tex

\newcommand{\PlotSingleImage}[1]{%
    \begin{tikzpicture}[scale=0.98]
        \begin{scope}
            \clip (0,0) -- (4.2,0) -- (4.2,4.2) -- (0,4.2) -- cycle;
            \path[fill overzoom image=images/point_correlation_design/#1] (0,0) rectangle (4.2cm,4.2cm);
        \end{scope}
        \draw (0,0) -- (4.2,0) -- (4.2,4.2) -- (0,4.2) -- cycle;
        \begin{scope}
        \end{scope}
    \end{tikzpicture}%
}

\newcommand{\PlotImageWithRectangleInset}[2]{
\begin{tikzpicture}[scale=0.98]
    \begin{scope}
        \clip (0.0,0.0) -- (4.2,0.0) -- (4.2,4.2) -- (0.0,4.2) -- cycle;
        \path[fill overzoom image=images/point_correlation_design/#1] (0,0) rectangle (4.2cm,4.2cm);
    \end{scope}
    \draw (0,0) -- (4.2,0) -- (4.2,4.2) -- (0,4.2) -- cycle;
    \begin{scope}
        \clip (2,0.0) -- (4.1,0.0) -- (4.1,1.3) -- (2,1.3) -- cycle;
        \path[fill overzoom image=images/point_correlation_design/#2] (2,0.0) rectangle (5.0cm,1.0cm);
    \end{scope}
\end{tikzpicture}
}

\newcommand{\PlotImageWithSquareInset}[2]{
\begin{tikzpicture}[scale=0.98]
    \begin{scope}
        \clip (0.0,0.0) -- (4.2,0.0) -- (4.2,4.2) -- (0.0,4.2) -- cycle;
        \path[fill overzoom image=images/point_correlation_design/#1] (0,0) rectangle (4.2cm,4.2cm);
    \end{scope}
    \draw (0,0) -- (4.2,0) -- (4.2,4.2) -- (0,4.2) -- cycle;
    \begin{scope}
        \clip (2.9,0.1) -- (4.1,0.1) -- (4.1,1.3) -- (2.9,1.3) -- cycle;
        \path[fill overzoom image=images/point_correlation_design/#2] (2.9,0.1) rectangle (4.1cm,1.3cm);
    \end{scope}
    \begin{scope}
        \draw[black,thick] (2.9,0.1) -- (4.1,0.1) -- (4.1,1.3) -- (2.9,1.3) -- cycle;
    \end{scope}
\end{tikzpicture}
}

\small
\hspace*{-2mm}
\begin{tabular}{c@{\;}c@{}}
\PlotSingleImage{adrian.png}
&
\PlotImageWithSquareInset{adrian_greengns_points_n33806_lowres.png}{greenspectrum_lowerexposure.PNG}
\\
(a) Input image & (b) \citet{zhou2012point}
\\
\PlotImageWithRectangleInset{adrian_cengizgreen_points_n33800_lowres.png}{adrian_greenpcf_pcf_n33800.png}
&
\PlotImageWithSquareInset{adrian_oursgreen_points_n33800_lowres.png}{green.PNG}
\\
(c) \citet{oztireli2012analysis} & (d) Ours (green)
\\
\PlotImageWithSquareInset{adrian_oursblue_points_n33800_v2_lowres.png}{blue.PNG}
&
\PlotImageWithSquareInset{adrian_oursbluegreen_points_n33800_varyingdotsize_v2_lowres.png}{adrian_oursbluegreen_v2.png}
\\
(e) Ours (blue) & (f) Ours (blue \& green)
\\
\end{tabular}

%% file: images/knownnoises.tex

\newcommand{\EditingOneImage}[1]{
    \begin{scope}
        \clip (0.05,0.05) -- (4.45,0.05) -- (4.45,4.45) -- (0.05,4.45) -- cycle;
        \path[fill overzoom image=images/ablation_study/known_noises/#1] (0,0) rectangle (4.5cm,4.5cm);
    \end{scope}
    \begin{scope}
    \end{scope}
}

\small
\hspace*{-2mm}
\begin{tabular}{c@{\;}c@{\;}c@{\;}c@{}}
Input points
\\
\begin{tikzpicture}[scale=1]
    \EditingOneImage{bnot_target_n1024_points.png}
\end{tikzpicture}
&
\begin{tikzpicture}[scale=1]
    \EditingOneImage{green_target_n1024_points.png}
\end{tikzpicture}
&
\begin{tikzpicture}[scale=1]
    \EditingOneImage{pink_target_n1024_points.png}
\end{tikzpicture}
&
\begin{tikzpicture}[scale=1]
    \EditingOneImage{step_target_n1024_points.png}
\end{tikzpicture}
\\
Colors
\\
\begin{tikzpicture}[scale=1]
    \EditingOneImage{bnot_gram_n1024_color.jpg}
\end{tikzpicture}
&
\begin{tikzpicture}[scale=1]
    \EditingOneImage{green_gram_n1024_color.jpg}
\end{tikzpicture}
&
\begin{tikzpicture}[scale=1]
    \EditingOneImage{pink_gram_n1024_color.jpg}
\end{tikzpicture}
&
\begin{tikzpicture}[scale=1]
    \EditingOneImage{step_gram_n1024_color.jpg}
\end{tikzpicture}
\\
Re-synthesized points
\\
\begin{tikzpicture}[scale=1]
    \EditingOneImage{bnot_gram_n1024_points.png}
\end{tikzpicture}
&
\begin{tikzpicture}[scale=1]
    \EditingOneImage{green_gram_n1024_points.png}
\end{tikzpicture}
&
\begin{tikzpicture}[scale=1]
    \EditingOneImage{pink_gram_n1024_points.png}
\end{tikzpicture}
&
\begin{tikzpicture}[scale=1]
    \EditingOneImage{step_gram_n1024_points.png}
\end{tikzpicture}
\end{tabular}

%% file: images/ablation_study/fig_ablation_study.tex

\newcommand{\PlotSingleImage}[1]{%
    \begin{tikzpicture}[scale=1.0]
        \begin{scope}
            \clip (0,0) -- (3.5,0) -- (3.5,3.5) -- (0,3.5) -- cycle;
            \path[fill overzoom image=images/ablation_study/#1] (0,0) rectangle (3.5cm,3.5cm);
        \end{scope}
        \draw (0,0) -- (3.5,0) -- (3.5,3.5) -- (0,3.5) -- cycle;
        \begin{scope}
        \end{scope}
    \end{tikzpicture}%
}

\small
\hspace*{-2mm}
\begin{tabular}{c@{\;}c@{\;}c@{\;}c@{\;}c@{}}
\PlotSingleImage{face5_00214_input_points_lowres.png}
&
\PlotSingleImage{face5_00214_edit00001_dencorr_kde.jpg}
&
\PlotSingleImage{face5_00214_points_kde_lowres.png}
&
\PlotSingleImage{face5_00214_edit00001_dencorr_ours.jpg}
&
\PlotSingleImage{face5_00214_points_ours_lowres.png}
\\
Input points & with KDE & Synthesized points (KDE) & Our network & Synthesized points
\\
\PlotSingleImage{face5_00166_input_points_lowres.png}
&
\PlotSingleImage{face5_00166_edit00001_dencorr_loss_wo_gan.jpg}
&
\PlotSingleImage{face5_00166_points_loss_wo_gan_lowres.png}
&
\PlotSingleImage{face5_00166_edit00001_dencorr_loss_w_gan.jpg}
&
\PlotSingleImage{face5_00166_points_loss_w_gan_lowres.png}
\\
Input points & without \networkLossGAN & Synthesized points & with \networkLossGAN & Synthesized points 
\end{tabular}

%% file: images/face_editing_existing_methods/fig_face_editing_existing_methods.tex

\newcommand{\PlotSingleImage}[1]{%
    \begin{tikzpicture}[scale=1.05]
        \begin{scope}
            \clip (0,0) -- (4.2,0) -- (4.2,4.2) -- (0,4.2) -- cycle;
            \path[fill overzoom image=images/face_editing_existing_methods/#1] (0,0) rectangle (4.2cm,4.2cm);
        \end{scope}
        \draw (0,0) -- (4.2,0) -- (4.2,4.2) -- (0,4.2) -- cycle;
        \begin{scope}
        \end{scope}
    \end{tikzpicture}%
}

\small
\hspace*{-2mm}
\begin{tabular}{c@{\;}c@{\;}c@{\;}c@{}}
Input points & Network output & Our editing & Synthesized points
\\
\PlotSingleImage{face5_01181_input_points_lowres.png}
&
\PlotSingleImage{face5_01181_edit00001_ours_dencorr.jpg}
&
\PlotSingleImage{face5_01181_edit00002_ours_dencorr.jpg}
&
\PlotSingleImage{face5_01181_edit00002_ours_points_lowres.png}
\\
\PlotSingleImage{face5_00112_edit00001_ours_points_lowres.png}
&
\PlotSingleImage{face5_00112_edit00001_ours_dencorr.jpg}
&
\PlotSingleImage{face5_00112_edit00002_ours_dencorr.jpg}
&
\PlotSingleImage{face5_00112_edit00002_ours_points_lowres.png}
\end{tabular}

%% file: paper.bbl

\begin{thebibliography}{93}


\ifx \showCODEN    \undefined \def \showCODEN     #1{\unskip}     \fi
\ifx \showDOI      \undefined \def \showDOI       #1{#1}\fi
\ifx \showISBNx    \undefined \def \showISBNx     #1{\unskip}     \fi
\ifx \showISBNxiii \undefined \def \showISBNxiii  #1{\unskip}     \fi
\ifx \showISSN     \undefined \def \showISSN      #1{\unskip}     \fi
\ifx \showLCCN     \undefined \def \showLCCN      #1{\unskip}     \fi
\ifx \shownote     \undefined \def \shownote      #1{#1}          \fi
\ifx \showarticletitle \undefined \def \showarticletitle #1{#1}   \fi
\ifx \showURL      \undefined \def \showURL       {\relax}        \fi
\providecommand\bibfield[2]{#2}
\providecommand\bibinfo[2]{#2}
\providecommand\natexlab[1]{#1}
\providecommand\showeprint[2][]{arXiv:#2}

\bibitem[\protect\citeauthoryear{Abdal, Qin, and Wonka}{Abdal
  et~al\mbox{.}}{2019}]%
        {abdal2019image2stylegan}
\bibfield{author}{\bibinfo{person}{Rameen Abdal}, \bibinfo{person}{Yipeng Qin},
  {and} \bibinfo{person}{Peter Wonka}.} \bibinfo{year}{2019}\natexlab{}.
\newblock \showarticletitle{Image2stylegan: How to embed images into the
  stylegan latent space?}. In \bibinfo{booktitle}{\emph{ICCV}}.
  \bibinfo{pages}{4432--4441}.
\newblock


\bibitem[\protect\citeauthoryear{Ahmed, Guo, Yan, Franceschia, Zhang, and
  Deussen}{Ahmed et~al\mbox{.}}{2017}]%
        {ahmed2017simple}
\bibfield{author}{\bibinfo{person}{Abdalla~GM Ahmed}, \bibinfo{person}{Jianwei
  Guo}, \bibinfo{person}{Dong-Ming Yan}, \bibinfo{person}{Jean-Yves
  Franceschia}, \bibinfo{person}{Xiaopeng Zhang}, {and} \bibinfo{person}{Oliver
  Deussen}.} \bibinfo{year}{2017}\natexlab{}.
\newblock \showarticletitle{A simple push-pull algorithm for blue-noise
  sampling}.
\newblock \bibinfo{journal}{\emph{IEEE Trans. Vis and Comp. Graph.}}
  \bibinfo{volume}{23}, \bibinfo{number}{12} (\bibinfo{year}{2017}).
\newblock


\bibitem[\protect\citeauthoryear{Ahmed, Huang, and Deussen}{Ahmed
  et~al\mbox{.}}{2015}]%
        {ahmed2015aa}
\bibfield{author}{\bibinfo{person}{Abdalla~GM Ahmed}, \bibinfo{person}{Hui
  Huang}, {and} \bibinfo{person}{Oliver Deussen}.}
  \bibinfo{year}{2015}\natexlab{}.
\newblock \showarticletitle{AA patterns for point sets with controlled spectral
  properties}.
\newblock \bibinfo{journal}{\emph{ACM Trans. Graph.}} \bibinfo{volume}{34},
  \bibinfo{number}{6} (\bibinfo{year}{2015}).
\newblock


\bibitem[\protect\citeauthoryear{Ahmed, Perrier, Coeurjolly, Ostromoukhov, Guo,
  Yan, Huang, and Deussen}{Ahmed et~al\mbox{.}}{2016}]%
        {ahmed2016low}
\bibfield{author}{\bibinfo{person}{Abdalla~GM Ahmed},
  \bibinfo{person}{H{\'e}l{\`e}ne Perrier}, \bibinfo{person}{David Coeurjolly},
  \bibinfo{person}{Victor Ostromoukhov}, \bibinfo{person}{Jianwei Guo},
  \bibinfo{person}{Dong-Ming Yan}, \bibinfo{person}{Hui Huang}, {and}
  \bibinfo{person}{Oliver Deussen}.} \bibinfo{year}{2016}\natexlab{}.
\newblock \showarticletitle{Low-discrepancy blue noise sampling}.
\newblock \bibinfo{journal}{\emph{ACM Trans. Graph.}} \bibinfo{volume}{35},
  \bibinfo{number}{6} (\bibinfo{year}{2016}).
\newblock


\bibitem[\protect\citeauthoryear{An, Tong, Denning, and Pellacini}{An
  et~al\mbox{.}}{2011}]%
        {an2011appwarp}
\bibfield{author}{\bibinfo{person}{Xiaobo An}, \bibinfo{person}{Xin Tong},
  \bibinfo{person}{Jonathan~D. Denning}, {and} \bibinfo{person}{Fabio
  Pellacini}.} \bibinfo{year}{2011}\natexlab{}.
\newblock \showarticletitle{AppWarp: Retargeting Measured Materials by
  Appearance-Space Warping}.
\newblock \bibinfo{journal}{\emph{ACM Trans. Graph.}} \bibinfo{volume}{30},
  \bibinfo{number}{6} (\bibinfo{year}{2011}), \bibinfo{pages}{1–10}.
\newblock
\showISSN{0730-0301}


\bibitem[\protect\citeauthoryear{Balzer, Schl{\"o}mer, and Deussen}{Balzer
  et~al\mbox{.}}{2009}]%
        {balzer2009capacity}
\bibfield{author}{\bibinfo{person}{Michael Balzer}, \bibinfo{person}{Thomas
  Schl{\"o}mer}, {and} \bibinfo{person}{Oliver Deussen}.}
  \bibinfo{year}{2009}\natexlab{}.
\newblock \showarticletitle{Capacity-constrained point distributions: a variant
  of Lloyd's method}.
\newblock \bibinfo{journal}{\emph{ACM Trans. Graph.}} \bibinfo{volume}{28},
  \bibinfo{number}{3} (\bibinfo{year}{2009}).
\newblock


\bibitem[\protect\citeauthoryear{Bertalmio, Sapiro, Caselles, and
  Ballester}{Bertalmio et~al\mbox{.}}{2000}]%
        {bertalmio2000inpainting}
\bibfield{author}{\bibinfo{person}{Marcelo Bertalmio},
  \bibinfo{person}{Guillermo Sapiro}, \bibinfo{person}{Vincent Caselles}, {and}
  \bibinfo{person}{Coloma Ballester}.} \bibinfo{year}{2000}\natexlab{}.
\newblock \showarticletitle{Image Inpainting}. In
  \bibinfo{booktitle}{\emph{Proc. SIGGRAPH}}. \bibinfo{pages}{417–424}.
\newblock


\bibitem[\protect\citeauthoryear{Büttner and Kosztra}{Büttner and
  Kosztra}{2017}]%
        {buettner2017european}
\bibfield{author}{\bibinfo{person}{György Büttner} {and}
  \bibinfo{person}{Barbara Kosztra}.} \bibinfo{year}{2017}\natexlab{}.
\newblock \bibinfo{booktitle}{\emph{CLC2018 Technical Guidelines}}.
\newblock \bibinfo{type}{{T}echnical {R}eport}. \bibinfo{institution}{European
  Environment Agency}.
\newblock


\bibitem[\protect\citeauthoryear{Chen, Ge, Wei, Wang, Wang, Wang, Fei, Qian,
  Yong, and Wang}{Chen et~al\mbox{.}}{2013}]%
        {chen2013bilateral}
\bibfield{author}{\bibinfo{person}{Jiating Chen}, \bibinfo{person}{Xiaoyin Ge},
  \bibinfo{person}{Li-Yi Wei}, \bibinfo{person}{Bin Wang},
  \bibinfo{person}{Yusu Wang}, \bibinfo{person}{Huamin Wang},
  \bibinfo{person}{Yun Fei}, \bibinfo{person}{Kang-Lai Qian},
  \bibinfo{person}{Jun-Hai Yong}, {and} \bibinfo{person}{Wenping Wang}.}
  \bibinfo{year}{2013}\natexlab{}.
\newblock \showarticletitle{Bilateral blue noise sampling}.
\newblock \bibinfo{journal}{\emph{ACM Trans. Graph.}} \bibinfo{volume}{32},
  \bibinfo{number}{6} (\bibinfo{year}{2013}), \bibinfo{pages}{1--11}.
\newblock


\bibitem[\protect\citeauthoryear{Chen, Paris, and Durand}{Chen
  et~al\mbox{.}}{2007}]%
        {chen2007real}
\bibfield{author}{\bibinfo{person}{Jiawen Chen}, \bibinfo{person}{Sylvain
  Paris}, {and} \bibinfo{person}{Fr{\'e}do Durand}.}
  \bibinfo{year}{2007}\natexlab{}.
\newblock \showarticletitle{Real-time edge-aware image processing with the
  bilateral grid}.
\newblock \bibinfo{journal}{\emph{ACM Trans. Graph.}} \bibinfo{volume}{26},
  \bibinfo{number}{3} (\bibinfo{year}{2007}), \bibinfo{pages}{103--113}.
\newblock


\bibitem[\protect\citeauthoryear{Choi, Uh, Yoo, and Ha}{Choi
  et~al\mbox{.}}{2020}]%
        {choi2020stargan}
\bibfield{author}{\bibinfo{person}{Yunjey Choi}, \bibinfo{person}{Youngjung
  Uh}, \bibinfo{person}{Jaejun Yoo}, {and} \bibinfo{person}{Jung-Woo Ha}.}
  \bibinfo{year}{2020}\natexlab{}.
\newblock \showarticletitle{Stargan v2: Diverse image synthesis for multiple
  domains}. In \bibinfo{booktitle}{\emph{Proceedings of the IEEE/CVF conference
  on computer vision and pattern recognition}}. \bibinfo{pages}{8188--8197}.
\newblock


\bibitem[\protect\citeauthoryear{ClipDrop}{ClipDrop}{2023}]%
        {ClipDrop}
\bibfield{author}{\bibinfo{person}{ClipDrop}.} \bibinfo{year}{2023}\natexlab{}.
\newblock \bibinfo{title}{Relight API}.
\newblock
\newblock
\urldef\tempurl%
\url{https://clipdrop.co/relight}
\showURL{%
\tempurl}


\bibitem[\protect\citeauthoryear{Cook}{Cook}{1986}]%
        {cook1986stochastic}
\bibfield{author}{\bibinfo{person}{Robert~L Cook}.}
  \bibinfo{year}{1986}\natexlab{}.
\newblock \showarticletitle{Stochastic sampling in computer graphics}.
\newblock \bibinfo{journal}{\emph{ACM Trans. Graph.}} \bibinfo{volume}{5},
  \bibinfo{number}{1} (\bibinfo{year}{1986}).
\newblock


\bibitem[\protect\citeauthoryear{De~Goes, Breeden, Ostromoukhov, and
  Desbrun}{De~Goes et~al\mbox{.}}{2012}]%
        {degoes2012blue}
\bibfield{author}{\bibinfo{person}{Fernando De~Goes},
  \bibinfo{person}{Katherine Breeden}, \bibinfo{person}{Victor Ostromoukhov},
  {and} \bibinfo{person}{Mathieu Desbrun}.} \bibinfo{year}{2012}\natexlab{}.
\newblock \showarticletitle{Blue noise through optimal transport}.
\newblock \bibinfo{journal}{\emph{ACM Trans. Graph.}} \bibinfo{volume}{31},
  \bibinfo{number}{6} (\bibinfo{year}{2012}).
\newblock


\bibitem[\protect\citeauthoryear{Deussen, Hiller, Van~Overveld, and
  Strothotte}{Deussen et~al\mbox{.}}{2000}]%
        {deussen2000floating}
\bibfield{author}{\bibinfo{person}{Oliver Deussen}, \bibinfo{person}{Stefan
  Hiller}, \bibinfo{person}{Cornelius Van~Overveld}, {and}
  \bibinfo{person}{Thomas Strothotte}.} \bibinfo{year}{2000}\natexlab{}.
\newblock \showarticletitle{Floating points: A method for computing stipple
  drawings}. In \bibinfo{booktitle}{\emph{Comp. Graph. Forum}},
  Vol.~\bibinfo{volume}{19}. \bibinfo{pages}{41--50}.
\newblock


\bibitem[\protect\citeauthoryear{Deussen and Isenberg}{Deussen and
  Isenberg}{2013}]%
        {deussen2013halftoning}
\bibfield{author}{\bibinfo{person}{Oliver Deussen} {and}
  \bibinfo{person}{Tobias Isenberg}.} \bibinfo{year}{2013}\natexlab{}.
\newblock \showarticletitle{Halftoning and Stippling}.
\newblock In \bibinfo{booktitle}{\emph{Image and Video-Based Artistic
  Stylisation}}, \bibfield{editor}{\bibinfo{person}{Paul Rosin} {and}
  \bibinfo{person}{John Collomosse}} (Eds.). \bibinfo{publisher}{Springer
  London}, \bibinfo{pages}{45--61}.
\newblock


\bibitem[\protect\citeauthoryear{Di~Renzo, Calabrese, and Pellacini}{Di~Renzo
  et~al\mbox{.}}{2014}]%
        {renzo2014appim}
\bibfield{author}{\bibinfo{person}{Francesco Di~Renzo},
  \bibinfo{person}{Claudio Calabrese}, {and} \bibinfo{person}{Fabio
  Pellacini}.} \bibinfo{year}{2014}\natexlab{}.
\newblock \showarticletitle{AppIm: Linear Spaces for Image-Based Appearance
  Editing}.
\newblock \bibinfo{journal}{\emph{ACM Trans. Graph.}} \bibinfo{volume}{33},
  \bibinfo{number}{6} (\bibinfo{year}{2014}).
\newblock
\showISSN{0730-0301}


\bibitem[\protect\citeauthoryear{Dipp\'{e} and Wold}{Dipp\'{e} and
  Wold}{1985}]%
        {dippe1985antialiasing}
\bibfield{author}{\bibinfo{person}{Mark A.~Z. Dipp\'{e}} {and}
  \bibinfo{person}{Erling~Henry Wold}.} \bibinfo{year}{1985}\natexlab{}.
\newblock \showarticletitle{Antialiasing through Stochastic Sampling}. In
  \bibinfo{booktitle}{\emph{SIGGRAPH}}. \bibinfo{pages}{69–78}.
\newblock


\bibitem[\protect\citeauthoryear{Duncan and Humphreys}{Duncan and
  Humphreys}{1989}]%
        {duncan1989visual}
\bibfield{author}{\bibinfo{person}{John Duncan} {and} \bibinfo{person}{Glyn~W
  Humphreys}.} \bibinfo{year}{1989}\natexlab{}.
\newblock \showarticletitle{Visual search and stimulus similarity.}
\newblock \bibinfo{journal}{\emph{Psychological review}} \bibinfo{volume}{96},
  \bibinfo{number}{3} (\bibinfo{year}{1989}), \bibinfo{pages}{433}.
\newblock


\bibitem[\protect\citeauthoryear{Durand}{Durand}{2011}]%
        {durand2011frequency}
\bibfield{author}{\bibinfo{person}{Fr\'edo Durand}.}
  \bibinfo{year}{2011}\natexlab{}.
\newblock \bibinfo{booktitle}{\emph{A frequency analysis of {Monte}-{Carlo} and
  other numerical integration schemes}}.
\newblock \bibinfo{type}{{T}echnical {R}eport} TR-2011-052.
  \bibinfo{institution}{MIT CSAIL}.
\newblock


\bibitem[\protect\citeauthoryear{Efros and Freeman}{Efros and Freeman}{2001}]%
        {efros2001image}
\bibfield{author}{\bibinfo{person}{Alexei~A. Efros} {and}
  \bibinfo{person}{William~T. Freeman}.} \bibinfo{year}{2001}\natexlab{}.
\newblock \showarticletitle{Image Quilting for Texture Synthesis and Transfer}.
  In \bibinfo{booktitle}{\emph{SIGGRAPH}}. \bibinfo{pages}{341–346}.
\newblock


\bibitem[\protect\citeauthoryear{Emilien, Vimont, Cani, Poulin, and
  Benes}{Emilien et~al\mbox{.}}{2015}]%
        {emilien2015worldbrush}
\bibfield{author}{\bibinfo{person}{Arnaud Emilien}, \bibinfo{person}{Ulysse
  Vimont}, \bibinfo{person}{Marie-Paule Cani}, \bibinfo{person}{Pierre Poulin},
  {and} \bibinfo{person}{Bedrich Benes}.} \bibinfo{year}{2015}\natexlab{}.
\newblock \showarticletitle{Worldbrush: Interactive example-based synthesis of
  procedural virtual worlds}.
\newblock \bibinfo{journal}{\emph{ACM Trans. Graph.}} \bibinfo{volume}{34},
  \bibinfo{number}{4} (\bibinfo{year}{2015}), \bibinfo{pages}{1--11}.
\newblock


\bibitem[\protect\citeauthoryear{Fairchild}{Fairchild}{2013}]%
        {fairchild2013color}
\bibfield{author}{\bibinfo{person}{Mark~D Fairchild}.}
  \bibinfo{year}{2013}\natexlab{}.
\newblock \bibinfo{booktitle}{\emph{Color appearance models}}.
\newblock \bibinfo{publisher}{John Wiley \& Sons}.
\newblock


\bibitem[\protect\citeauthoryear{Fattal}{Fattal}{2011}]%
        {fattal2011blue}
\bibfield{author}{\bibinfo{person}{Raanan Fattal}.}
  \bibinfo{year}{2011}\natexlab{}.
\newblock \showarticletitle{Blue-noise point sampling using kernel density
  model}.
\newblock \bibinfo{journal}{\emph{ACM Trans. Graph.}} \bibinfo{volume}{30},
  \bibinfo{number}{4} (\bibinfo{year}{2011}).
\newblock


\bibitem[\protect\citeauthoryear{Gatys, Ecker, and Bethge}{Gatys
  et~al\mbox{.}}{2016}]%
        {gatys2016styletransfer}
\bibfield{author}{\bibinfo{person}{Leon~A. Gatys},
  \bibinfo{person}{Alexander~S. Ecker}, {and} \bibinfo{person}{Matthias
  Bethge}.} \bibinfo{year}{2016}\natexlab{}.
\newblock \showarticletitle{Image Style Transfer Using Convolutional Neural
  Networks}. In \bibinfo{booktitle}{\emph{CVPR}}. \bibinfo{pages}{2414--2423}.
\newblock


\bibitem[\protect\citeauthoryear{Guerrero, Bernstein, Li, and Mitra}{Guerrero
  et~al\mbox{.}}{2016}]%
        {guerrero2016patex}
\bibfield{author}{\bibinfo{person}{Paul Guerrero}, \bibinfo{person}{Gilbert
  Bernstein}, \bibinfo{person}{Wilmot Li}, {and} \bibinfo{person}{Niloy~J.
  Mitra}.} \bibinfo{year}{2016}\natexlab{}.
\newblock \showarticletitle{PATEX: Exploring Pattern Variations}.
\newblock \bibinfo{journal}{\emph{ACM Trans. Graph.}} \bibinfo{volume}{35},
  \bibinfo{number}{4} (\bibinfo{year}{2016}).
\newblock
\showISSN{0730-0301}


\bibitem[\protect\citeauthoryear{Heck, Schl{\"o}mer, and Deussen}{Heck
  et~al\mbox{.}}{2013}]%
        {heck2013blue}
\bibfield{author}{\bibinfo{person}{Daniel Heck}, \bibinfo{person}{Thomas
  Schl{\"o}mer}, {and} \bibinfo{person}{Oliver Deussen}.}
  \bibinfo{year}{2013}\natexlab{}.
\newblock \showarticletitle{Blue noise sampling with controlled aliasing}.
\newblock \bibinfo{journal}{\emph{ACM Trans. Graph. (Proc. SIGGRAPH)}}
  \bibinfo{volume}{32}, \bibinfo{number}{3} (\bibinfo{year}{2013}).
\newblock


\bibitem[\protect\citeauthoryear{Henzler, Mitra, , and Ritschel}{Henzler
  et~al\mbox{.}}{2019}]%
        {henzler2020neuraltexture}
\bibfield{author}{\bibinfo{person}{Philipp Henzler}, \bibinfo{person}{Niloy~J
  Mitra}, \bibinfo{person}{}, {and} \bibinfo{person}{Tobias Ritschel}.}
  \bibinfo{year}{2019}\natexlab{}.
\newblock \showarticletitle{Learning a Neural 3D Texture Space from 2D
  Exemplars}. In \bibinfo{booktitle}{\emph{CVPR}}.
\newblock


\bibitem[\protect\citeauthoryear{Hermosilla, Ritschel, V\'{a}zquez, Vinacua,
  and Ropinski}{Hermosilla et~al\mbox{.}}{2018}]%
        {hermosilla2018montecarlo}
\bibfield{author}{\bibinfo{person}{Pedro Hermosilla}, \bibinfo{person}{Tobias
  Ritschel}, \bibinfo{person}{Pere-Pau V\'{a}zquez}, \bibinfo{person}{\`{A}lvar
  Vinacua}, {and} \bibinfo{person}{Timo Ropinski}.}
  \bibinfo{year}{2018}\natexlab{}.
\newblock \showarticletitle{Monte Carlo Convolution for Learning on
  Non-uniformly Sampled Point Clouds}.
\newblock \bibinfo{journal}{\emph{ACM Trans. Graph (Proc. SIGGRAPH Asia)}}
  \bibinfo{volume}{37}, \bibinfo{number}{5} (\bibinfo{year}{2018}).
\newblock


\bibitem[\protect\citeauthoryear{Hertzmann, Jacobs, Oliver, CNOurless, and
  Salesin}{Hertzmann et~al\mbox{.}}{2001}]%
        {hertzmann2001image}
\bibfield{author}{\bibinfo{person}{Aaron Hertzmann}, \bibinfo{person}{Charles~E
  Jacobs}, \bibinfo{person}{Nuria Oliver}, \bibinfo{person}{Brian CNOurless},
  {and} \bibinfo{person}{David~H Salesin}.} \bibinfo{year}{2001}\natexlab{}.
\newblock \showarticletitle{Image analogies}. In
  \bibinfo{booktitle}{\emph{SIGGRAPH}}. \bibinfo{pages}{327--340}.
\newblock


\bibitem[\protect\citeauthoryear{Hsu, Wei, You, and Zhang}{Hsu
  et~al\mbox{.}}{2020}]%
        {hsu2020autocomplete}
\bibfield{author}{\bibinfo{person}{Chen-Yuan Hsu}, \bibinfo{person}{Li-Yi Wei},
  \bibinfo{person}{Lihua You}, {and} \bibinfo{person}{Jian~Jun Zhang}.}
  \bibinfo{year}{2020}\natexlab{}.
\newblock \showarticletitle{Autocomplete element fields}. In
  \bibinfo{booktitle}{\emph{Proc. CHI}}. \bibinfo{pages}{1--13}.
\newblock


\bibitem[\protect\citeauthoryear{Huang, Memari, Seidel, and Singh}{Huang
  et~al\mbox{.}}{2022}]%
        {huang2022point}
\bibfield{author}{\bibinfo{person}{Xingchang Huang}, \bibinfo{person}{Pooran
  Memari}, \bibinfo{person}{Hans-Peter Seidel}, {and} \bibinfo{person}{Gurprit
  Singh}.} \bibinfo{year}{2022}\natexlab{}.
\newblock \showarticletitle{Point-Pattern Synthesis using Gabor and Random
  Filters}. In \bibinfo{booktitle}{\emph{Comp. Graph. Forum}},
  Vol.~\bibinfo{volume}{41}. \bibinfo{pages}{169--179}.
\newblock


\bibitem[\protect\citeauthoryear{Isola, Zhu, Zhou, and Efros}{Isola
  et~al\mbox{.}}{2017}]%
        {isola2017imagetoimage}
\bibfield{author}{\bibinfo{person}{Phillip Isola}, \bibinfo{person}{Jun-Yan
  Zhu}, \bibinfo{person}{Tinghui Zhou}, {and} \bibinfo{person}{Alexei~A
  Efros}.} \bibinfo{year}{2017}\natexlab{}.
\newblock \showarticletitle{Image-to-Image Translation with Conditional
  Adversarial Networks}.
\newblock \bibinfo{journal}{\emph{CVPR}} (\bibinfo{year}{2017}).
\newblock


\bibitem[\protect\citeauthoryear{Jarabo, Masia, Bousseau, Pellacini, and
  Gutierrez}{Jarabo et~al\mbox{.}}{2014}]%
        {jarabo2014how}
\bibfield{author}{\bibinfo{person}{Adrian Jarabo}, \bibinfo{person}{Belen
  Masia}, \bibinfo{person}{Adrien Bousseau}, \bibinfo{person}{Fabio Pellacini},
  {and} \bibinfo{person}{Diego Gutierrez}.} \bibinfo{year}{2014}\natexlab{}.
\newblock \showarticletitle{How Do People Edit Light Fields?}
\newblock \bibinfo{journal}{\emph{ACM Trans. Graph.}} \bibinfo{volume}{33},
  \bibinfo{number}{4} (\bibinfo{year}{2014}).
\newblock
\showISSN{0730-0301}


\bibitem[\protect\citeauthoryear{Jensen}{Jensen}{2001}]%
        {jensen2001realistic}
\bibfield{author}{\bibinfo{person}{Henrik~Wann Jensen}.}
  \bibinfo{year}{2001}\natexlab{}.
\newblock \bibinfo{booktitle}{\emph{Realistic image synthesis using photon
  mapping}}.
\newblock \bibinfo{publisher}{AK Peters/crc Press}.
\newblock


\bibitem[\protect\citeauthoryear{Kailkhura, Thiagarajan, Bremer, and
  Varshney}{Kailkhura et~al\mbox{.}}{2016}]%
        {kailkhura2016stair}
\bibfield{author}{\bibinfo{person}{Bhavya Kailkhura},
  \bibinfo{person}{Jayaraman~J Thiagarajan}, \bibinfo{person}{Peer-Timo
  Bremer}, {and} \bibinfo{person}{Pramod~K Varshney}.}
  \bibinfo{year}{2016}\natexlab{}.
\newblock \showarticletitle{Stair blue noise sampling}.
\newblock \bibinfo{journal}{\emph{ACM Trans. Graph.}} \bibinfo{volume}{35},
  \bibinfo{number}{6} (\bibinfo{year}{2016}).
\newblock


\bibitem[\protect\citeauthoryear{Kapp, Gain, Gu{\'e}rin, Galin, and
  Peytavie}{Kapp et~al\mbox{.}}{2020}]%
        {kapp2020data}
\bibfield{author}{\bibinfo{person}{Konrad Kapp}, \bibinfo{person}{James Gain},
  \bibinfo{person}{Eric Gu{\'e}rin}, \bibinfo{person}{Eric Galin}, {and}
  \bibinfo{person}{Adrien Peytavie}.} \bibinfo{year}{2020}\natexlab{}.
\newblock \showarticletitle{Data-driven authoring of large-scale ecosystems}.
\newblock \bibinfo{journal}{\emph{ACM Trans. Graph.}} \bibinfo{volume}{39},
  \bibinfo{number}{6} (\bibinfo{year}{2020}), \bibinfo{pages}{1--14}.
\newblock


\bibitem[\protect\citeauthoryear{Kim, Maciejewski, Isenberg, Andrews, Chen,
  Sousa, and Ebert}{Kim et~al\mbox{.}}{2009}]%
        {kim2009stippling}
\bibfield{author}{\bibinfo{person}{Sung~Ye Kim}, \bibinfo{person}{Ross
  Maciejewski}, \bibinfo{person}{Tobias Isenberg}, \bibinfo{person}{William~M.
  Andrews}, \bibinfo{person}{Wei Chen}, \bibinfo{person}{Mario~Costa Sousa},
  {and} \bibinfo{person}{David~S. Ebert}.} \bibinfo{year}{2009}\natexlab{}.
\newblock \showarticletitle{Stippling by Example}. In
  \bibinfo{booktitle}{\emph{Proc. NPAR}}. \bibinfo{pages}{41–50}.
\newblock


\bibitem[\protect\citeauthoryear{Kingma and Ba}{Kingma and Ba}{2014}]%
        {kingma2014adam}
\bibfield{author}{\bibinfo{person}{Diederik~P Kingma} {and}
  \bibinfo{person}{Jimmy Ba}.} \bibinfo{year}{2014}\natexlab{}.
\newblock \showarticletitle{Adam: A method for stochastic optimization}.
\newblock \bibinfo{journal}{\emph{arXiv preprint arXiv:1412.6980}}
  (\bibinfo{year}{2014}).
\newblock


\bibitem[\protect\citeauthoryear{Knutsson and Westin}{Knutsson and
  Westin}{1993}]%
        {knutsson1993normalized}
\bibfield{author}{\bibinfo{person}{Hans Knutsson} {and} \bibinfo{person}{C-F
  Westin}.} \bibinfo{year}{1993}\natexlab{}.
\newblock \showarticletitle{Normalized and differential convolution}. In
  \bibinfo{booktitle}{\emph{CVPR}}. \bibinfo{pages}{515--523}.
\newblock


\bibitem[\protect\citeauthoryear{Kopf, Cohen-Or, Deussen, and Lischinski}{Kopf
  et~al\mbox{.}}{2006}]%
        {kopf2006recursive}
\bibfield{author}{\bibinfo{person}{Johannes Kopf}, \bibinfo{person}{Daniel
  Cohen-Or}, \bibinfo{person}{Oliver Deussen}, {and} \bibinfo{person}{Dani
  Lischinski}.} \bibinfo{year}{2006}\natexlab{}.
\newblock \showarticletitle{Recursive Wang tiles for real-time blue noise}.
\newblock \bibinfo{journal}{\emph{ACM Trans. Graph. (Proc. SIGGRAPH)}}
  \bibinfo{volume}{25}, \bibinfo{number}{3} (\bibinfo{year}{2006}).
\newblock


\bibitem[\protect\citeauthoryear{Lagae and Dutre}{Lagae and Dutre}{2008}]%
        {lagae2008}
\bibfield{author}{\bibinfo{person}{Ares Lagae} {and} \bibinfo{person}{Philip
  Dutre}.} \bibinfo{year}{2008}\natexlab{}.
\newblock \showarticletitle{A Comparison of Methods for Generating Poisson Disk
  Distributions}.
\newblock \bibinfo{journal}{\emph{Comp. Graph. Forum}} \bibinfo{volume}{27},
  \bibinfo{number}{1} (\bibinfo{year}{2008}).
\newblock


\bibitem[\protect\citeauthoryear{Lau, Arce, and Gallagher}{Lau
  et~al\mbox{.}}{1999}]%
        {lau99digital}
\bibfield{author}{\bibinfo{person}{Daniel~L. Lau}, \bibinfo{person}{Gonzalo~R.
  Arce}, {and} \bibinfo{person}{Neal~C. Gallagher}.}
  \bibinfo{year}{1999}\natexlab{}.
\newblock \showarticletitle{Digital halftoning by means of green-noise masks}.
\newblock \bibinfo{journal}{\emph{J OSA}} \bibinfo{volume}{16},
  \bibinfo{number}{7} (\bibinfo{year}{1999}), \bibinfo{pages}{1575--1586}.
\newblock


\bibitem[\protect\citeauthoryear{Lee, Liu, Wu, and Luo}{Lee
  et~al\mbox{.}}{2020}]%
        {CelebAMask-HQ}
\bibfield{author}{\bibinfo{person}{Cheng-Han Lee}, \bibinfo{person}{Ziwei Liu},
  \bibinfo{person}{Lingyun Wu}, {and} \bibinfo{person}{Ping Luo}.}
  \bibinfo{year}{2020}\natexlab{}.
\newblock \showarticletitle{MaskGAN: Towards Diverse and Interactive Facial
  Image Manipulation}. In \bibinfo{booktitle}{\emph{CVPR}}.
\newblock


\bibitem[\protect\citeauthoryear{Leimk\"{u}hler, Singh, Myszkowski, Seidel, and
  Ritschel}{Leimk\"{u}hler et~al\mbox{.}}{2019}]%
        {leimkuhler2019deep}
\bibfield{author}{\bibinfo{person}{Thomas Leimk\"{u}hler},
  \bibinfo{person}{Gurprit Singh}, \bibinfo{person}{Karol Myszkowski},
  \bibinfo{person}{Hans-Peter Seidel}, {and} \bibinfo{person}{Tobias
  Ritschel}.} \bibinfo{year}{2019}\natexlab{}.
\newblock \showarticletitle{Deep Point Correlation Design}.
\newblock \bibinfo{journal}{\emph{ACM Trans. Graph.}} \bibinfo{volume}{38},
  \bibinfo{number}{6} (\bibinfo{year}{2019}).
\newblock
\showISSN{0730-0301}


\bibitem[\protect\citeauthoryear{Lindow, Baum, and Hege}{Lindow
  et~al\mbox{.}}{2012}]%
        {lindow2012perceptually}
\bibfield{author}{\bibinfo{person}{Norbert Lindow}, \bibinfo{person}{Daniel
  Baum}, {and} \bibinfo{person}{Hans-Christian Hege}.}
  \bibinfo{year}{2012}\natexlab{}.
\newblock \showarticletitle{Perceptually linear parameter variations}. In
  \bibinfo{booktitle}{\emph{Comp. Graph. Forum}}, Vol.~\bibinfo{volume}{31}.
  \bibinfo{pages}{535--544}.
\newblock


\bibitem[\protect\citeauthoryear{Liu, Wang, L{\'e}vy, Sun, Yan, Lu, and
  Yang}{Liu et~al\mbox{.}}{2009}]%
        {liu2009centroidal}
\bibfield{author}{\bibinfo{person}{Yang Liu}, \bibinfo{person}{Wenping Wang},
  \bibinfo{person}{Bruno L{\'e}vy}, \bibinfo{person}{Feng Sun},
  \bibinfo{person}{Dong-Ming Yan}, \bibinfo{person}{Lin Lu}, {and}
  \bibinfo{person}{Chenglei Yang}.} \bibinfo{year}{2009}\natexlab{}.
\newblock \showarticletitle{On centroidal Voronoi tessellation -- energy
  smoothness and fast computation}.
\newblock \bibinfo{journal}{\emph{ACM Trans. Graph.}} \bibinfo{volume}{28},
  \bibinfo{number}{4} (\bibinfo{year}{2009}).
\newblock


\bibitem[\protect\citeauthoryear{Lloyd}{Lloyd}{1982}]%
        {lloyd1982least}
\bibfield{author}{\bibinfo{person}{Stuart Lloyd}.}
  \bibinfo{year}{1982}\natexlab{}.
\newblock \showarticletitle{Least squares quantization in {PCM}}.
\newblock \bibinfo{journal}{\emph{IEEE Trans Inform. Theory}}
  \bibinfo{volume}{28}, \bibinfo{number}{2} (\bibinfo{year}{1982}).
\newblock


\bibitem[\protect\citeauthoryear{Ma, Wei, and Tong}{Ma et~al\mbox{.}}{2011}]%
        {ma2011discrete}
\bibfield{author}{\bibinfo{person}{Chongyang Ma}, \bibinfo{person}{Li-Yi Wei},
  {and} \bibinfo{person}{Xin Tong}.} \bibinfo{year}{2011}\natexlab{}.
\newblock \showarticletitle{Discrete Element Textures}.
\newblock \bibinfo{journal}{\emph{ACM Trans. Graph.}} \bibinfo{volume}{30},
  \bibinfo{number}{4} (\bibinfo{year}{2011}).
\newblock
\showISSN{0730-0301}


\bibitem[\protect\citeauthoryear{Mart{\'\i}n, Arroyo, Rodr{\'\i}guez, and
  Isenberg}{Mart{\'\i}n et~al\mbox{.}}{2017}]%
        {martin2017survey}
\bibfield{author}{\bibinfo{person}{Domingo Mart{\'\i}n},
  \bibinfo{person}{Germ{\'a}n Arroyo}, \bibinfo{person}{Alejandro
  Rodr{\'\i}guez}, {and} \bibinfo{person}{Tobias Isenberg}.}
  \bibinfo{year}{2017}\natexlab{}.
\newblock \showarticletitle{A survey of digital stippling}.
\newblock \bibinfo{journal}{\emph{Computers \& Graphics}}  \bibinfo{volume}{67}
  (\bibinfo{year}{2017}), \bibinfo{pages}{24--44}.
\newblock
\showISSN{0097-8493}


\bibitem[\protect\citeauthoryear{Nguyen, Ritschel, and Seidel}{Nguyen
  et~al\mbox{.}}{2015}]%
        {nguyen2015data}
\bibfield{author}{\bibinfo{person}{Chuong~H Nguyen}, \bibinfo{person}{Tobias
  Ritschel}, {and} \bibinfo{person}{Hans-Peter Seidel}.}
  \bibinfo{year}{2015}\natexlab{}.
\newblock \showarticletitle{Data-driven color manifolds}.
\newblock \bibinfo{journal}{\emph{ACM Trans. Graph.}} \bibinfo{volume}{34},
  \bibinfo{number}{2} (\bibinfo{year}{2015}), \bibinfo{pages}{1--9}.
\newblock


\bibitem[\protect\citeauthoryear{Ostromoukhov, Donohue, and
  Jodoin}{Ostromoukhov et~al\mbox{.}}{2004}]%
        {ostromoukhov2004fast}
\bibfield{author}{\bibinfo{person}{Victor Ostromoukhov},
  \bibinfo{person}{Charles Donohue}, {and} \bibinfo{person}{Pierre-Marc
  Jodoin}.} \bibinfo{year}{2004}\natexlab{}.
\newblock \showarticletitle{Fast hierarchical importance sampling with blue
  noise properties}.
\newblock \bibinfo{journal}{\emph{ACM Trans. Graph.}} \bibinfo{volume}{23},
  \bibinfo{number}{3} (\bibinfo{year}{2004}).
\newblock


\bibitem[\protect\citeauthoryear{{\"O}ztireli and Gross}{{\"O}ztireli and
  Gross}{2012}]%
        {oztireli2012analysis}
\bibfield{author}{\bibinfo{person}{A~Cengiz {\"O}ztireli} {and}
  \bibinfo{person}{Markus Gross}.} \bibinfo{year}{2012}\natexlab{}.
\newblock \showarticletitle{Analysis and synthesis of point distributions based
  on pair correlation}.
\newblock \bibinfo{journal}{\emph{ACM Trans. Graph.}} \bibinfo{volume}{31},
  \bibinfo{number}{6} (\bibinfo{year}{2012}).
\newblock


\bibitem[\protect\citeauthoryear{Paszke, Gross, Chintala, Chanan, Yang, DeVito,
  Lin, Desmaison, Antiga, and Lerer}{Paszke et~al\mbox{.}}{2017}]%
        {paszke2017automatic}
\bibfield{author}{\bibinfo{person}{Adam Paszke}, \bibinfo{person}{Sam Gross},
  \bibinfo{person}{Soumith Chintala}, \bibinfo{person}{Gregory Chanan},
  \bibinfo{person}{Edward Yang}, \bibinfo{person}{Zachary DeVito},
  \bibinfo{person}{Zeming Lin}, \bibinfo{person}{Alban Desmaison},
  \bibinfo{person}{Luca Antiga}, {and} \bibinfo{person}{Adam Lerer}.}
  \bibinfo{year}{2017}\natexlab{}.
\newblock \showarticletitle{Automatic differentiation in pytorch}.
\newblock  (\bibinfo{year}{2017}).
\newblock


\bibitem[\protect\citeauthoryear{Pellacini}{Pellacini}{2010}]%
        {pellacini2010envylight}
\bibfield{author}{\bibinfo{person}{Fabio Pellacini}.}
  \bibinfo{year}{2010}\natexlab{}.
\newblock \showarticletitle{envyLight: An Interface for Editing Natural
  Illumination}.
\newblock \bibinfo{journal}{\emph{ACM Trans. Graph. (SIGGRAPH)}}
  (\bibinfo{year}{2010}).
\newblock


\bibitem[\protect\citeauthoryear{Pellacini, Ferwerda, and Greenberg}{Pellacini
  et~al\mbox{.}}{2000}]%
        {pellacini2000toward}
\bibfield{author}{\bibinfo{person}{Fabio Pellacini}, \bibinfo{person}{James~A
  Ferwerda}, {and} \bibinfo{person}{Donald~P Greenberg}.}
  \bibinfo{year}{2000}\natexlab{}.
\newblock \showarticletitle{Toward a psychophysically-based light reflection
  model for image synthesis}. In \bibinfo{booktitle}{\emph{SIGGRAPH}}.
  \bibinfo{pages}{55--64}.
\newblock


\bibitem[\protect\citeauthoryear{Pellacini and Lawrence}{Pellacini and
  Lawrence}{2007}]%
        {pellacini2007appwand}
\bibfield{author}{\bibinfo{person}{Fabio Pellacini} {and}
  \bibinfo{person}{Jason Lawrence}.} \bibinfo{year}{2007}\natexlab{}.
\newblock \showarticletitle{AppWand: Editing Measured Materials Using
  Appearance-Driven Optimization}.
\newblock \bibinfo{journal}{\emph{ACM Trans. Graph.}} \bibinfo{volume}{26},
  \bibinfo{number}{3} (\bibinfo{year}{2007}), \bibinfo{pages}{54–64}.
\newblock
\showISSN{0730-0301}


\bibitem[\protect\citeauthoryear{Petschnigg, Szeliski, Agrawala, Cohen, Hoppe,
  and Toyama}{Petschnigg et~al\mbox{.}}{2004}]%
        {petschnigg2004digital}
\bibfield{author}{\bibinfo{person}{Georg Petschnigg}, \bibinfo{person}{Richard
  Szeliski}, \bibinfo{person}{Maneesh Agrawala}, \bibinfo{person}{Michael
  Cohen}, \bibinfo{person}{Hugues Hoppe}, {and} \bibinfo{person}{Kentaro
  Toyama}.} \bibinfo{year}{2004}\natexlab{}.
\newblock \showarticletitle{Digital photography with flash and no-flash image
  pairs}.
\newblock \bibinfo{journal}{\emph{ACM Trans. Graph.}} \bibinfo{volume}{23},
  \bibinfo{number}{3} (\bibinfo{year}{2004}), \bibinfo{pages}{664--672}.
\newblock


\bibitem[\protect\citeauthoryear{Piovar\v{c}i, Levin, Kaufman, and
  Didyk}{Piovar\v{c}i et~al\mbox{.}}{2018}]%
        {piovarci2018modelling}
\bibfield{author}{\bibinfo{person}{Michal Piovar\v{c}i},
  \bibinfo{person}{David~I.W. Levin}, \bibinfo{person}{Danny Kaufman}, {and}
  \bibinfo{person}{Piotr Didyk}.} \bibinfo{year}{2018}\natexlab{}.
\newblock \showarticletitle{Perception-Aware Modeling and Fabrication of
  Digital Drawing Tools}.
\newblock \bibinfo{journal}{\emph{ACM Trans. Graph. (SIGGRAPH)}}
  \bibinfo{volume}{37}, \bibinfo{number}{4} (\bibinfo{year}{2018}).
\newblock


\bibitem[\protect\citeauthoryear{Pols, Van~der Kamp, and Plomp}{Pols
  et~al\mbox{.}}{1969}]%
        {pols1969perceptual}
\bibfield{author}{\bibinfo{person}{Louis~CW Pols}, \bibinfo{person}{LJ~Th
  Van~der Kamp}, {and} \bibinfo{person}{Reinier Plomp}.}
  \bibinfo{year}{1969}\natexlab{}.
\newblock \showarticletitle{Perceptual and physical space of vowel sounds}.
\newblock \bibinfo{journal}{\emph{J ASA}} \bibinfo{volume}{46},
  \bibinfo{number}{2B} (\bibinfo{year}{1969}), \bibinfo{pages}{458--467}.
\newblock


\bibitem[\protect\citeauthoryear{Portilla and Simoncelli}{Portilla and
  Simoncelli}{2000}]%
        {portilla2000parametric}
\bibfield{author}{\bibinfo{person}{Javier Portilla} {and}
  \bibinfo{person}{Eero~P Simoncelli}.} \bibinfo{year}{2000}\natexlab{}.
\newblock \showarticletitle{A parametric texture model based on joint
  statistics of complex wavelet coefficients}.
\newblock \bibinfo{journal}{\emph{Int. J Computer Vision}}
  \bibinfo{volume}{40} (\bibinfo{year}{2000}), \bibinfo{pages}{49--70}.
\newblock


\bibitem[\protect\citeauthoryear{Qi, Su, Mo, and Guibas}{Qi
  et~al\mbox{.}}{2017}]%
        {qi2017pointnet}
\bibfield{author}{\bibinfo{person}{Charles~R Qi}, \bibinfo{person}{Hao Su},
  \bibinfo{person}{Kaichun Mo}, {and} \bibinfo{person}{Leonidas~J Guibas}.}
  \bibinfo{year}{2017}\natexlab{}.
\newblock \showarticletitle{Pointnet: Deep learning on point sets for {3D}
  classification and segmentation}.
\newblock \bibinfo{journal}{\emph{CVPR}} (\bibinfo{year}{2017}).
\newblock


\bibitem[\protect\citeauthoryear{Qin, Chen, He, and Chen}{Qin
  et~al\mbox{.}}{2017}]%
        {qin2017wasserstein}
\bibfield{author}{\bibinfo{person}{Hongxing Qin}, \bibinfo{person}{Yi Chen},
  \bibinfo{person}{Jinlong He}, {and} \bibinfo{person}{Baoquan Chen}.}
  \bibinfo{year}{2017}\natexlab{}.
\newblock \showarticletitle{Wasserstein Blue Noise Sampling}.
\newblock \bibinfo{journal}{\emph{ACM Trans. Graph.}} \bibinfo{volume}{36},
  \bibinfo{number}{5} (\bibinfo{year}{2017}).
\newblock
\showISSN{0730-0301}


\bibitem[\protect\citeauthoryear{Reddy, Guerrero, Fisher, Li, and Mitra}{Reddy
  et~al\mbox{.}}{2020}]%
        {reddy2020discovering}
\bibfield{author}{\bibinfo{person}{Pradyumna Reddy}, \bibinfo{person}{Paul
  Guerrero}, \bibinfo{person}{Matt Fisher}, \bibinfo{person}{Wilmot Li}, {and}
  \bibinfo{person}{Niloy~J. Mitra}.} \bibinfo{year}{2020}\natexlab{}.
\newblock \showarticletitle{Discovering Pattern Structure Using Differentiable
  Compositing}.
\newblock \bibinfo{journal}{\emph{ACM Trans. Graph.}} \bibinfo{volume}{39},
  \bibinfo{number}{6} (\bibinfo{year}{2020}).
\newblock
\showISSN{0730-0301}


\bibitem[\protect\citeauthoryear{Reinert, Ritschel, and Seidel}{Reinert
  et~al\mbox{.}}{2013}]%
        {reinert2013interactive}
\bibfield{author}{\bibinfo{person}{Bernhard Reinert}, \bibinfo{person}{Tobias
  Ritschel}, {and} \bibinfo{person}{Hans-Peter Seidel}.}
  \bibinfo{year}{2013}\natexlab{}.
\newblock \showarticletitle{Interactive By-Example Design of Artistic Packing
  Layouts}.
\newblock \bibinfo{journal}{\emph{ACM Trans. Graph.}} \bibinfo{volume}{32},
  \bibinfo{number}{6} (\bibinfo{year}{2013}).
\newblock
\showISSN{0730-0301}


\bibitem[\protect\citeauthoryear{Rosin and Collomosse}{Rosin and
  Collomosse}{2012}]%
        {paul2012image}
\bibfield{author}{\bibinfo{person}{Paul Rosin} {and} \bibinfo{person}{John
  Collomosse}.} \bibinfo{year}{2012}\natexlab{}.
\newblock \bibinfo{booktitle}{\emph{Image and Video-Based Artistic
  Stylisation}}.
\newblock \bibinfo{publisher}{Springer Publishing Company, Incorporated}.
\newblock


\bibitem[\protect\citeauthoryear{Rott~Shaham, Dekel, and Michaeli}{Rott~Shaham
  et~al\mbox{.}}{2019}]%
        {rottshaham2019singan}
\bibfield{author}{\bibinfo{person}{Tamar Rott~Shaham}, \bibinfo{person}{Tali
  Dekel}, {and} \bibinfo{person}{Tomer Michaeli}.}
  \bibinfo{year}{2019}\natexlab{}.
\newblock \showarticletitle{SinGAN: Learning a Generative Model from a Single
  Natural Image}. In \bibinfo{booktitle}{\emph{ICCV}}.
\newblock


\bibitem[\protect\citeauthoryear{Roveri, Öztireli, and Gross}{Roveri
  et~al\mbox{.}}{2017}]%
        {roveri2017general}
\bibfield{author}{\bibinfo{person}{Riccardo Roveri}, \bibinfo{person}{A.~Cengiz
  Öztireli}, {and} \bibinfo{person}{Markus Gross}.}
  \bibinfo{year}{2017}\natexlab{}.
\newblock \showarticletitle{General Point Sampling with Adaptive Density and
  Correlations}.
\newblock \bibinfo{journal}{\emph{Comp. Graph. Forum}} \bibinfo{volume}{36},
  \bibinfo{number}{2} (\bibinfo{year}{2017}), \bibinfo{pages}{107--117}.
\newblock


\bibitem[\protect\citeauthoryear{Sala\"{u}n, Georgiev, Seidel, and
  Singh}{Sala\"{u}n et~al\mbox{.}}{2022}]%
        {salaun2022scalable}
\bibfield{author}{\bibinfo{person}{Corentin Sala\"{u}n},
  \bibinfo{person}{Iliyan Georgiev}, \bibinfo{person}{Hans-Peter Seidel}, {and}
  \bibinfo{person}{Gurprit Singh}.} \bibinfo{year}{2022}\natexlab{}.
\newblock \showarticletitle{Scalable Multi-Class Sampling via Filtered Sliced
  Optimal Transport}.
\newblock \bibinfo{journal}{\emph{ACM Trans. Graph. (SIGGRAPH)}}
  \bibinfo{volume}{41}, \bibinfo{number}{6} (\bibinfo{year}{2022}).
\newblock
\showISSN{0730-0301}


\bibitem[\protect\citeauthoryear{Schmaltz, Gwosdek, Bruhn, and
  Weickert}{Schmaltz et~al\mbox{.}}{2010}]%
        {schmaltz2010electrostatic}
\bibfield{author}{\bibinfo{person}{Christian Schmaltz}, \bibinfo{person}{Pascal
  Gwosdek}, \bibinfo{person}{Andres Bruhn}, {and} \bibinfo{person}{Joachim
  Weickert}.} \bibinfo{year}{2010}\natexlab{}.
\newblock \showarticletitle{Electrostatic Halftoning}.
\newblock \bibinfo{journal}{\emph{Comp. Graph. Forum}} (\bibinfo{year}{2010}).
\newblock


\bibitem[\protect\citeauthoryear{Schulz, Kwan, Becher, Baumgartner, Reina,
  Deussen, and Weiskopf}{Schulz et~al\mbox{.}}{2021}]%
        {schulz2021multiclass}
\bibfield{author}{\bibinfo{person}{Christoph Schulz},
  \bibinfo{person}{Kin~Chung Kwan}, \bibinfo{person}{Michael Becher},
  \bibinfo{person}{Daniel Baumgartner}, \bibinfo{person}{Guido Reina},
  \bibinfo{person}{Oliver Deussen}, {and} \bibinfo{person}{Daniel Weiskopf}.}
  \bibinfo{year}{2021}\natexlab{}.
\newblock \showarticletitle{Multi-Class Inverted Stippling}.
\newblock \bibinfo{journal}{\emph{ACM Trans. Graph.}} \bibinfo{volume}{40},
  \bibinfo{number}{6} (\bibinfo{year}{2021}).
\newblock
\showISSN{0730-0301}


\bibitem[\protect\citeauthoryear{Secord}{Secord}{2002}]%
        {secord2002weighted}
\bibfield{author}{\bibinfo{person}{Adrian Secord}.}
  \bibinfo{year}{2002}\natexlab{}.
\newblock \showarticletitle{Weighted {Voronoi} stippling}. In
  \bibinfo{booktitle}{\emph{Proc. NPAR}}.
\newblock


\bibitem[\protect\citeauthoryear{Sendik and Cohen-Or}{Sendik and
  Cohen-Or}{2017}]%
        {sendik2017deep}
\bibfield{author}{\bibinfo{person}{Omry Sendik} {and} \bibinfo{person}{Daniel
  Cohen-Or}.} \bibinfo{year}{2017}\natexlab{}.
\newblock \showarticletitle{Deep Correlations for Texture Synthesis}.
\newblock \bibinfo{journal}{\emph{ACM Trans. Graph.}} \bibinfo{volume}{36},
  \bibinfo{number}{5} (\bibinfo{year}{2017}).
\newblock
\showISSN{0730-0301}


\bibitem[\protect\citeauthoryear{Simoncelli and Olshausen}{Simoncelli and
  Olshausen}{2001}]%
        {simoncelli2001natural}
\bibfield{author}{\bibinfo{person}{Eero~P Simoncelli} {and}
  \bibinfo{person}{Bruno~A Olshausen}.} \bibinfo{year}{2001}\natexlab{}.
\newblock \showarticletitle{Natural image statistics and neural
  representation}.
\newblock \bibinfo{journal}{\emph{Ann. Review Neuroscience}}
  \bibinfo{volume}{24}, \bibinfo{number}{1} (\bibinfo{year}{2001}).
\newblock


\bibitem[\protect\citeauthoryear{Simonyan and Zisserman}{Simonyan and
  Zisserman}{2015}]%
        {simonyan2014vgg19}
\bibfield{author}{\bibinfo{person}{Karen Simonyan} {and}
  \bibinfo{person}{Andrew Zisserman}.} \bibinfo{year}{2015}\natexlab{}.
\newblock \showarticletitle{Very Deep Convolutional Networks for Large-Scale
  Image Recognition}. In \bibinfo{booktitle}{\emph{Proc. ICLR}},
  \bibfield{editor}{\bibinfo{person}{Yoshua Bengio} {and} \bibinfo{person}{Yann
  LeCun}} (Eds.).
\newblock


\bibitem[\protect\citeauthoryear{Singh, Oztireli, Ahmed, Coeurjolly, Subr,
  Deussen, Ostromoukhov, Ramamoorthi, and Jarosz}{Singh et~al\mbox{.}}{2019}]%
        {singh19analysis}
\bibfield{author}{\bibinfo{person}{Gurprit Singh}, \bibinfo{person}{Cengiz
  Oztireli}, \bibinfo{person}{Abdalla~G.M. Ahmed}, \bibinfo{person}{David
  Coeurjolly}, \bibinfo{person}{Kartic Subr}, \bibinfo{person}{Oliver Deussen},
  \bibinfo{person}{Victor Ostromoukhov}, \bibinfo{person}{Ravi Ramamoorthi},
  {and} \bibinfo{person}{Wojciech Jarosz}.} \bibinfo{year}{2019}\natexlab{}.
\newblock \showarticletitle{Analysis of Sample Correlations for Monte Carlo
  Rendering}.
\newblock \bibinfo{journal}{\emph{Comp. Graph Form. (Proc. EGSR)}}
  \bibinfo{volume}{38}, \bibinfo{number}{2} (\bibinfo{year}{2019}).
\newblock


\bibitem[\protect\citeauthoryear{Sochorov{\'a} and
  Jamri{\v{s}}ka}{Sochorov{\'a} and Jamri{\v{s}}ka}{2021}]%
        {sochorova2021practical}
\bibfield{author}{\bibinfo{person}{{\v{S}}{\'a}rka Sochorov{\'a}} {and}
  \bibinfo{person}{Ond{\v{r}}ej Jamri{\v{s}}ka}.}
  \bibinfo{year}{2021}\natexlab{}.
\newblock \showarticletitle{Practical pigment mixing for digital painting}.
\newblock \bibinfo{journal}{\emph{ACM Trans. Graph.}} \bibinfo{volume}{40},
  \bibinfo{number}{6} (\bibinfo{year}{2021}), \bibinfo{pages}{1--11}.
\newblock


\bibitem[\protect\citeauthoryear{Spicker, Hahn, Lindemeier, Saupe, and
  Deussen}{Spicker et~al\mbox{.}}{2017}]%
        {spicker2017quantifying}
\bibfield{author}{\bibinfo{person}{Marc Spicker}, \bibinfo{person}{Franz Hahn},
  \bibinfo{person}{Thomas Lindemeier}, \bibinfo{person}{Dietmar Saupe}, {and}
  \bibinfo{person}{Oliver Deussen}.} \bibinfo{year}{2017}\natexlab{}.
\newblock \showarticletitle{Quantifying Visual Abstraction Quality for Stipple
  Drawings}. In \bibinfo{booktitle}{\emph{Proc. NPAR}}.
\newblock


\bibitem[\protect\citeauthoryear{Strassmann}{Strassmann}{1986}]%
        {strassmann1986hairy}
\bibfield{author}{\bibinfo{person}{Steve Strassmann}.}
  \bibinfo{year}{1986}\natexlab{}.
\newblock \showarticletitle{Hairy brushes}.
\newblock \bibinfo{journal}{\emph{SIGGRAPH}} \bibinfo{volume}{20},
  \bibinfo{number}{4} (\bibinfo{year}{1986}), \bibinfo{pages}{225--232}.
\newblock


\bibitem[\protect\citeauthoryear{Sun, Barron, Tsai, Xu, Yu, Fyffe, Rhemann,
  Busch, Debevec, and Ramamoorthi}{Sun et~al\mbox{.}}{2019}]%
        {sun2019single}
\bibfield{author}{\bibinfo{person}{Tiancheng Sun}, \bibinfo{person}{Jonathan~T.
  Barron}, \bibinfo{person}{Yun-Ta Tsai}, \bibinfo{person}{Zexiang Xu},
  \bibinfo{person}{Xueming Yu}, \bibinfo{person}{Graham Fyffe},
  \bibinfo{person}{Christoph Rhemann}, \bibinfo{person}{Jay Busch},
  \bibinfo{person}{Paul Debevec}, {and} \bibinfo{person}{Ravi Ramamoorthi}.}
  \bibinfo{year}{2019}\natexlab{}.
\newblock \showarticletitle{Single Image Portrait Relighting}.
\newblock \bibinfo{journal}{\emph{ACM Trans. Graph.}} \bibinfo{volume}{38},
  \bibinfo{number}{4} (\bibinfo{year}{2019}).
\newblock
\showISSN{0730-0301}


\bibitem[\protect\citeauthoryear{Tu, Lischinski, and Huang}{Tu
  et~al\mbox{.}}{2019}]%
        {tu2019pointpattern}
\bibfield{author}{\bibinfo{person}{Peihan Tu}, \bibinfo{person}{Dani
  Lischinski}, {and} \bibinfo{person}{Hui Huang}.}
  \bibinfo{year}{2019}\natexlab{}.
\newblock \showarticletitle{Point Pattern Synthesis via Irregular Convolution}.
\newblock \bibinfo{journal}{\emph{Comp. Graph. Forum}}  \bibinfo{volume}{38}
  (\bibinfo{year}{2019}).
\newblock


\bibitem[\protect\citeauthoryear{Ulichney}{Ulichney}{1987}]%
        {ulichney1987digital}
\bibfield{author}{\bibinfo{person}{Robert Ulichney}.}
  \bibinfo{year}{1987}\natexlab{}.
\newblock \bibinfo{booktitle}{\emph{Digital Halftoning}}.
\newblock \bibinfo{publisher}{MIT Press}.
\newblock


\bibitem[\protect\citeauthoryear{Wachtel, Pilleboue, Coeurjolly, Breeden,
  Singh, Cathelin, De~Goes, Desbrun, and Ostromoukhov}{Wachtel
  et~al\mbox{.}}{2014}]%
        {wachtel2014fast}
\bibfield{author}{\bibinfo{person}{Florent Wachtel}, \bibinfo{person}{Adrien
  Pilleboue}, \bibinfo{person}{David Coeurjolly}, \bibinfo{person}{Katherine
  Breeden}, \bibinfo{person}{Gurprit Singh}, \bibinfo{person}{Ga{\"e}l
  Cathelin}, \bibinfo{person}{Fernando De~Goes}, \bibinfo{person}{Mathieu
  Desbrun}, {and} \bibinfo{person}{Victor Ostromoukhov}.}
  \bibinfo{year}{2014}\natexlab{}.
\newblock \showarticletitle{Fast tile-based adaptive sampling with
  user-specified Fourier spectra}.
\newblock \bibinfo{journal}{\emph{ACM Trans. Graph.}} \bibinfo{volume}{33},
  \bibinfo{number}{4} (\bibinfo{year}{2014}).
\newblock


\bibitem[\protect\citeauthoryear{Wei}{Wei}{2010}]%
        {wei2010multi}
\bibfield{author}{\bibinfo{person}{Li-Yi Wei}.}
  \bibinfo{year}{2010}\natexlab{}.
\newblock \showarticletitle{Multi-class blue noise sampling}.
\newblock \bibinfo{journal}{\emph{ACM Trans. Graph.}} \bibinfo{volume}{29},
  \bibinfo{number}{4} (\bibinfo{year}{2010}).
\newblock


\bibitem[\protect\citeauthoryear{Wei and Wang}{Wei and Wang}{2011}]%
        {wei2011differential}
\bibfield{author}{\bibinfo{person}{Li-Yi Wei} {and} \bibinfo{person}{Rui
  Wang}.} \bibinfo{year}{2011}\natexlab{}.
\newblock \showarticletitle{Differential domain analysis for non-uniform
  sampling}.
\newblock \bibinfo{journal}{\emph{ACM Trans. Graph.}} \bibinfo{volume}{30},
  \bibinfo{number}{4} (\bibinfo{year}{2011}).
\newblock


\bibitem[\protect\citeauthoryear{Wills, Agarwal, Kriegman, and Belongie}{Wills
  et~al\mbox{.}}{2009}]%
        {wills2009toward}
\bibfield{author}{\bibinfo{person}{Josh Wills}, \bibinfo{person}{Sameer
  Agarwal}, \bibinfo{person}{David Kriegman}, {and} \bibinfo{person}{Serge
  Belongie}.} \bibinfo{year}{2009}\natexlab{}.
\newblock \showarticletitle{Toward a perceptual space for gloss}.
\newblock \bibinfo{journal}{\emph{ACM Trans. Graph.}} \bibinfo{volume}{28},
  \bibinfo{number}{4} (\bibinfo{year}{2009}), \bibinfo{pages}{1--15}.
\newblock


\bibitem[\protect\citeauthoryear{Xu, Fan, Yuan, and Singh}{Xu
  et~al\mbox{.}}{2020}]%
        {xu2020ladybird}
\bibfield{author}{\bibinfo{person}{Yifan Xu}, \bibinfo{person}{Tianqi Fan},
  \bibinfo{person}{Yi Yuan}, {and} \bibinfo{person}{Gurprit Singh}.}
  \bibinfo{year}{2020}\natexlab{}.
\newblock \showarticletitle{Ladybird: Quasi-Monte Carlo Sampling for Deep
  Implicit Field Based 3D Reconstruction with Symmetry}.
\newblock \bibinfo{journal}{\emph{ECCV}} (\bibinfo{year}{2020}),
  \bibinfo{pages}{248--263}.
\newblock


\bibitem[\protect\citeauthoryear{Yan, Guo, Wang, Zhang, and Wonka}{Yan
  et~al\mbox{.}}{2015}]%
        {yan2015survey}
\bibfield{author}{\bibinfo{person}{Dong-Ming Yan}, \bibinfo{person}{Jian-Wei
  Guo}, \bibinfo{person}{Bin Wang}, \bibinfo{person}{Xiao-Peng Zhang}, {and}
  \bibinfo{person}{Peter Wonka}.} \bibinfo{year}{2015}\natexlab{}.
\newblock \showarticletitle{A survey of blue-noise sampling and its
  applications}.
\newblock \bibinfo{journal}{\emph{Journal of Comp. Sci. and Tech.}}
  \bibinfo{volume}{30}, \bibinfo{number}{3} (\bibinfo{year}{2015}).
\newblock


\bibitem[\protect\citeauthoryear{Yellott}{Yellott}{1983}]%
        {yellott1983spectral}
\bibfield{author}{\bibinfo{person}{John~I Yellott}.}
  \bibinfo{year}{1983}\natexlab{}.
\newblock \showarticletitle{Spectral consequences of photoreceptor sampling in
  the rhesus retina}.
\newblock \bibinfo{journal}{\emph{Science}} \bibinfo{volume}{221},
  \bibinfo{number}{4608} (\bibinfo{year}{1983}).
\newblock


\bibitem[\protect\citeauthoryear{Yu, Seff, Zhang, Song, Funkhouser, and
  Xiao}{Yu et~al\mbox{.}}{2015}]%
        {yu2015lsun}
\bibfield{author}{\bibinfo{person}{Fisher Yu}, \bibinfo{person}{Ari Seff},
  \bibinfo{person}{Yinda Zhang}, \bibinfo{person}{Shuran Song},
  \bibinfo{person}{Thomas Funkhouser}, {and} \bibinfo{person}{Jianxiong Xiao}.}
  \bibinfo{year}{2015}\natexlab{}.
\newblock \showarticletitle{Lsun: Construction of a large-scale image dataset
  using deep learning with humans in the loop}.
\newblock \bibinfo{journal}{\emph{arXiv preprint arXiv:1506.03365}}
  (\bibinfo{year}{2015}).
\newblock


\bibitem[\protect\citeauthoryear{Zhang, \"{O}ztireli, Mandt, and Salvi}{Zhang
  et~al\mbox{.}}{2019}]%
        {zhang2019active}
\bibfield{author}{\bibinfo{person}{Cheng Zhang}, \bibinfo{person}{Cengiz
  \"{O}ztireli}, \bibinfo{person}{Stephan Mandt}, {and}
  \bibinfo{person}{Giampiero Salvi}.} \bibinfo{year}{2019}\natexlab{}.
\newblock \showarticletitle{Active Mini-Batch Sampling Using Repulsive Point
  Processes}.
\newblock \bibinfo{journal}{\emph{AAAI}}.
\newblock


\bibitem[\protect\citeauthoryear{Zhou, Huang, Wei, and Wang}{Zhou
  et~al\mbox{.}}{2012}]%
        {zhou2012point}
\bibfield{author}{\bibinfo{person}{Yahan Zhou}, \bibinfo{person}{Haibin Huang},
  \bibinfo{person}{Li-Yi Wei}, {and} \bibinfo{person}{Rui Wang}.}
  \bibinfo{year}{2012}\natexlab{}.
\newblock \showarticletitle{Point sampling with general noise spectrum}.
\newblock \bibinfo{journal}{\emph{ACM Trans. Graph.}} \bibinfo{volume}{31},
  \bibinfo{number}{4} (\bibinfo{year}{2012}).
\newblock


\bibitem[\protect\citeauthoryear{Zhou, Zhu, Bai, Lischinski, Cohen-Or, and
  Huang}{Zhou et~al\mbox{.}}{2018}]%
        {zhou2018nonstationary}
\bibfield{author}{\bibinfo{person}{Yang Zhou}, \bibinfo{person}{Zhen Zhu},
  \bibinfo{person}{Xiang Bai}, \bibinfo{person}{Dani Lischinski},
  \bibinfo{person}{Daniel Cohen-Or}, {and} \bibinfo{person}{Hui Huang}.}
  \bibinfo{year}{2018}\natexlab{}.
\newblock \showarticletitle{Non-Stationary Texture Synthesis by Adversarial
  Expansion}.
\newblock \bibinfo{journal}{\emph{ACM Trans. Graph.}} \bibinfo{volume}{37},
  \bibinfo{number}{4} (\bibinfo{year}{2018}).
\newblock
\showISSN{0730-0301}


\end{thebibliography}



\begin{thebibliography}{11}


\ifx \showCODEN    \undefined \def \showCODEN     #1{\unskip}     \fi
\ifx \showDOI      \undefined \def \showDOI       #1{#1}\fi
\ifx \showISBNx    \undefined \def \showISBNx     #1{\unskip}     \fi
\ifx \showISBNxiii \undefined \def \showISBNxiii  #1{\unskip}     \fi
\ifx \showISSN     \undefined \def \showISSN      #1{\unskip}     \fi
\ifx \showLCCN     \undefined \def \showLCCN      #1{\unskip}     \fi
\ifx \shownote     \undefined \def \shownote      #1{#1}          \fi
\ifx \showarticletitle \undefined \def \showarticletitle #1{#1}   \fi
\ifx \showURL      \undefined \def \showURL       {\relax}        \fi
\providecommand\bibfield[2]{#2}
\providecommand\bibinfo[2]{#2}
\providecommand\natexlab[1]{#1}
\providecommand\showeprint[2][]{arXiv:#2}

\bibitem[\protect\citeauthoryear{Büttner and Kosztra}{Büttner and
  Kosztra}{2017}]%
        {buettner2017european}
\bibfield{author}{\bibinfo{person}{György Büttner} {and}
  \bibinfo{person}{Barbara Kosztra}.} \bibinfo{year}{2017}\natexlab{}.
\newblock \bibinfo{booktitle}{\emph{CLC2018 Technical Guidelines}}.
\newblock \bibinfo{type}{{T}echnical {R}eport}. \bibinfo{institution}{European
  Environment Agency}.
\newblock


\bibitem[\protect\citeauthoryear{Choi, Uh, Yoo, and Ha}{Choi
  et~al\mbox{.}}{2020}]%
        {choi2020stargan}
\bibfield{author}{\bibinfo{person}{Yunjey Choi}, \bibinfo{person}{Youngjung
  Uh}, \bibinfo{person}{Jaejun Yoo}, {and} \bibinfo{person}{Jung-Woo Ha}.}
  \bibinfo{year}{2020}\natexlab{}.
\newblock \showarticletitle{Stargan v2: Diverse image synthesis for multiple
  domains}. In \bibinfo{booktitle}{\emph{Proceedings of the IEEE/CVF conference
  on computer vision and pattern recognition}}. \bibinfo{pages}{8188--8197}.
\newblock


\bibitem[\protect\citeauthoryear{Isola, Zhu, Zhou, and Efros}{Isola
  et~al\mbox{.}}{2017}]%
        {isola2017imagetoimage}
\bibfield{author}{\bibinfo{person}{Phillip Isola}, \bibinfo{person}{Jun-Yan
  Zhu}, \bibinfo{person}{Tinghui Zhou}, {and} \bibinfo{person}{Alexei~A
  Efros}.} \bibinfo{year}{2017}\natexlab{}.
\newblock \showarticletitle{Image-to-Image Translation with Conditional
  Adversarial Networks}.
\newblock \bibinfo{journal}{\emph{CVPR}} (\bibinfo{year}{2017}).
\newblock


\bibitem[\protect\citeauthoryear{Kingma and Ba}{Kingma and Ba}{2014}]%
        {kingma2014adam}
\bibfield{author}{\bibinfo{person}{Diederik~P Kingma} {and}
  \bibinfo{person}{Jimmy Ba}.} \bibinfo{year}{2014}\natexlab{}.
\newblock \showarticletitle{Adam: A method for stochastic optimization}.
\newblock \bibinfo{journal}{\emph{arXiv preprint arXiv:1412.6980}}
  (\bibinfo{year}{2014}).
\newblock


\bibitem[\protect\citeauthoryear{Lee, Liu, Wu, and Luo}{Lee
  et~al\mbox{.}}{2020}]%
        {CelebAMask-HQ}
\bibfield{author}{\bibinfo{person}{Cheng-Han Lee}, \bibinfo{person}{Ziwei Liu},
  \bibinfo{person}{Lingyun Wu}, {and} \bibinfo{person}{Ping Luo}.}
  \bibinfo{year}{2020}\natexlab{}.
\newblock \showarticletitle{MaskGAN: Towards Diverse and Interactive Facial
  Image Manipulation}. In \bibinfo{booktitle}{\emph{CVPR}}.
\newblock


\bibitem[\protect\citeauthoryear{Leimk\"{u}hler, Singh, Myszkowski, Seidel, and
  Ritschel}{Leimk\"{u}hler et~al\mbox{.}}{2019}]%
        {leimkuhler2019deep}
\bibfield{author}{\bibinfo{person}{Thomas Leimk\"{u}hler},
  \bibinfo{person}{Gurprit Singh}, \bibinfo{person}{Karol Myszkowski},
  \bibinfo{person}{Hans-Peter Seidel}, {and} \bibinfo{person}{Tobias
  Ritschel}.} \bibinfo{year}{2019}\natexlab{}.
\newblock \showarticletitle{Deep Point Correlation Design}.
\newblock \bibinfo{journal}{\emph{ACM Trans. Graph.}} \bibinfo{volume}{38},
  \bibinfo{number}{6} (\bibinfo{year}{2019}).
\newblock
\showISSN{0730-0301}


\bibitem[\protect\citeauthoryear{{\"O}ztireli and Gross}{{\"O}ztireli and
  Gross}{2012}]%
        {oztireli2012analysis}
\bibfield{author}{\bibinfo{person}{A~Cengiz {\"O}ztireli} {and}
  \bibinfo{person}{Markus Gross}.} \bibinfo{year}{2012}\natexlab{}.
\newblock \showarticletitle{Analysis and synthesis of point distributions based
  on pair correlation}.
\newblock \bibinfo{journal}{\emph{ACM Trans. Graph.}} \bibinfo{volume}{31},
  \bibinfo{number}{6} (\bibinfo{year}{2012}).
\newblock


\bibitem[\protect\citeauthoryear{Sala\"{u}n, Georgiev, Seidel, and
  Singh}{Sala\"{u}n et~al\mbox{.}}{2022}]%
        {salaun2022scalable}
\bibfield{author}{\bibinfo{person}{Corentin Sala\"{u}n},
  \bibinfo{person}{Iliyan Georgiev}, \bibinfo{person}{Hans-Peter Seidel}, {and}
  \bibinfo{person}{Gurprit Singh}.} \bibinfo{year}{2022}\natexlab{}.
\newblock \showarticletitle{Scalable Multi-Class Sampling via Filtered Sliced
  Optimal Transport}.
\newblock \bibinfo{journal}{\emph{ACM Trans. Graph. (SIGGRAPH)}}
  \bibinfo{volume}{41}, \bibinfo{number}{6} (\bibinfo{year}{2022}).
\newblock
\showISSN{0730-0301}


\bibitem[\protect\citeauthoryear{Simonyan and Zisserman}{Simonyan and
  Zisserman}{2014}]%
        {simonyan2014very}
\bibfield{author}{\bibinfo{person}{Karen Simonyan} {and}
  \bibinfo{person}{Andrew Zisserman}.} \bibinfo{year}{2014}\natexlab{}.
\newblock \showarticletitle{Very deep convolutional networks for large-scale
  image recognition}.
\newblock \bibinfo{journal}{\emph{arXiv preprint arXiv:1409.1556}}
  (\bibinfo{year}{2014}).
\newblock


\bibitem[\protect\citeauthoryear{Yu, Seff, Zhang, Song, Funkhouser, and
  Xiao}{Yu et~al\mbox{.}}{2015}]%
        {yu2015lsun}
\bibfield{author}{\bibinfo{person}{Fisher Yu}, \bibinfo{person}{Ari Seff},
  \bibinfo{person}{Yinda Zhang}, \bibinfo{person}{Shuran Song},
  \bibinfo{person}{Thomas Funkhouser}, {and} \bibinfo{person}{Jianxiong Xiao}.}
  \bibinfo{year}{2015}\natexlab{}.
\newblock \showarticletitle{Lsun: Construction of a large-scale image dataset
  using deep learning with humans in the loop}.
\newblock \bibinfo{journal}{\emph{arXiv preprint arXiv:1506.03365}}
  (\bibinfo{year}{2015}).
\newblock


\bibitem[\protect\citeauthoryear{Zhou, Huang, Wei, and Wang}{Zhou
  et~al\mbox{.}}{2012}]%
        {zhou2012point}
\bibfield{author}{\bibinfo{person}{Yahan Zhou}, \bibinfo{person}{Haibin Huang},
  \bibinfo{person}{Li-Yi Wei}, {and} \bibinfo{person}{Rui Wang}.}
  \bibinfo{year}{2012}\natexlab{}.
\newblock \showarticletitle{Point sampling with general noise spectrum}.
\newblock \bibinfo{journal}{\emph{ACM Trans. Graph.}} \bibinfo{volume}{31},
  \bibinfo{number}{4} (\bibinfo{year}{2012}).
\newblock


\end{thebibliography}
